\documentclass{jfm}
\usepackage{bm}

\usepackage{amsmath}
\usepackage{amssymb}
\usepackage{amsbsy}
\usepackage{graphicx}
\usepackage{subcaption}
\usepackage{mathtools}
\usepackage[export]{adjustbox}
\usepackage{dashbox}
\usepackage{array}
\usepackage{color}
\usepackage[dvipsnames]{xcolor}
\newcolumntype{L}[1]{>{\raggedright\let\newline\\\arraybackslash\hspace{0pt}}m{#1}}
\newcolumntype{C}[1]{>{\centering\let\newline\\\arraybackslash\hspace{0pt}}m{#1}}
\newcolumntype{R}[1]{>{\raggedleft\let\newline\\\arraybackslash\hspace{0pt}}m{#1}}

\begin{document}

\newtheorem{lemma}{Lemma}
\newtheorem{corollary}{Corollary}

\shorttitle{Recurrent Neural Nets for Stability Predictions} 
\shortauthor{Chizfahm and Jaiman} 

\title{Deep Learning for Stability Analysis of a Freely Vibrating Sphere at Moderate Reynolds Number}

\author
{
	A. Chizfahm\aff{1}
	R. Jaiman\aff{1}
	\corresp{\email{rjaiman@mech.ubc.ca}},
}

\affiliation
{
	\aff{1}
	Department of Mechanical Engineering, The University of British Columbia, Canada}

\maketitle

\begin{abstract}
In this paper, we present a deep learning-based reduced-order model (DL-ROM) for the stability prediction of unsteady 3D fluid-structure interaction systems. The proposed DL-ROM has the format of a nonlinear state-space model and employs a recurrent neural network with long short-term memory (LSTM). We consider a canonical fluid-structure system of an elastically-mounted sphere coupled with the incompressible fluid flow in a state-space format.
We develop a nonlinear data-driven coupling for predicting the unsteady forces and the vortex-induced vibration (VIV) lock-in of the freely vibrating sphere in a transverse direction. We design an input-output relationship as a temporal sequence of force and displacement datasets for a low-dimensional approximation of the fluid-structure system. Based on the prior knowledge of the VIV lock-in process, the input function contains a range of frequencies and amplitudes, which enables an efficient DL-ROM without the need for a massive training dataset for the low-dimensional modeling. Once trained, the network provides a nonlinear mapping of input-output dynamics that can predict the coupled fluid-structure dynamics for a longer horizon via the feedback process. By integrating the LSTM network with the eigensystem realization algorithm (ERA), we construct a data-driven state-space model for the reduced-order stability analysis.
We investigate the underlying mechanism and stability characteristics of VIV of a freely vibrating sphere via an eigenvalue selection process. To understand the frequency lock-in mechanism, we study the eigenvalue trajectories for a range of the reduced oscillation frequencies and the mass ratios. Consistent with the full-order simulations, the frequency lock-in branches are accurately captured by the combined LSTM-ERA procedure. The proposed DL-ROM aligns with the development of physics-based digital twin of engineering systems involving moving boundaries and fluid-structure interactions.
\end{abstract}

\section{Introduction}
Predictions and control of the spatial-temporal dynamics of fluid-structure systems are crucial in various engineering disciplines ranging from marine/offshore, aerospace to biomedical and energy harvesting. The two-way coupling between fluid and structure exhibits rich flow dynamics such as wake-body interaction and vortex-induced vibrations \citep{sarpkaya2004,williamson2004vortex, jaimancomputational}.
While the physics models based on coupled nonlinear partial differential equations are readily available, the analytical solutions of these differential equations are intractable. Numerical simulations are central for modeling such complex fluid-structure interactions. Using powerful numerical algorithms and large-scale computing resources, high-fidelity simulations can provide accurate predictions and a vast amount of physical insight \citep{jaiman2016stable,liu2016interaction,miyanawala2019decomposition}.
However, such simulations are very expensive for extensive parametric analysis and stability predictions for emerging technologies such as digital twins \citep{tuegel2011}.
%
This work is motivated by the need of making coupled physics simulations efficient for the digital twin technology whereby multi-query analysis, design optimization and control can be achieved through the recent advances in nonlinear model reduction and deep learning algorithms. We consider a prototypical fluid-structure interaction problem of an elastically mounted three-dimensional bluff body undergoing vortex-induced vibration and frequency lock-in phenomenon \citep{sareen2018vortex,rajamuni2018transverse,chizfahm2021transverse}.
During the frequency lock-in/synchronization for a certain range of physical parameters, the vibrating body undergoes a nonlinear coupled flow-structural instability with self-sustained oscillations.  Effective real-time control strategies are required to handle these oscillations and the undesired effects such as noise and structural failure.

In this work, we are interested in the development of a data-driven reduced-order model that can learn the dynamical system well enough to efficiently predict the fluid-structure stability using full-order or measurement data. During the lock-in, the vibration response is limited by nonlinearity either from fluid or structure hence a nonlinear reduced-order model is desired.
A wide majority of reduced-order models (ROMs) are projection-based as they provide low-dimensional representations of an underlying high-dimensional system \citep{lall2002,benner2015survey,rowley2017model}. The basic assumption of these models is that a lower-order representation of a higher-order model may exist and can be identified efficiently within reasonable accuracy. Using Galerkin-type projections of the full-order system onto a small subspace spanned by mode vectors, one can construct the mode vectors or the optimal subspace via proper orthogonal decomposition \citep{berkooz1993proper},
balanced truncation \citep{moore1981principal} or dynamic mode decomposition (DMD) \citep{schmid2010dynamic}.
Conventional POD/Galerkin models capture the most energetic modes based on a user-defined energy norm, whereby
low-energy features may be crucial to the dynamics of the underlying problem. While the POD extracts modes from snapshots of the primary system, the balanced truncation method derives the modes by collecting snapshots of both the primary and the adjoint
systems. This feature of the balanced truncation method allows identification of the modes that are dynamically important.
This algorithm is usually referred to as the balanced proper orthogonal decomposition (BPOD) and provides two sets of modes, namely primal
and adjoint modes \citep{rowley2017model}.
While these linear projection-based ROMs enjoy numerous attractive properties to construct an input-output model for control and parametric analysis, these methods are difficult to generalize for high-dimensional nonlinear systems.
Petrov-Galerkin projections or Koopman operators can be incorporated to introduce the nonlinearity \citep{rowley2017model}. Through a nonlinear combination of the POD modes, one can also use the discrete empirical interpolation method (DEIM; \cite{chaturantabut2009discrete}). The DEIM method relies on the additional POD basis to enrich the low-rank approximation of the nonlinear terms.

An alternative approach to projection-based ROM is based on the system identification via input-output dynamical relationship \citep{zadeh1956identification}.
Using input and output data, system identification methods attempt to build mathematical models of dynamical systems
and consider the original system as a black box \citep{ljung1999system}. In a system identification process, one needs to collect the data, identify any model structures,
estimate the parameters of the model structure, and then validate the model. While the nonlinear (NARX) models rely on the static inference function (i.e., the regression vector)
between input and output data, the nonlinear state-space models provide a general nonlinear dynamical system form whereby the
information in the state can sustain longer in the horizon by the feedback process.
One of the popular system identification methods is the eigensystem realization algorithm (ERA) introduced by \cite{juang1985eigensystem}.
The ERA is a non-intrusive linear state-space model and generates a minimal realization that follows the evolution of the system output when it is subjected to an impulse input. In the theoretical study by \cite{ma2011reduced}, the authors demonstrated the equivalence of the ERA-based model reduction with the BPOD method
for which the observability and controllability gramians are the same.
The ERA-based methodology is primarily data-driven and is used to analyze the stability of dynamical systems.
In the context of flow-induced vibration and control problems, \cite{yao2017model}, \cite{yao2017feedback}, \cite{bukka2020stability} and \cite{chizfahm2021transverse} explored
the ERA-based model reduction. A unified description of frequency lock-in for 2D elastically mounted cylinders has been provided by \cite{yao2017model} along with the
generalized stability properties of the fluid-structure system as functions of Reynolds number, mass ration and the geometry of the bluff body.
While \cite{bukka2020stability} studied the passive suppression mechanism, \cite{yao2017feedback} proposed a feedback control via ERA-based ROM.
\cite{chizfahm2021transverse} applied the ERA-based to three-dimensional geometry of the sphere and provided physical insight on the VIV stability properties.
A more general idea of ROM in nonlinear data-driven dynamics can be constructed using ERA, DMD and their variants.
In that regard, Koopman theory has got a lot of attention for which it is mainly used for linear control (\cite{korda2018linear}, \cite{peitz2019koopman}) and modal decomposition (\cite{liu2018decomposition}, \cite{mezic2013analysis}). An interesting connection between DMD and Koopman theory is presented by \cite{korda2018convergence}. \cite{budivsic2012applied} investigated the applications of Koopman theory for nonlinear dynamical systems.
A harmonic balance (HB) technique for the reduced-order computation of vortex-induced vibration is presented by \cite{yao2016harmonic}. Although the HB model is not robust and general in predicting all the complex nonlinear dynamics of VIV, it appears to be quite effective to extract the basic features of wake dynamics and response characteristics in the lock-in range. Another challenge for the HB-based ROM procedure is to handle the inherently chaotic behavior of vortex-induced vibration at a high Reynolds number.

In recent years, deep learning techniques have gained significant attention in the fluid mechanics community \citep{miyanawala2017efficient,wang2017physics,mohan2018deep,wang2018model}.
Deep learning is a sub-field of machine learning that refers to the use of highly multilayered neural networks to analyze a complicated dataset
in order to predict certain characteristics in the dataset \citep{lecun2015deep,goodfellow2016deep}.
Deep neural nets provide parametric nonlinear function approximations that can fit datasets to learn functions from input vectors to output vectors. This process generates a low-dimensional subspace to represent the underlying behavior of the system. Recently, convolutional neural networks (CNNs) were utilized to develop al nonlinear modal decomposition method, which performed superior to the traditional POD \citep{miyanawala2019decomposition,murata2019nonlinear}. A recent study performed by \cite{bukka2021assessment} presents a review of deep learning-based reduced-order models for the prediction of unsteady fluid flow where these hybrid models rely on recurrent neural networks (RNNs) to evolve low-dimensional states of unsteady fluid flow.
ML has been used to learn the hidden governing equations of fluid flow directly from field data in various cases in the literature (\cite{brunton2016discovering}, \cite{rudy2019data}, \cite{raissi2019deep}, \cite{champion2019data}, \cite{choudhury2018developing}, \cite{long2019pde}). These deep neural network types have been recently adopted in fluid mechanics by \cite{bukka2021assessment,maulik2020time}. Several other works that employ deep learning-based models for nonlinear dynamical systems by building on the mathematical framework of Koopman theory, are investigated by \cite{yeo2019deep}, \cite{otto2019linearly} and \cite{lusch2018deep}.

Predicting the evolution of the parameter of interest can be categorized under sequence modeling problems in machine learning. While neural networks need persistence and retention of information dealing with sequence prediction problems, they vary from other canonical learning problems in machine learning \citep{lipton2015critical}.
Traditional neural networks lack a mechanism for information persistence and retention \citep{cho2014learning,sutskever2014sequence}.
Recurrent neural networks (RNN) alleviate information retention during training and inference. RNNs contain recursive hidden states and learn functions from an input sequence to an output sequence. The internal state of the model is preserved and propagated by adding a new dimension in recurrent neural networks. Despite the success of RNN, several stability issues have been observed. The most prevalent is the vanishing gradient as they are unable to learn long-term dependencies in the data. In the present study, long short-term memory networks are employed to address the issue of long-term dependence in the unsteady dynamical data \citep{hochreiter1997long}.
In contrast to feedforward nets such as the NARX model, the LSTM networks retain the information for a relatively long horizon.  During prediction, the LSTM net utilizes the information based on its relevance and context. For example, they can be presented one observation at a time from a sequence and can learn relevant features using previous observations.
Notably, both deep learning and system identification techniques attempt to address the same fundamental problem i.e., the construction of inference models from observable data. There are many connections between the two techniques as illuminated by \cite{ljung2020deep}. In the present work, we will employ the RNN-LSTM methodology for the DL-based model reduction of the nonlinear fluid-structure system.

In this work, we present for the first time a complete data-driven stability analysis via deep learning model for the flow past a freely vibrating sphere at moderate Reynolds numbers. We study the underlying mechanism of transverse flow-induced vibration by exploiting a unified description of frequency lock-in for an elastically mounted sphere.
We propose a methodology that connects the DL-based model reduction with the ERA to capture just enough dynamics to extract the stability properties
of the fluid-structure systems. The resulting reduced-order model is nonlinear and makes use of data from full-order numerical simulations \citep{chizfahm2021transverse,rajamuni2018transverse} to identify the dynamics relevant to the input-output map of the dynamical system.
It is of particular interest to provide a generalized description of these frequency lock-in regimes at both low and high Reynolds numbers where the wake is laminar ($Re=300$) and turbulent ($Re=2\,000$) respectively via our proposed model reduction technique and the eigenvalue selection process. The results from the ROM are compared with the FOM simulations based
on the incompressible Navier-Stokes equations. Using nonlinear force and motion data, we employ the DL-based ROM integrated with ERA to predict the stability through the eigenvalue distribution in the complex plane.
The present study is based on the following questions pertaining to nonlinear stability predictions of the coupled fluid-structure system. (i) Can we characterize the frequency lock-in regimes of a transversely vibrating sphere by utilizing the proposed nonlinear DL-based ROM methodology? (ii) How can we train the network in order to enhance its performance based on the general underlying physics of the problem? (iii) Once the network is trained, can we employ the constructed DL-based ROM to perform parameter space exploration for a range of mass ratios ($m^*$) and reduced velocities ($U^*$)? (iv) Does nonlinear DL-based ROM integrated with ERA allow stability prediction at high Reynolds number ($Re$) where the wake is turbulent?
In this article, we attempt to answer these questions via our DL-based model reduction procedure which can be useful for the development of emerging digital twin technologies requiring real-time control and structural health monitoring.

This article is organized as follows.
Section \ref{NM} deals with the methodology to construct the DL-based ROM via RNN-LSTM as a system identification technique, the state-space formulation for the model reduction, and the idea of stability predictions for the wake flow and VIV. Section \ref{VandV} describes the VIV problem set-up and presents the numerical verification of the DL-based ROM model.
A systematic analysis of the frequency lock-in mechanism and the effects of mass ratio ($m^*$) and Reynolds number ($Re$) are provided in Section \ref{RandD}.
Finally, concluding remarks are provided in Section \ref{Conclu}.

\section{Numerical Methodology}
\label{NM}
In this section, we first provide an overview of forward and inverse representations of the nonlinear dynamical system. For the sake of completeness, we briefly summarize our high-dimensional FOM to simulate the fluid-structure interaction using the incompressible Navier-Stokes equations and the rigid body dynamics. We next introduce our DL-based model reduction for the nonlinear fluid-structure system of an elastically-mounted bluff body coupled with the incompressible Navier-Stokes equations. We finally provide our methodology to integrate ERA with the constructed DL-based ROM for the stability prediction.

\subsection{Full-Order vs. Reduced-Order State-Space Model}
\label{FwdvsInv}
This section starts by describing the full-order equations of a coupled fluid-structure interaction, followed by a brief description of reduced-order modeling.

\subsubsection{Full-order Fluid-structure Model}
\label{FOM}
In this section, we present a brief description of the coupled fluid-structure solver.
The coupled equations comprise the initial-boundary value problems of the fluid and the structure, which are complemented by 
the traction and velocity continuity conditions at the fluid-structure interface. The deformation of the structure is driven by the traction force exerted by 
the fluid at the fluid-structure interface.
The structural equations are conventionally formulated in Lagrangian coordinates on a mesh that moves along with the material, while
the fluid equations are formulated in Eulerian coordinates, where the mesh serves as a fixed reference for the fluid motion.
The spatially filtered Navier-Stokes equations in an  arbitrary Lagrangian-Eulerian (ALE) framework for an incompressible isothermal flow are given as
\begin{align} \label{NS}
\rho^\mathrm{f}\frac{\partial \bar{\boldsymbol{u}}^\mathrm{f}}{\partial t}\bigg|_{\hat{{x}}^\mathrm{f}} + \rho^\mathrm{f}(\bar{\boldsymbol{u}}^\mathrm{f} - {\boldsymbol{u}^\mathrm{m}})\cdot\nabla\bar{\boldsymbol{u}}^\mathrm{f} &= \nabla\cdot \bar{\boldsymbol{\sigma}}^\mathrm{f} + \boldsymbol{b}^\mathrm{f} \ \mathrm{on\ \Omega^\mathrm{f}(t)}, \\
\frac{\partial \rho^\mathrm{f} }{\partial t} + \nabla\cdot\bar{\boldsymbol{u}}^\mathrm{f} &= 0\ \mathrm{on\ \Omega^\mathrm{f}(t)}, 
\end{align}
where $\bar{\boldsymbol{u}}^\mathrm{f} = \bar{\boldsymbol{u}}^\mathrm{f}(\boldsymbol{x}^\mathrm{f},t)$ and $\boldsymbol{u}^\mathrm{m}=\boldsymbol{u}^\mathrm{m}(\boldsymbol{x}^\mathrm{f},t)$ represent the fluid and mesh velocities defined for each spatial point $\boldsymbol{x}^\mathrm{f} \in \Omega^\mathrm{f}(t)$ respectively. The fluid density is denoted by $\rho^\mathrm{f}$ and $\boldsymbol{b}^\mathrm{f}$ represents the body force acting on the fluid and $\bar{\boldsymbol{\sigma}}^\mathrm{f}$ is the Cauchy stress tensor for a Newtonian fluid which is given by
$\bar{\boldsymbol{\sigma}}^\mathrm{f} = -\bar{p}\boldsymbol{I} + \mu^\mathrm{f}( \nabla\bar{\boldsymbol{u}}^\mathrm{f} + (\nabla\bar{\boldsymbol{u}}^\mathrm{f})^T)$,
where $\bar{p}$ is the filtered fluid pressure, $\boldsymbol{I}$ denotes the second-order identity tensor, $\mu^\mathrm{f}$ represents the dynamic viscosity of the fluid. The partial derivative with respect to the ALE referential coordinate $\hat{x}^\mathrm{f}$ is kept fixed in Eq. (\ref{NS}). 

A rigid-body structure submerged in the fluid may undergo flow-induced vibrations due to unsteady fluid forces if the body is mounted elastically. 
To simulate the translational motion of a three-dimensional rigid body about its centre of mass, the Lagrangian motion along the Cartesian axes is given by 
\begin{align}  
\boldsymbol{m}\frac{d \boldsymbol{u}^\mathrm{s}}{d t} + \boldsymbol{c}\cdot\boldsymbol{u}^\mathrm{s} + \boldsymbol{k}(\boldsymbol{\varphi}^\mathrm{s}(t)-\boldsymbol{\varphi}^\mathrm{s}(0)) &= \boldsymbol{f}^\mathrm{s},\ \mathrm{on}\ \Omega^\mathrm{s}, \\
\boldsymbol{u}^\mathrm{s}=\frac{\partial \boldsymbol{\varphi}^\mathrm{s}}{\partial t},
\label{rigid}
\end{align} 
where $\boldsymbol{u}^\mathrm{s}$ denotes the velocity of the immersed rigid body in the fluid domain and $\boldsymbol{\varphi}^\mathrm{s}$ represents the position of the centre of the rigid body at time $t$. $\boldsymbol{m}$, $\boldsymbol{c}$ and $\boldsymbol{k}$ are mass, damping and stiffness coefficient matrices respectively for the translational motions, and $\boldsymbol{f}^\mathrm{s}$ is the fluid force applied on the rigid body. 
The kinematic and dynamic equilibrium for the coupled fluid-structure interaction problem are satisfied at the fluid-structure interface $\Gamma^\mathrm{fs}$. Mathematically, these relations can be written as 
\begin{align}
{\bar{\boldsymbol{u}}}^\mathrm{f}(t) &= \boldsymbol{u}^\mathrm{s}(t),\\
\int_{\Gamma^\mathrm{fs}} \bar{\boldsymbol{\sigma}}^\mathrm{f}&\cdot \boldsymbol{n} \mathrm{d\Gamma} +  \boldsymbol{f}^\mathrm{s}  = 0
\end{align}
Here, $\boldsymbol{n}$ is the outer normal to the fluid-body interface $\Gamma^\mathrm{fs}$ in the reference configuration. 
The body-fitted ALE formulation restricts the fluid velocity to be exactly equal to the velocity of the body along the interface. 
The immersed body motion is governed by the fluid forces which include the integration of pressure and shear stress effects on the body surface.
The coupling algorithm between the fluid and the rigid-body structural equations is based on the nonlinear iterative force correction (NIFC) scheme \citep{jaiman2016stable}.
The coupled partial differential equations for the full order model are discretized via the stabilized finite element procedure \citep{jaimancomputational}. 
The FOM approach can also be termed a forward problem for the solution of a physical system. Given the model differential equations with appropriate initial/boundary conditions, 
full-order (high-dimensional) solutions can be generated by solving the forward problem.
 
From a state space dynamical system perspective, the forward problem for a general nonlinear dynamics of a system can be written in a discrete form as 
\begin{align}
\begin{split}
\mathbf{x}(t+1) &= \mathbf{F}(\mathbf{x}(t),\mathbf{u}(t)) ,\\
\mathbf{y}(t) &= \mathbf{H}(\mathbf{x}(t)),
\label{nonlinear_dyn_d}
\end{split}
\end{align}
where $\mathbf{x} \in \mathbb{R}^{M}$ is the state vector for a coupled FSI domain with a total of $M$ variables in the system. 
For our fluid-body system in the current study, the state vector involves the fluid velocity and the pressure as $\mathbf{x} = \{\boldsymbol{u}^\mathrm{f}, p^\mathrm{f} \}$ and the structural velocity includes the three translational degrees-of-freedom.
Note that the pressure $p^{f}$ can be written as some function of density $\rho^{f}$ via the state law of a fluid flow. The right-hand side term $\textbf{F}$ represents a dynamic model and can be associated with a vector-valued differential operator describing Eqs. (\ref{NS})-(\ref{rigid}).
The resultant spatial-temporal dynamics of the fluid-structure interaction are driven by the inputs $\mathbf{u}$ such as the modeling parameters and boundary conditions.
$\mathbf{y}$ represents the observable quantities of interest, and $\mathbf{H}$ is a nonlinear function that maps the state and input of the system to the quantities of interest.
The goal of the forward problem is to compute the functions $\mathbf{F}$ and $\mathbf{H}$ via principles of conservation and numerical discretizations.
Next, we briefly review data-driven reduced-order modeling based on traditional projection and deep learning-based recurrent neural nets.

\subsubsection{Nonlinear DL-based Model Reduction}
From the perspective of a data-driven approach, the idea is to build the functions $\mathbf{F}$ and $\mathbf{H}$  using projection-based model reduction, system identification or machine learning method as the inverse problem.
The inverse problems begin with the available data and aim at estimating the parameters in the model. As a general way, a model (${\mathcal{M}(\theta)}$) can be considered as a predictor of the next output based on prior input-output data. The output data may be written as $\hat{\mathbf{y}}(t|\theta)$, where $\theta$ is a parameter that spans a parameter set according to the model ${\mathcal{M}(\theta)}$. The inference in this inverse problem is purely based on the dynamical data. A general nonlinear state-space (NLSS) model can be written as
\begin{subequations}
	\begin{align}
	\mathbf{x}(t+1) &= \boldsymbol{\mathcal{F}}(\mathbf{x}(t),\mathbf{y}(t),\mathbf{u}(t),\theta),
	\label{NLSS_a}
	\end{align}
	\begin{align}
	\hat{\mathbf{y}}(t|\theta) &= \boldsymbol{\mathcal{H}}(\mathbf{x}(t),\theta),
	\label{NLSS_b}
	\end{align}
	\label{NLSS}
\end{subequations}
where $ \boldsymbol{\mathcal{F}}$ and $\boldsymbol{\mathcal{H}}$ can be parametrized by $\theta$ in many different ways according to the model structure.
The inverse problem aims to construct transfer functions $ \boldsymbol{\mathcal{F}}$ and $\boldsymbol{\mathcal{H}}$ as close approximations to the functions $\mathbf{F}$ and $\mathbf{H}$. 
The estimation of the parameters is essentially done by minimizing the fit between observed outputs $\mathbf{y}(t)$ and predicted model outputs $\hat{\mathbf{y}}(t|\theta)$, which can be expressed as
\begin{align}
\min_{\theta}\sum_{t=1}^{N} \left\lVert \mathbf{y}(t)-\hat{\mathbf{y}}(t|\theta) \right\rVert^2,
\end{align}
where $N$ is the number of data points in the data set. 
In practice, the ground truth data are obtained via forward numerical simulations, experiments or field measurements. The ground truth data are obtained from numerical simulations in the current work. The functions $\boldsymbol{\mathcal{F}}$ and $\boldsymbol{\mathcal{H}}$ represent the approximate (surrogate) reduced models that generate the predicted data. We investigate the application of the inverse problem on the canonical problem of the flow past a sphere. The next section presents the mathematical formulation of nonlinear DL-based ROM.

\subsubsection{Nonlinear System Identification via LSTM}
\label{LSTM}
Here we provide a brief description of the RNN-LSTM which is a well-established architecture in deep learning. 
In particular, we intend to illustrate the connection between DL-based ROM using RNN-LSTM with nonlinear system identification.
The RNN-LSTM based ROM model can be considered an extension of the ERA-based system identification technique for predicting the unsteady forces and the VIV lock-in, as
presented earlier by \cite{yao2017model}, \cite{bukka2020stability} and \cite{chizfahm2021transverse}.  
The goal of system identification is to construct mathematical models of dynamical systems based on pure input and output signals (Eq. \ref{NLSS}). 
Our DL-based model reduction approach aims to combine the features of the RNN-LSTM and eigenvalue realization algorithm, 
thereby developing a new method of model reduction for nonlinear systems.

 To address the vanishing gradient problem, LSTM employs gating mechanisms during the dataflow. A single LSTM layer consists of four gates and two states to boost the recurrent computation. In particular, the $\mathrm{l^{th}}$ layer has input gate $\mathrm{i^{(l)}_t}$, forget gate $\mathrm{f^{(l)}_t}$, cell gate $\mathrm{\tilde{c}^{(l)}_t}$, output gate $\mathrm{o^{(l)}_t}$, cell state $\mathrm{c^{(l)}_t}$, and hidden state $\mathrm{h^{(l)}_t}$. The mapping from the input $\mathrm{\tilde{u}^{(l)}_t}$ to the output $\mathrm{\tilde{y}^{(l)}_t}=\mathrm{h^{(l)}_t}$ of the $\mathrm{l^{th}}$ layer is as follows: 
\begin{align}
\begin{split}
\mathrm{i^{(l)}_t}&=\sigma(\mathrm{{W}^{(l)}_{i}}\cdot[\mathrm{{{h}}_{t-1}^{(l)}},{\mathrm{{\tilde{u}}^{(l)}_{t}}}]+\mathrm{{b}_{i}^{(l)}}),\\
\mathrm{f^{(l)}_t}&=\sigma(\mathrm{{W}^{(l)}_{f}}\cdot[\mathrm{{{h}}_{t-1}^{(l)}},{{\mathrm{\tilde{u}}}^{(l)}_{t}}]+\mathrm{{b}_{f}^{(l)}}),\\
\mathrm{\tilde{c}^{(l)}_t}&=\mathrm{tanh}(\mathrm{{W}^{(l)}_{c}}\cdot[\mathrm{{h}_{t-1}^{(l)}},\mathrm{{\tilde{u}}^{(l)}_{t}}]+\mathrm{{b}_{c}^{(l)}}),\\
\mathrm{o^{(l)}_t}&=\sigma(\mathrm{{W}^{(l)}_{o}}\cdot[\mathrm{{h}_{t-1}^{(l)}},{\mathrm{{\tilde{u}}}^{(l)}_{t}}]+\mathrm{{b}_{o}^{(l)}}),\\
\mathrm{c^{(l)}_t}&=\mathrm{f^{(l)}_t*c^{(l)}_{t-1}}+\mathrm{i^{(l)}_t*\tilde{c}^{(l)}_t},\\
\mathrm{h^{(l)}_t}&=\mathrm{o^{(l)}_t*\mathrm{tanh}(c^{(l)}_t)},
\end{split}
\label{LSTM_Structure}
\end{align}
where $\sigma(.)$ is a sigmoid function operating element-wise and ($*$) is the element-wise product. The weight matrices $\mathbf{W}$ and the bias vectors $\mathbf{b}$ include the parameters of the LSTM layers. The \textit{Number of Units} in an LSTM layer is the dimension of the hidden and the cell states and defines the dimensions of the matrices $\mathbf{W}$ and $\mathbf{b}$. 
The gates layout and the data control are the main reasons for the robustness of LSTMs with regard to the vanishing gradient problem that plagues traditional RNNs. 
As a result, LSTMs can serve as a powerful tool for time-series predictions of fluid-structure interaction.

\begin{figure}
	\centering
	\begin{subfigure}{1\textwidth}
		\centering
		\includegraphics[trim={0cm 16cm 15cm 1cm},clip,scale=0.45]{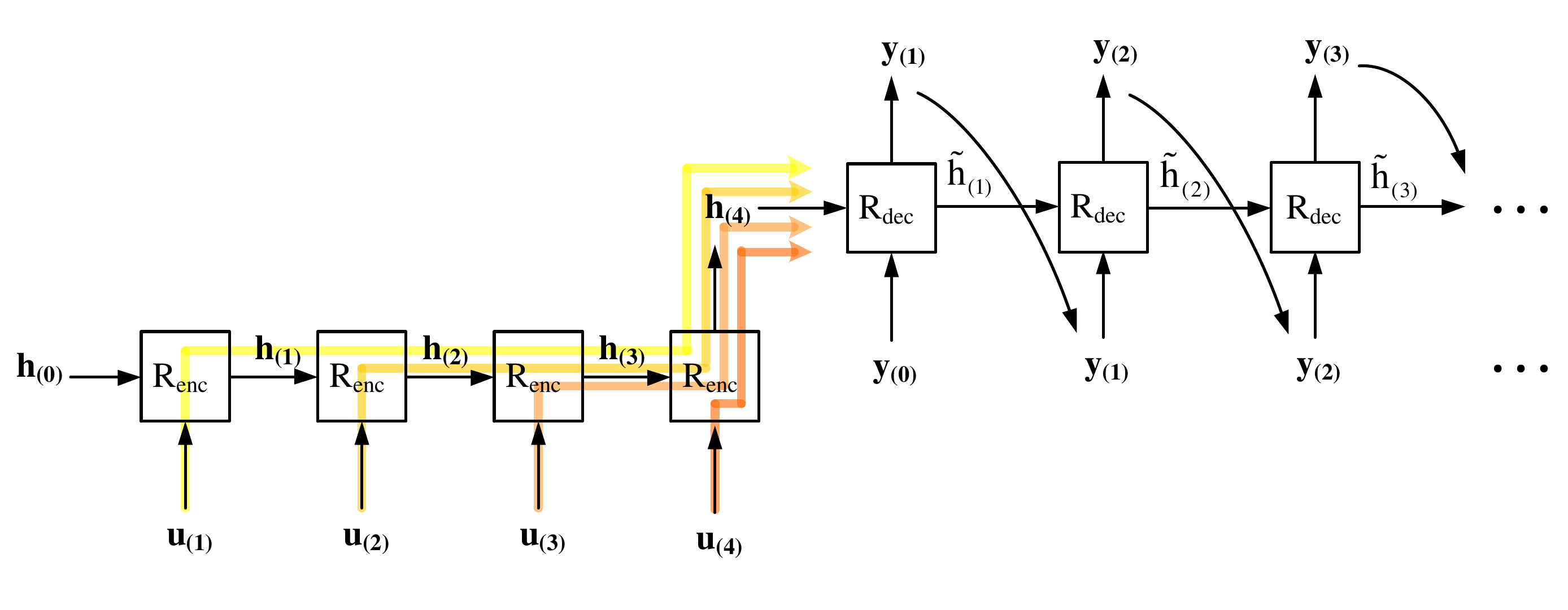}
	\end{subfigure}
	\caption{
		A typical example of a sequence-to-sequence architecture with an RNN encoder $\mathrm{R_{enc}}$ and RNN decoder $\mathrm{R_{dec}}$. }
	\label{LSTM_Cell}
\end{figure}

The LSTM layers are connected by taking the output of a layer as the input to the next layer: $\mathrm{\tilde{u}^{(l)}}=\mathrm{h^{(l-1)}}$ for $\mathrm{l = 2, \dots, L}$. The output of the last LSTM layer is the input to a fully connected linear layer: $\mathrm{y_t}=\mathrm{W^{(l)}_{fc}h^{(L)}_t+b_{fc}}$. It is important to mention that an LSTM network is a NLSS model (Eq. (\ref{NLSS})), since a state vector $\mathrm{x(t)}$ that collects the cell and hidden states in all layers $\mathrm{x^{(l)}_t=[(c^{(l)}_t)^T\ (h^{(l)}_t)^T]^T}$ and $\mathrm{x(t)=[(x^{(1)}_t)^T\ \dots \ (x^{(L)}_t)^T]^T}$, input $\mathrm{u(t)=\tilde{u}^{(l)}}$, output $\mathrm{\hat{y}(t|\theta) = y_t}$, and a parameter vector $\theta$ that collects all the elements in the weight matrices $\mathbf{W}$ and the bias vectors $\mathbf{b}$. 
More explicitly, the output $\mathrm{y_t}$ in Eq. (\ref{NLSS_b}) is a function of $\mathrm{h^{L}_t}$, which is part of the state vector $\mathrm{x(t)}$. Moreover, for the state equation, Eq. (\ref{NLSS_a}), we can observe from Eq.~ (\ref{LSTM_Structure}) that $\mathrm{x^{(l)}_{t+1}}$ depends on the states in the same layer one time step earlier $\mathrm{x^{(l)}_{t}}$ and on the input $\mathrm{\tilde{u}^{(l)}_{t+1}}$ to the layer. For $\mathrm{l = L, L-1, \ \dots \  2}$, this input $\mathrm{\tilde{u}^{(l)}_{t+1}=h^{(l-1)}_{t+1}}$ is part of the state of the previous layer $\mathrm{x^{(l-1)}_{t+1}}$, which in turn depends on the states of that layer one time step earlier $\mathrm{x^{(l-1)}_{t}}$, and on the input $\mathrm{\tilde{u}^{(l-1)}_{t+1}}$. Ultimately, $\mathrm{x^{(l)}_{t+1}}$ depends on the states of all the previous layers one time step earlier $\mathrm{x^{(l-1)}_{t}\ \dots \ x^{(1)}_{t}}$, which is part of $\mathrm{x(t)}$, and on the input $\mathrm{\tilde{u}^{(1)}_{t+1}}$ to the first layer, which is $\mathrm{u(t)}$. A schematic of a closed-loop recurrent neural network is depicted in figure \ref{LSTM_Cell}. The decoder is infused by the final summary vector $\mathrm{h^{(l)}_{T}}$ generated by the encoder. The decoder then proceeds in an autoregressive fashion for the prediction.
 At each step of $\mathrm{R_{dec}}$, the input is generated by the predicted output from the previous step. The so-called summary vector or the bottleneck vector provides an emphasis to 
store all relevant information in the input sequence.  The summary vector $\mathrm{h^{(l)}_{T}}$ is illustrated with the transparent lines in figure \ref{LSTM_Cell}. 
Further details about the dataflow of the LSTM can found in \cite{bronstein2021geometric}.

The work procedure becomes quite similar when the deep learning model structures ($\mathrm{\hat{y}(t|\theta)}$) are integrated with the model objects of system identification. Once the model is estimated, it can be validated and used for simulation and prediction. A benchmark example of a nonlinear dynamical system is provided in the appendix to evaluate the performance of the LSTM network toolbox for system identification.
In the next section, we present the basic state-space formulation and DL-based model reduction for the coupled fluid-structure system. We then provide our methodology for the training process of the RNN-LSTM as a nonlinear system identification method for fluid-structure interaction problem in Section \ref{VandV}.

\subsubsection{Coupled Nonlinear Fluid-Structure Formulation}
\label{Coupled Formulation}
Of particular interest in this work is the rigid-body equation for the freely vibrating sphere in a flowing stream. The sphere is mounted on a spring system in a cross-flow direction, which allows the sphere to vibrate through unsteady lift comprising of the pressure and shear stresses.
We adopt a partitioned coupled formulation for the reduced fluid-structure problem.  The LSTM-based ROM for the fluid sub-system is integrated with the linear structural model in a partitioned manner. The non-dimensional structural equation for a transversely vibrating sphere with 1-DOF can be expressed as
\begin{align}
{\ddot{A}_y}+4\zeta \pi F_s {\dot{A}_y} + (2\pi F_s)^2 {A_y}=\frac{1}{m^*}{C}_y
\end{align}
where $A_y$ is the transverse displacement and $C_y$ is the normalized transverse force acting on the structural body due to fluid traction, defined as $C_y={f}^\mathrm{s}_y/(\frac{1}{2}\rho U^2 S)$. $m^*$ and $\zeta$ are the mass ratio (i.e. ratio of the mass of the structural body to the mass of the displaced fluid) and the damping coefficient respectively. $F_s$ is the reduced natural frequency of the structure, defined as $F_s=f_nD/U=1/U^*$, where $U^*$ is the reduced velocity which is an alternative parameter to characterize the frequency lock-in phenomenon. The definitions of the key parameters can be found in Table \ref{Parameters}. The non-dimensional structural equation can be presented into a state-space formulation as 
\begin{align}
\bold{\dot{x}_s}=\bold{A_s}\bold{x_s}+\bold{B_s}{C}_y
\label{ROM_Structural}
\end{align}
where the state matrices and vectors are
\begin{align}
\bold{A_s}=
\begin{bmatrix}
0 & 1  \\
-(2\pi F_s)^2 & -4\zeta \pi F_s &  \\
\end{bmatrix}
\bold{B_s}= 
\begin{bmatrix}
0  \\
a_s/m^*  \\
\end{bmatrix}
\bold{x_s}= 
\begin{bmatrix}
{A_y}  \\
{\dot{A}_y}  \\
\end{bmatrix}
\end{align}
The characteristic length scale factor $a_s$ is related to the geometry of the body which is $a_s=3/8$ for a spherical body. The above formulation can be transformed into discrete state-space form as follows, 
\begin{align}\label{ROM_Structural_d}
\begin{split}
\bold{x_s}{(t+1)}&=\bold{A_{sd}}\bold{x_s}{(t)}+\bold{B_{sd}}{C}_{y}(t)\\
A_{y}(t)&=\bold{C_{sd}}\bold{x_s}{(t)}
\end{split}
\end{align}
where the state matrices are $\bold{A_{sd}}=e^{\bold{A_s}\Delta t}$, $\bold{B_{sd}}=\bold{A_s}^{-1}(e^{\bold{A_s}\Delta t}-I)\bold{B_s}$ at discrete times $t=k\Delta t;\ k=0, 1, 2 \hdots$ with a constant sampling time $\Delta t$. $I$ is the identity matrix and $\bold{C_{sd}}=[1 \ 0]$. 
The fluid ROM is derived by the RNN-LSTM method as described in Eq. (\ref{NLSS}) and Eq. (\ref{LSTM_Structure}) through the input-output dynamics.
The input for the ROM is the transverse displacement $A_y$, and the output is the normalized transverse force $C_y$. The DL-based ROM with the single input and single output (SISO) can be reformulated as
\begin{align}
\hat{C}_y(t|\theta) &= \boldsymbol{\mathcal{H}}(C_y(t),A_y(t),\theta),
\label{ROM_Fluid}
\end{align}

In this work, we aim to characterize a complex dynamical relation between the transverse amplitude $A_y$ and the transverse force $C_y$. A state-space relationship between the transverse force and the amplitude is constructed directly from the NS equations subject to random prescribed motion. Therefore, now we proceed to the formulation of the coupled fluid-structure system. By substituting Eq. (\ref{ROM_Fluid}) into Eq. (\ref{ROM_Structural}), the resultant ROM can be expressed as 
\begin{align}\label{ROM_FSI}
\begin{split}
\bold{{x}_{fs}}{(t+1)}&=
\begin{bmatrix}
\bold{A_{s}}& \bold{B_{s}}  \\
\boldsymbol{\mathcal{H}_s} & \boldsymbol{\mathcal{H}_f} &  \\
\end{bmatrix}
\bold{x_{fs}}{(t)}=\bold{\boldsymbol{\mathcal{A}}_{fs}}\bold{x_{fs}}{(t)}
\end{split}
\end{align}
where $\boldsymbol{\mathcal{H}_s}$ and $\boldsymbol{\mathcal{H}_f}$ are nonlinear functions that can be parametrized according to the black-box DL-based model structure, $\bold{\boldsymbol{\mathcal{A}}_{fs}}$ denotes the coupled nonlinear fluid-structure matrix in the discrete state-space form, and $\bold{x_{fs}}=[\bold{x_{s}}\ C_y]^T$. The input-output dynamics of the full-order system is emulated through the DL-based ROM. 

\subsection{Stability Analysis via DL-based ROM Integrated with ERA}
\label{DL_ERA}

\begin{figure}
	\centering
	
	\begin{subfigure}{1\textwidth}
		\centering
		\includegraphics[trim={18cm 8cm 8cm 6cm},clip,scale=0.38]{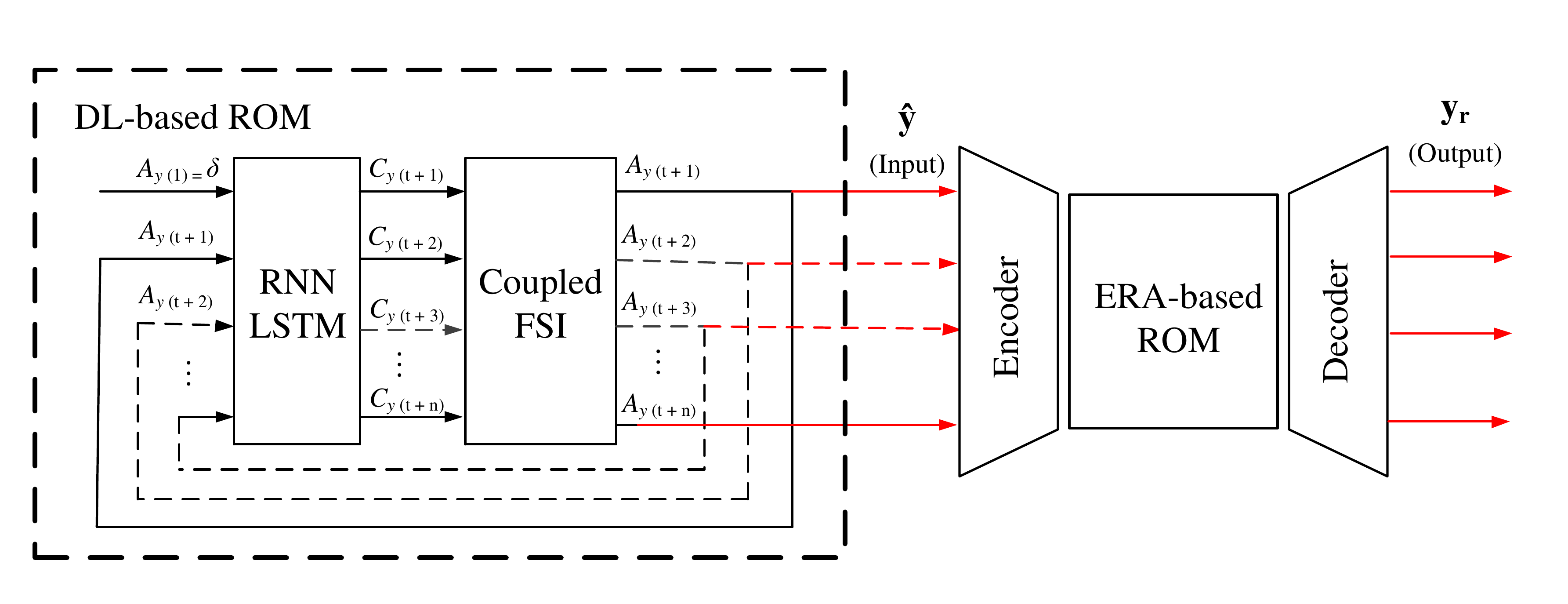}
	\end{subfigure}
	\caption{Schematic of proposed hybrid RNN-LSTM framework with ERA for stability prediction of coupled fluid-structure system.}
	\label{Schematic_Process}
\end{figure}

Here, we present the methodology to integrate DL-based ROM with ERA system identification for stability analysis. Figure \ref{Schematic_Process} shows the schematic of the overall process, where the predictive RNN-LSTM as a nonlinear DL-based ROM is integrated with the ERA-based ROM to provide a linear approximation of the nonlinear model for the stability prediction. Later in this section, we present the eigenvalue selection process as a requirement to interpret the results for the stability prediction.

\subsubsection{Linear Approximation via ERA}
\label{ERA}
We assume that the dynamic system of interest can be modeled as
\begin{align}\label{ROM_FSI_NLin}
\bold{x_{fs}}{(t+1)}=\bold{\boldsymbol{\mathcal{A}}_{fs}}\bold{x_{fs}}{(t)}+\eta_t,\ \ \ t = 0, 1, 2, \dots. 
\end{align}
where $\bold{x_{fs}}{(t)}$ presents the state of the system at time $t$. The function $\bold{\boldsymbol{\mathcal{A}}_{fs}}: \mathbb{R}^n \rightarrow \mathbb{R}^n$ maps the state at time $t$ to the state at time $t + 1$, and $\eta_t$ is a small perturbation. The perturbation $\eta_t$ might include modeling errors, such as slowly changing operating conditions (unmodeled dynamics) or discretization errors as investigated by \cite{erichson2019physics}. If $\eta_t$ is small, the dynamics simply specify that the state $\bold{x_{fs}}{(t+1)}$ depends only on the value of the previous state $\bold{x_{fs}}{(t)}$. Hence, through the transformation $\bold{\boldsymbol{\mathcal{A}}_{fs}}$, the state $\bold{x_{fs}}{(t)}$ provides all information needed for predicting the future state at $\bold{x_{fs}}{(t+1)}$.

To perform the stability analysis, we use a linear time-invariant approximation of the nonlinear system as of the following form
\begin{align}\label{ROM_FSI_Lin}
\bold{x_{fs}}{(t+1)}=\bold{{A}_{fs}}\bold{x_{fs}}{(t)}+\eta_t,\ \ \ t = 0, 1, 2, \dots. 
\end{align}
where $\bold{{A}_{fs}}: \mathbb{R}^n \rightarrow \mathbb{R}^n$ presents a linear map. Linear models provide a reasonably good approximation for the underlying mechanism in many applications. Using the linear dynamics, the stability of the origin can be checked by utilizing an eigenvalue analysis. 

For this purpose, we use the ERA method as a system identification technique that makes a reduced model through a linear projection of the original system on the most observable and controllable subspaces. The sample data from the DL-based ROM is passed through ERA to generate a low dimensional linear space. The generated latent space from ERA is designed to have a lower dimension than the original input and output space, thus, achieving compression of the data as an autoencoder-decoder-like framework. The ERA-based ROM is then used for the stability prediction of the coupled FSI system using eigenvalue analysis.
However, despite the simplicity of this procedure, it often turns out to be a challenge to find a reliable estimate for the coupled system $\bold{{A}_{fs}}$.
Therefore, we aim to formulate the basic state-space formulation to estimate the desired output of the coupled nonlinear system using the eigensystem realization algorithm.
The state-space form at discrete times $t=k\Delta t;\ k=0, 1, 2, ...,$ with a constant sampling time $\Delta t$ is represented as follows:

\begin{align}\label{ERA_ROM}
\begin{split}
\bold{x_{r}}{(t+1)}=\bold{A}_r\bold{x_{r}}{(t)}+\bold{B}_r\bold{u_{r}}{(t)},\\
\bold{{y}_r}=\bold{C}_r\bold{x_{r}}{(t)}+\bold{D}_r\bold{u_{r}}{(t)}&
\end{split}
\end{align}
where $\bold{x}_r$ is an $n_r$-dimensional state vector, $\bold{u}$ is a $q$-dimensional input vector and $\bold{{y}_r}$ is a $p$-dimensional output vector. The system matrices are ($\bold{A}_r,\ \bold{B}_r,\ \bold{C}_r,\ \bold{D}_r$), where the matrix $\bold{A}_r$ characterizes the dynamics of the system and $\bold{B}_r,\ \bold{C}_r$ and $\bold{D}_r$ are the input, output and feed-through matrices. The statement of system realization is to construct the system matrices ($\bold{A}_r,\ \bold{B}_r,\ \bold{C}_r,\ \bold{D}_r$) such that the vector $\bold{{y}_r}$ is reproduced by the state-space model. In a discrete-time setting, the state-space realization matrices ($\bold{A}_r,\ \bold{B}_r,\ \bold{C}_r,\ \bold{D}_r$) of the dynamical system are constructed by the eigensystem realization algorithm (ERA), in which only the impulse response function (IRF) of the DL-based ROM is required for the system realization. 

Based on the impulse response, the block Hankel matrix can be constructed as
\begin{align}\label{Hankel}
\bold{H}(t-1)
=
\begin{bmatrix}
\mathrm{y}_{t+1} & \mathrm{y}_{t+2} & \hdots & \mathrm{y}_{t+l} \\
\mathrm{y}_{t+2} & \mathrm{y}_{t+3} & \hdots & \mathrm{y}_{t+l+1} \\
\vdots & \vdots & \ddots & \vdots \\
\mathrm{y}_{t+r} & \mathrm{y}_{t+r+1} & \hdots & \mathrm{y}_{t+(l+r-1)} \\
\end{bmatrix}
\end{align}
where $\mathrm{y}(t)$ is the desired output from the DL-based ROM, and $t$ denotes the time step. The singular value decomposition of the Hankel matrix at $t=1$ gives, 
\begin{align}
\bold{H}(0)
=
\bold{U}\boldsymbol{\Sigma}\bold{V}^*
=
\begin{bmatrix}
\bold{U}_r & \bold{U}_\mathrm{tr}  \\
\end{bmatrix}
\begin{bmatrix}
\boldsymbol{\Sigma}_r & 0  \\
0 & \boldsymbol{\Sigma}_\mathrm{tr} &  \\
\end{bmatrix}
\begin{bmatrix}
\bold{V}_r^*  \\
\bold{V}_\mathrm{tr}^* \\
\end{bmatrix}
\end{align}
where the diagonal matrices $\boldsymbol{\Sigma}_r$ and $\boldsymbol{\Sigma}_\mathrm{tr}$ are the reduced and truncated Hankel singular values (HSVs) respectively. The dominant eigenvalues are sorting such that $\sigma_1\geq \hdots \geq \sigma_n \geq 0$. By truncating the dynamically less significant states, we estimate $\bold{H}(0)=\bold{U}_r\boldsymbol{\Sigma}_r\bold{V}_r^*$. The reduced system is then defined as
\begin{align}
\begin{split}
\bold{A}_r&=\boldsymbol{\Sigma}_{r}^{-1/2}\bold{U}_{r}^*\bold{H}(1)\bold{V}_{r}\boldsymbol{\Sigma}_{r}^{-1/2}\\
\bold{B}_r&=\boldsymbol{\Sigma}_{r}^{1/2}\bold{V}_{r}^*\bold{E_q}\\
\bold{C}_r&=\bold{E_p^*}\bold{U}_{r}^*\boldsymbol{\Sigma}_{r}^{1/2}\\
\bold{D}_r&=\mathrm{y}_1
\end{split}
\end{align}
where, $\bold{E_q^*}=[\bold{I}_q \ 0]_{q\times N}$ and $\bold{E_p^*}=[\bold{I}_p \ 0]_{p\times M}$ as $N=l\times q$ and $M=r\times p$. $\bold{I}_q$ and $\bold{I}_p$ are the identity matrices. 

The input for the ERA-based ROM is a tiny perturbation of amplitude and the output is the temporal evolution of the transverse displacement $A_y$.
The continuous-time eigenvalues and eigenvectors of $\bold{A_{r}}$, denoted by ($\lambda, \bold{\hat{x}}$), characterize the temporal behaviour by the complex scalar ($\lambda$), and the spatial structure by the complex vector field ($\bold{\hat{x}}$). The stability analysis can be established by tracing the trajectory of the complex eigenvalue in the complex plane, whereby $\bold{\hat{x}}$ provides the spatial global modes of the ERA-based ROM. Through the eigenvalues of the linear ERA-based approximation of the nonlinear model, the growth rate and the frequency of the corresponding global modes are identified by $\mathrm{Re}(\lambda)$ and $\mathrm{Im}(\lambda/2\pi)$,  respectively.

\subsubsection{Eigenvalue Selection Process}
\label{Eigenvalue Selection}
In this section, we present the eigenvalue selection process for stability prediction. We aim to identify the unstable region corresponding to VIV lock-in for the coupled FSI system.
In section \ref{ERA}, the ERA methodology is applied to construct a linear approximation of the nonlinear DL-based ROM.
Consistent with the previous literature by \cite{yao2017model}, we utilize the methodology of variation of unstable modes to classify the distinct eigenvalue trajectories of the fluid-structure system. In their work, the fluid-structure mode is considered as a structural mode (SM) if the eigenvalue of the coupled linearized system comes close to the natural frequency of the structural body in a vacuum. The ERA method is applied to the desired output from the coupled nonlinear DL-based ROM, which is the transverse displacement ($A_y$). We can plot the eigenvalue distribution corresponding to the transverse displacement ($A_y$), which is equivalent to structural mode, by selecting the most unstable mode.
The stability prediction can be performed based on the growth rate and the frequency of the most unstable mode.
Based on the methodology described in the previous sections, the process of constructing the ROM for the coupled FSI system contains the following vital components:

\begin{itemize}
	\item The first step involves the extraction of the amplitude response ($A_y$) from the nonlinear DL-based ROM. By imposing a tiny amplitude perturbation ($\delta =A_y/D= 10^{-3}$), the equilibrium position is perturbed and we can assess the growth of the disturbance and the stability of the system. The sensitivity of the unstable system to the initial condition is confirmed by comparing the response subject to two small perturbations with $\delta = 10^{-4}$ and $\delta = 10^{-3}$. Small amplitude perturbation is applied such that the transverse force ($C_y$) evolves for a relatively long time. An adequate number of cycles is required to capture the dynamics of the system to ensure that the unstable modes start to dominate the essential dynamics of the input-output relationship. However, an excessively long simulation time should be avoided that leads to an increase in the error norm associated with the RNN-LSTM.
	\item The second step is to construct the linear approximation through the ERA methodology. This procedure consists of the formation of the Hankel matrix by stacking the transverse displacement ($A_y$) at each timestamp, applying the SVD and evaluating the reduced system state matrices. The dimensions of the Hankel matrix can be determined by examining the convergence of the unstable eigenvalues. The linear ROM is calculated through Eq.~ (\ref{ERA_ROM}), where $\bold{A}_r$ characterizes the dynamics of the system.
	\item In the final step, the eigenvalues of the matrix $\bold{A}_r$ are calculated and the most unstable mode is selected. In this study the continuous eigenvalues, $\lambda = \mathrm{log}(\mathrm{eig}( \bold{A_r} ))/\Delta t$ are considered for the stability analysis. The eigenspectrum can be constructed by plotting the root loci of selected modes as a function of the reduced natural frequency ($F_s$). The position of the eigenvalues on this spectrum provides information about the stability of the coupled system.
\end{itemize}

The above process with the three main components forms a general procedure to be followed while constructing the nonlinear DL-based ROM integrated with ERA for the stability prediction of the VIV lock-in. We aim to identify the lock-in regions and to predict the unseen dynamics by utilizing the eigenvalue selection process on an amplitude response of the sphere as the desired output.

\section{Problem Setup and Hyperparameter Analysis}
\label{VandV}
This section first deals with the problem setup of the FOM for flow past a sphere at a uniform flow field. The numerical verification of the DL-based ROM integrated with ERA is presented. We assess the generality of the DL-based ROM by evaluating the error associated with the trained model and by testing the model on different test datasets. Later, we perform hyper-parameter sensitivity analysis for the training process to assess the robustness of the model. Finally, we check the reliability of the ERA for the stability prediction.

\subsection{Full-order Problem Setup}
While the NLSS (Eq. \ref{NLSS}) formulation is general for a fluid-structure system, as a prototypical problem, we consider the transverse motion of a fully submerged three-dimensional canonical geometry of a sphere exposed to a uniform flow field. The sphere is mounted on a spring-damper system in the cross-flow direction and the unsteady fluid force makes the sphere vibrate in the transverse direction. 

\begin{figure}
	\centering
	
	\begin{subfigure}{1\textwidth}
		\centering
		\includegraphics[trim={0 0.5cm 0.5cm 0},clip,scale=0.25]{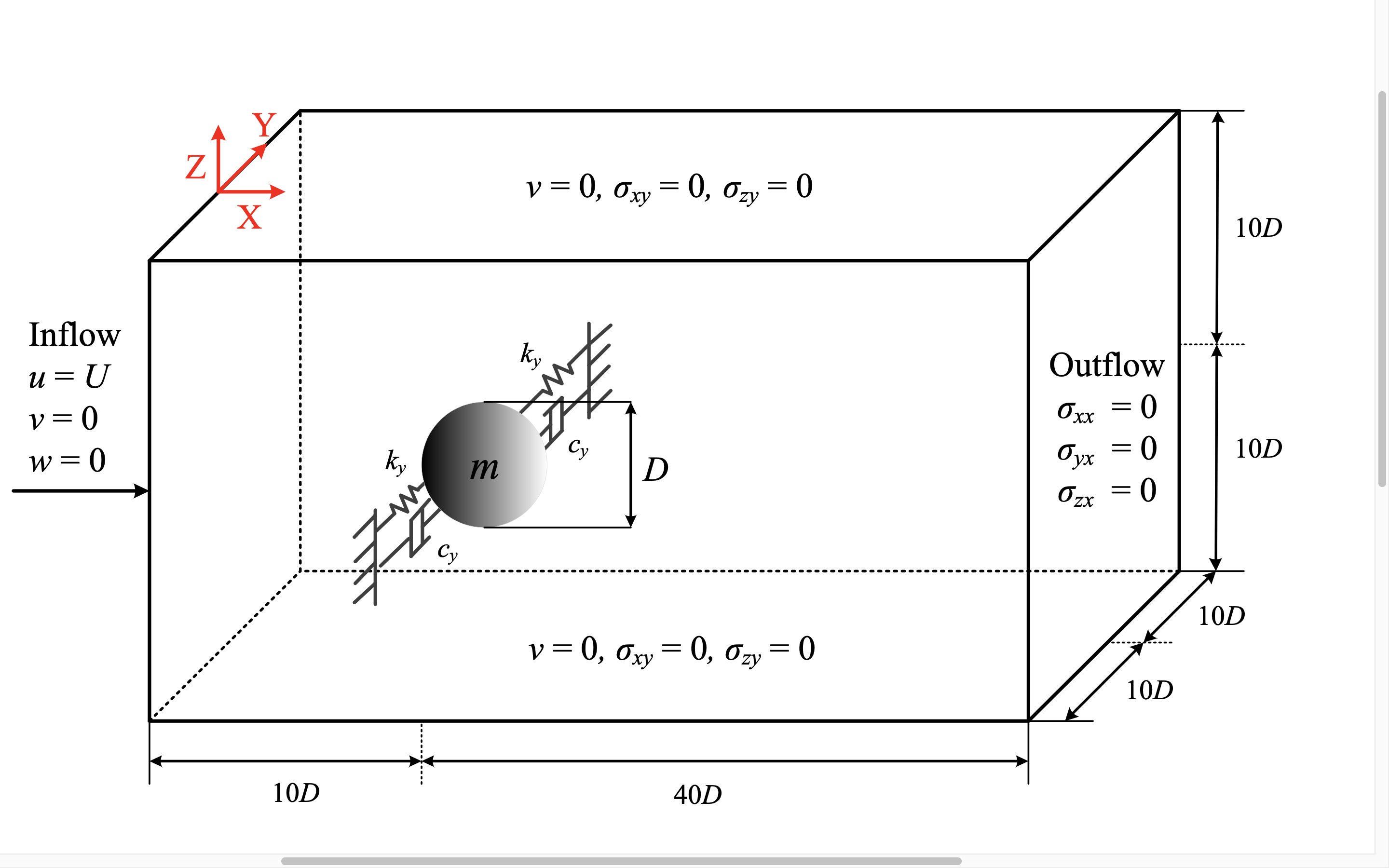}
	\end{subfigure}
	\caption{Schematic and associated boundary conditions of the fluid flow past a 1-DOF fully submerged elastically mounted sphere. }
	\label{domain}
\end{figure}

Figure \ref{domain} shows a schematic of the problem setup used in our simulation study for a sphere, both stationary and elastically mounted cases. A three-dimensional computational domain of the size ($50 \times 20 \times 20$)$D$ with a sphere of diameter $D$ placed at an offset of $10D$ from the inflow surface is considered, which is sufficient enough to reduce the effects of artificial boundary conditions around the fluid domain. The origin of the coordinate system is fixed at the center of the sphere. We consider the $x$-axis as the streamwise flow direction, the $y$-axis in the transverse direction, and the $z$-axis represents the vertical direction. While the streamwise motion corresponds to the freestream ($x$-direction), the transverse motion is parallel to the $y$-direction. A uniform freestream flow with velocity $U$ is along the $x$-axis. At the inlet boundary, a stream of water enters into the domain with velocity $(u, v, w)=(U, 0, 0)$ where $u$, $v$ and $w$ denote the streamwise, transverse and vertical velocities in $x, y$ and $z$ directions, respectively. The sphere is elastically mounted on springs with a stiffness value of $k$ and linear dampers with a damping value of $c$ in the transverse direction. The damping coefficient $\zeta$ is set to zero in the present work. We have considered the slip-wall boundary condition along the top, bottom and side surfaces, in addition to the Dirichlet and traction-free Neumann boundary conditions along the inflow and outflow boundaries, respectively.

\begin{table}
	\begin{center}
		\begin{tabular}{L{6cm}L{6cm}}
			Parameter &  Definition  \\
			\hline
			\raggedright Reynolds number  &  $Re={\rho ^\mathrm{f} U D}/{\mu ^\mathrm{f}}$    \\[5pt]
			\raggedright Reduced velocity  &  $U^*={U}/{f_n D}$    \\[5pt]
			\raggedright Mass ratio  &  ${m}^*={m}/{m_d}$    \\[5pt]
			\raggedright Damping ratio  &  $\zeta=c/2\sqrt{mk}$    \\[5pt]
			\raggedright Non-dimensional amplitude  &  ${A}^*_\mathrm{rms}=\sqrt{2}{A_\mathrm{rms}/D}$    \\[5pt]
			\raggedright Normalized horizontal force  &  $C_x={f}^{\mathrm{s}}_x/(\frac{1}{2}\rho U^2 S)$    \\[5pt]
			\raggedright Normalized transverse force  &  $C_y={f}^{\mathrm{s}}_y/(\frac{1}{2}\rho U^2 S)$    \\[5pt]
			\raggedright Normalized vertical force  &  $C_z={f}^{\mathrm{s}}_z/(\frac{1}{2}\rho U^2 S)$    \\[5pt]
			\raggedright Normalized frequency  &  $f^*=f/f_n$    \\[5pt]
		\end{tabular}
		\caption{Definition of the non-dimensional parameters and post-processing quantities.}
		\label{Parameters}
	\end{center}
\end{table}

The definitions of some relevant important non-dimensional parameters are summarized in Table \ref{Parameters}. 
The non-dimensional amplitude response $A^*$ is defined as ${A^*=A/D}$ and $f^*$ denotes the normalized frequency and $f_n=\frac{1}{2\pi}\sqrt{\frac{k}{m}}$ is the natural frequency of the spring-mass system in vacuum, where $m$ is the mass of the sphere and $k$ is the spring stiffness.  The mass ratio is given by ${m}^*={m}/{m_d}$, where $m$ is the mass of sphere and $m_d$ is the mass of displaced fluid. The value of $U$ is considered to be $1$ throughout the numerical study. The Reynolds number $Re$ and the reduced velocity $U^*$ are varied by changing the values of dynamic viscosity $\mu ^\mathrm{f}$ and the natural frequency $f_n$. 

The normalized forces are evaluated from the fluid traction, acting on the structural body, where $C_x$ is the normalized drag force, $C_y$ and $C_z$ are the normalized transverse and vertical forces in $y$ and $z$ directions, respectively. 
\begin{align}
C_x=\frac{1}{\frac{1}{2}\rho U^2 S}\int_{\Gamma} (\bar{\boldsymbol{\sigma}}^\mathrm{f}\cdot \boldsymbol{n})\cdot \boldsymbol{n_x} \mathrm{d\Gamma} \\
C_y=\frac{1}{\frac{1}{2}\rho U^2 S}\int_{\Gamma} (\bar{\boldsymbol{\sigma}}^\mathrm{f}\cdot \boldsymbol{n})\cdot \boldsymbol{n_y} \mathrm{d\Gamma} \\
C_z=\frac{1}{\frac{1}{2}\rho U^2 S}\int_{\Gamma} (\bar{\boldsymbol{\sigma}}^\mathrm{f}\cdot \boldsymbol{n})\cdot \boldsymbol{n_z} \mathrm{d\Gamma} 
\end{align}
where $\boldsymbol{n_x}$, $\boldsymbol{n_y}$ and $\boldsymbol{n_z}$ are the Cartesian components of the unit normal $\boldsymbol{n}$ to the sphere surface, and $S$ is the relevant surface area which is defined as $S=\pi D^2/4$. The normalized lift coefficient is obtained as ${C}_{L}=\sqrt{{C_y}^2+{C_z}^2}$. A comprehensive mesh convergence study along with the validation of the solver by comparing with the experimental and available numerical data are provided in \cite{chizfahm2021transverse} and \cite{chizfahm2021data}. In the current study we use the same mesh configuration used in our aforementioned studies. 

\subsection{RNN-LSTM Training Procedure}

\begin{figure}
	\centering
	
	\begin{subfigure}{1\textwidth}
		\centering
		\includegraphics[trim={0.5cm 0 0 0},clip,scale=0.25]{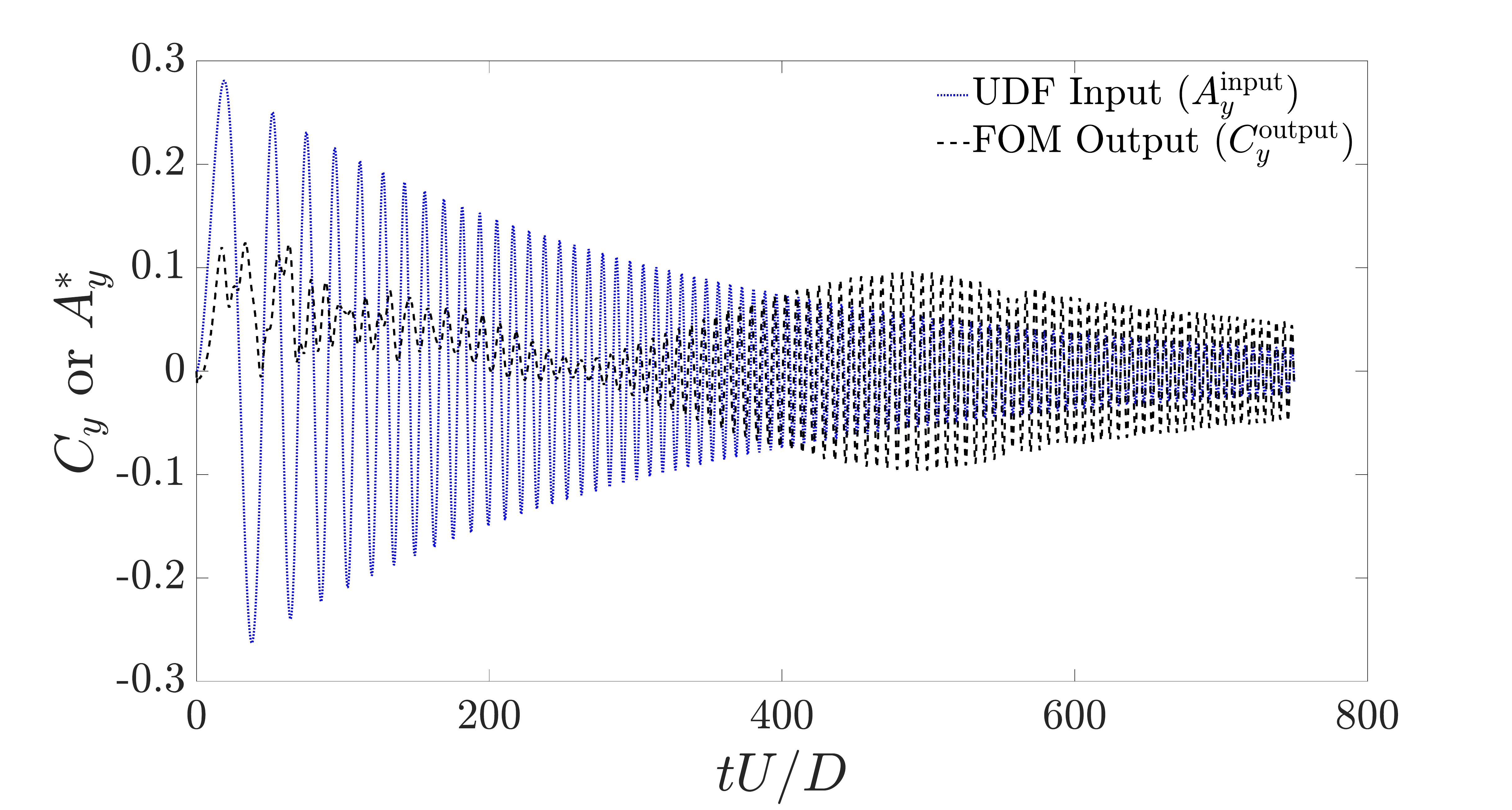}
		\caption{}
	\end{subfigure}
	\begin{subfigure}{1\textwidth}
		\centering
		\includegraphics[trim={0.5cm 0 0 0},clip,scale=0.25]{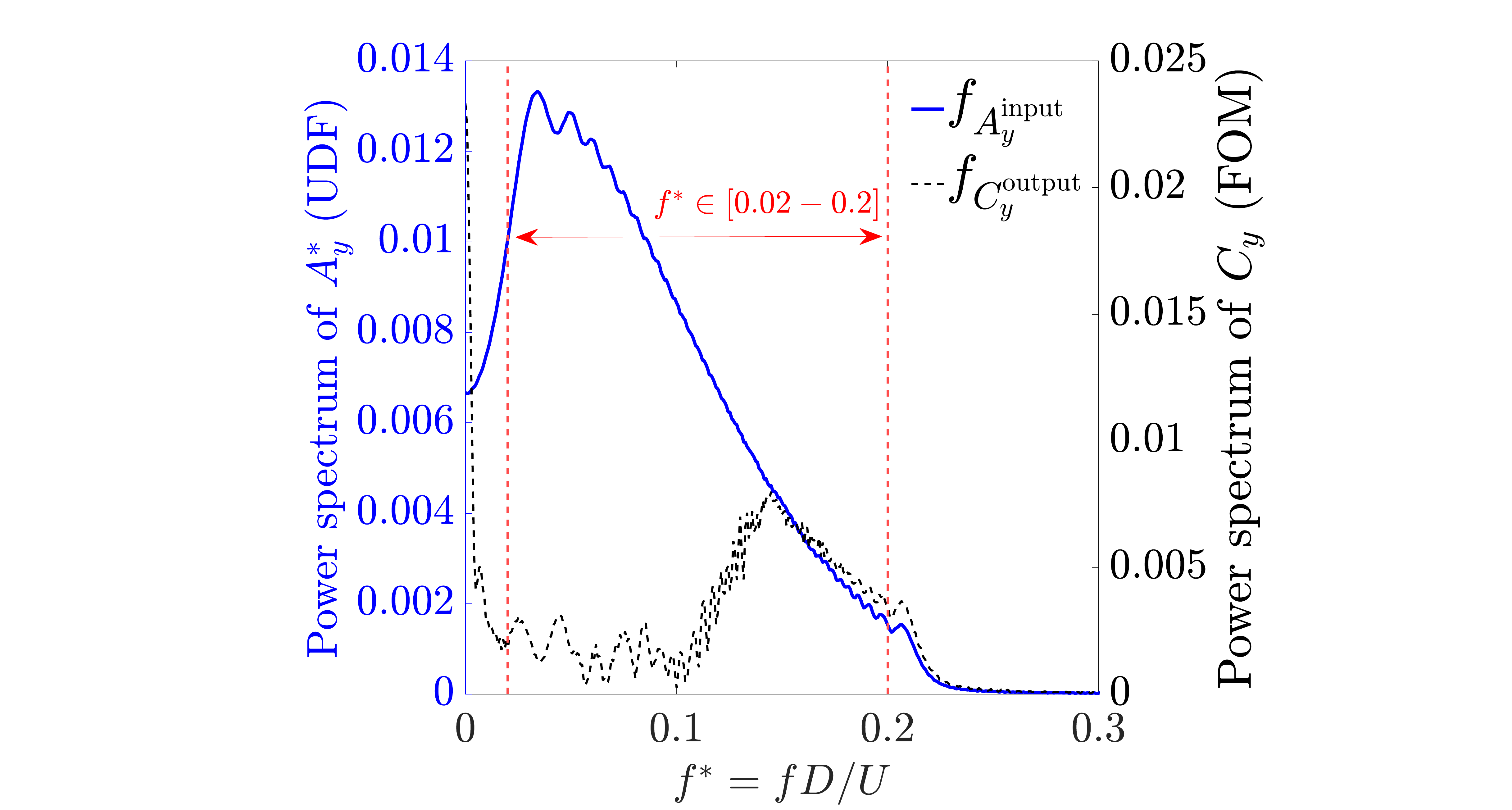}
		\caption{}
	\end{subfigure}~
	\caption{Input-output training dataset for RNN-LSTM for a sphere exposed to uniform flow at $Re = 300$: (a) Time history of the forced displacement ($A_y^{\mathrm{input}}$) via UDF (Eq. \ref{UDF}: $\alpha=0.3$, $\beta=3.5\times10^{-3}$, $\omega(t)=(1.42\times10^{-2})t^{0.6}$, $\phi=0$, and $tU/D\in[0-750]$), and the normalized transverse force ($C_y^{\mathrm{output}}$) from FOM as the output response, (b) Corresponding frequency spectrum of input-output dataset. } 
	\label{IO}
\end{figure}

The formulation of the LSTM network as a well-established standard architecture along with its general nonlinear state-space (NLSS) form (Eq. \ref{NLSS}) are provided in sections \ref{FwdvsInv} and \ref{LSTM}.
While the NLSS formulation is applicable to any fluid-structure system, we choose the transverse motion of a fully submerged three-dimensional canonical geometry of a sphere exposed to a uniform flow field as a prototype problem (see figure \ref{domain}).
The main objective is to identify the VIV lock-in states for this coupled FSI problem for a range of parameter space (i.e.,  Reynolds number, mass ratio, reduced velocity and damping ratio).
To this aim, we need to construct a ROM model which is based on pure input-output data to make a relationship between system observables (i.e. the displacement of the bluff body and the corresponding hydrodynamic force applied on the bluff body).
We train a network using the RNN-LSTM framework as a system identification toolbox to extract temporal feature relationships from high-fidelity numerical solutions.
Since VIV is all about frequency lock-in, we need to generate a large training dataset with useful information across a wide range of frequencies where we expect VIV to occur to get a reliable ROM.
Therefore, to acquire an efficient training dataset from FOM, we provide our methodology to impose a user-defined arbitrary input function (UDF) as a sequence of forced displacement ($A_{y}^\mathrm{input}$) to the bluff body that contains a range of frequencies and amplitudes with a general form as follows 
\begin{align}
	A_{y}^\mathrm{input}=\alpha e^{-\beta t}\mathrm{sin}(\omega(t) t + \phi)
	\label{UDF}
\end{align}
The normalized transverse force ($C_{y}^\mathrm{output}$) is recorded for every time step ($\Delta t = 0.15$) as the output response from the FOM. Figure \ref{IO} shows the input-output dataset from FOM, where a set of $5\,000$ responses are stacked resulting in a total simulation time of $tU/D=750$.

It should be noted that the parameters of the input function should be defined properly based on the underlying physics of the problem.
For instance, if the parameter related to the amplitude of the input function ($\alpha$) is extremely small, it may have a minimal effect on the hydrodynamic output response and the input-output relationship cannot be realized properly. On the other hand, extremely large ($\alpha$) may lead to divergence of the output response.
The amplitude decay/growth rate parameter ($\beta$) is another important parameter to consider the effect of change in the amplitude in the process of learning and to avoid large impulsive displacements at the higher frequency domain.
$\omega(t)$ defines the rate of change of frequency in the input function. For our FSI problem setup, we have adjusted $\omega(t)$ to have a minimum of 40 time steps ($\Delta t$) per fluctuation cycle at the highest frequency range. Here we considered a range of non-dimensional frequencies $f^*\in[0.02-0.2]$ where we expect VIV lock-in to happen. Based on the insight gained from the underlying physics of the problem we know that at higher frequency ratios (i.e. $f^*>0.2$) we do not expect high amplitude vibrations corresponding to VIV and the fluctuations are most probably negligible.
The RNN-LSTM model is only dependant on the Reynolds number and the geometry of the bluff body. The system dynamic associated with the RNN-LSTM model is decoupled and there is a partitioned coupling between the fluid and structural solver. Using this decoupled FSI idea for the relationship between the displacement and the hydrodynamic force, we next verify the methodology for the coupled FSI cycle.

\subsection{Verification of DL-based ROM Integrated with ERA} 
\begin{figure}
	\centering
	
	\begin{subfigure}{1\textwidth}
		\centering
		\includegraphics[trim={0.5cm 0 0 0},clip,scale=0.2]{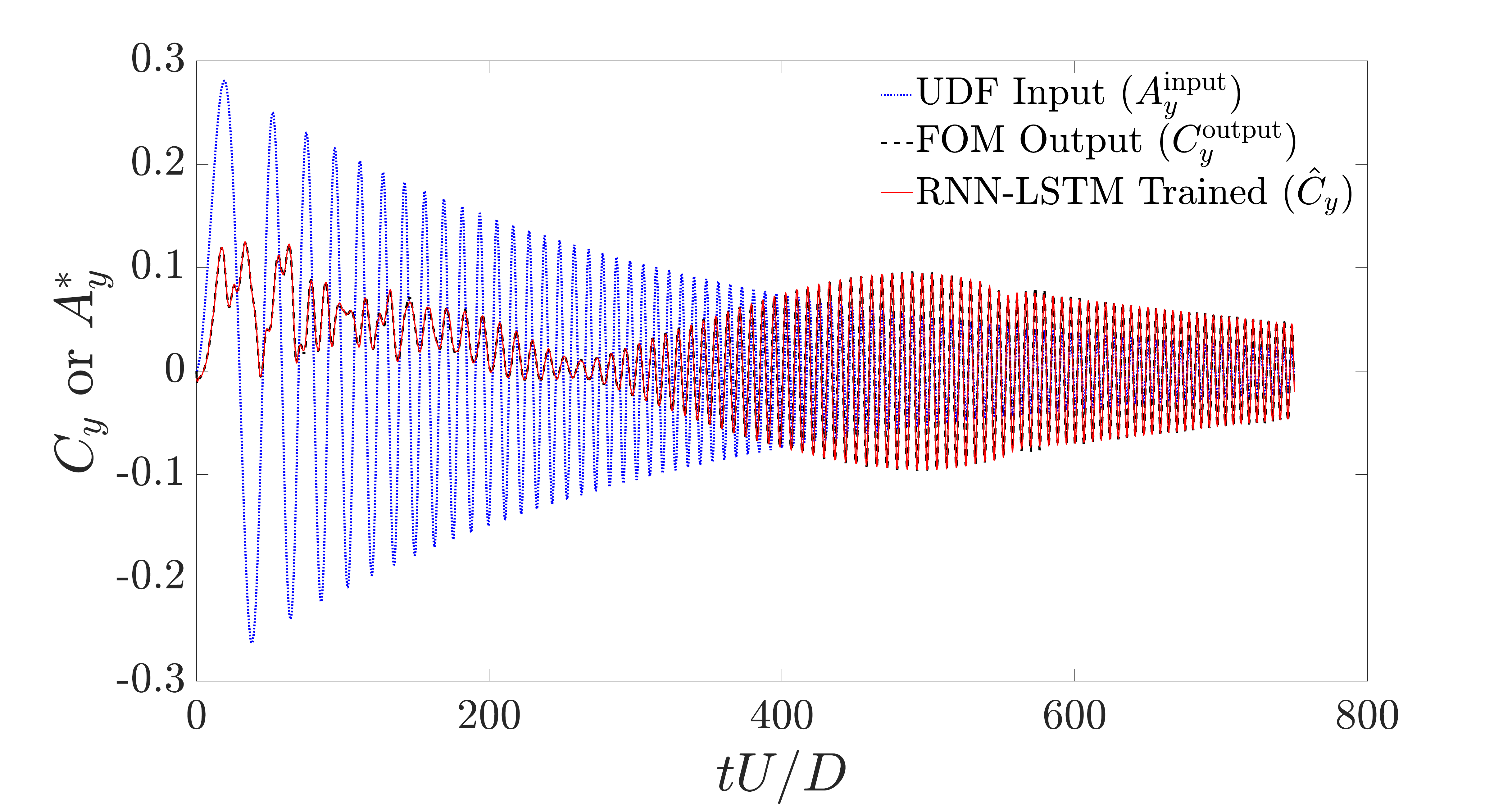}
	\end{subfigure}
	
	\begin{subfigure}{1\textwidth}
		\centering
		\includegraphics[trim={0.5cm 0 0 0},clip,scale=0.2]{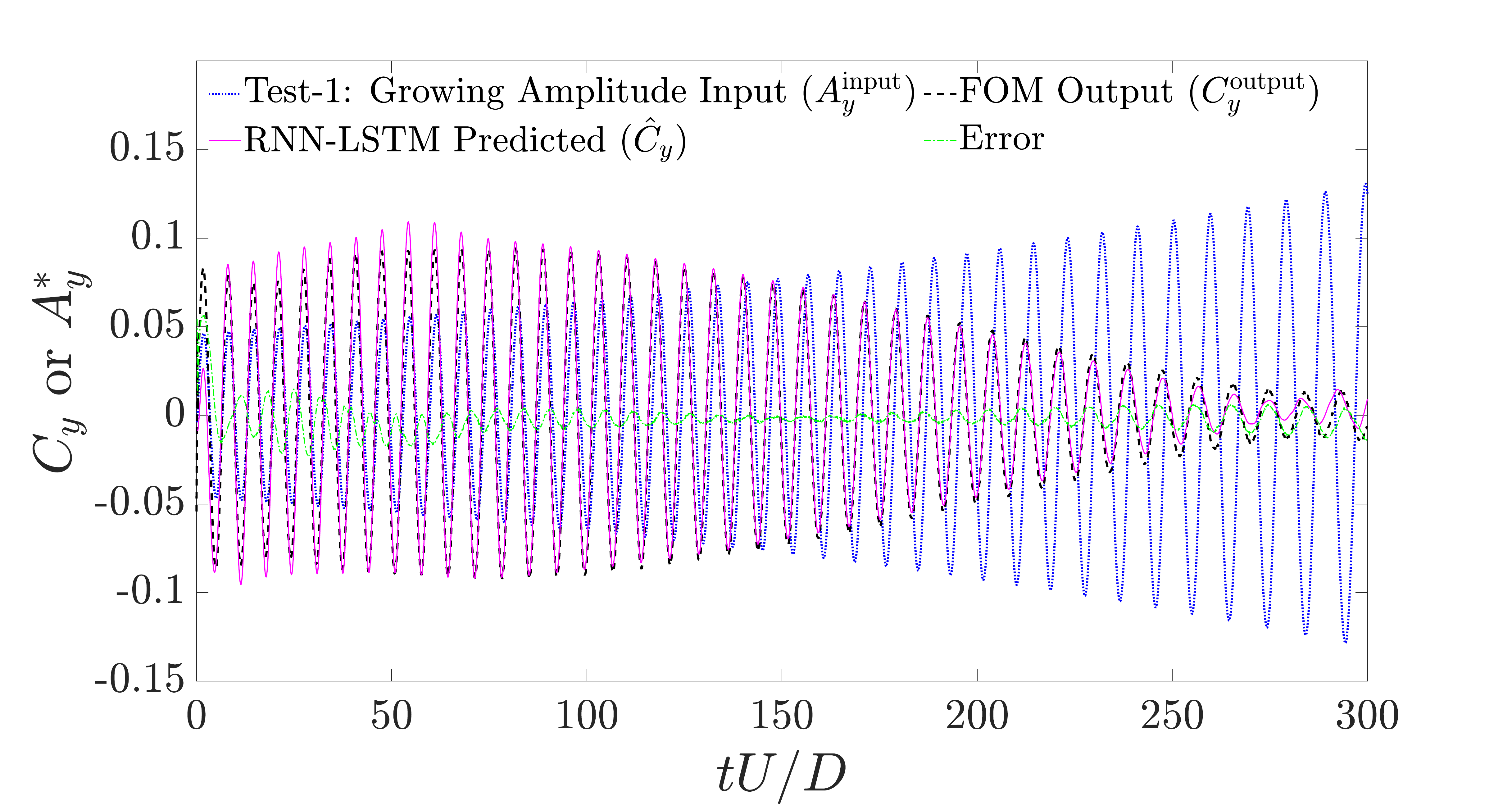}
	\end{subfigure}
	
	\begin{subfigure}{1\textwidth}
		\centering
		\includegraphics[trim={0.5cm 0 0 0},clip,scale=0.2]{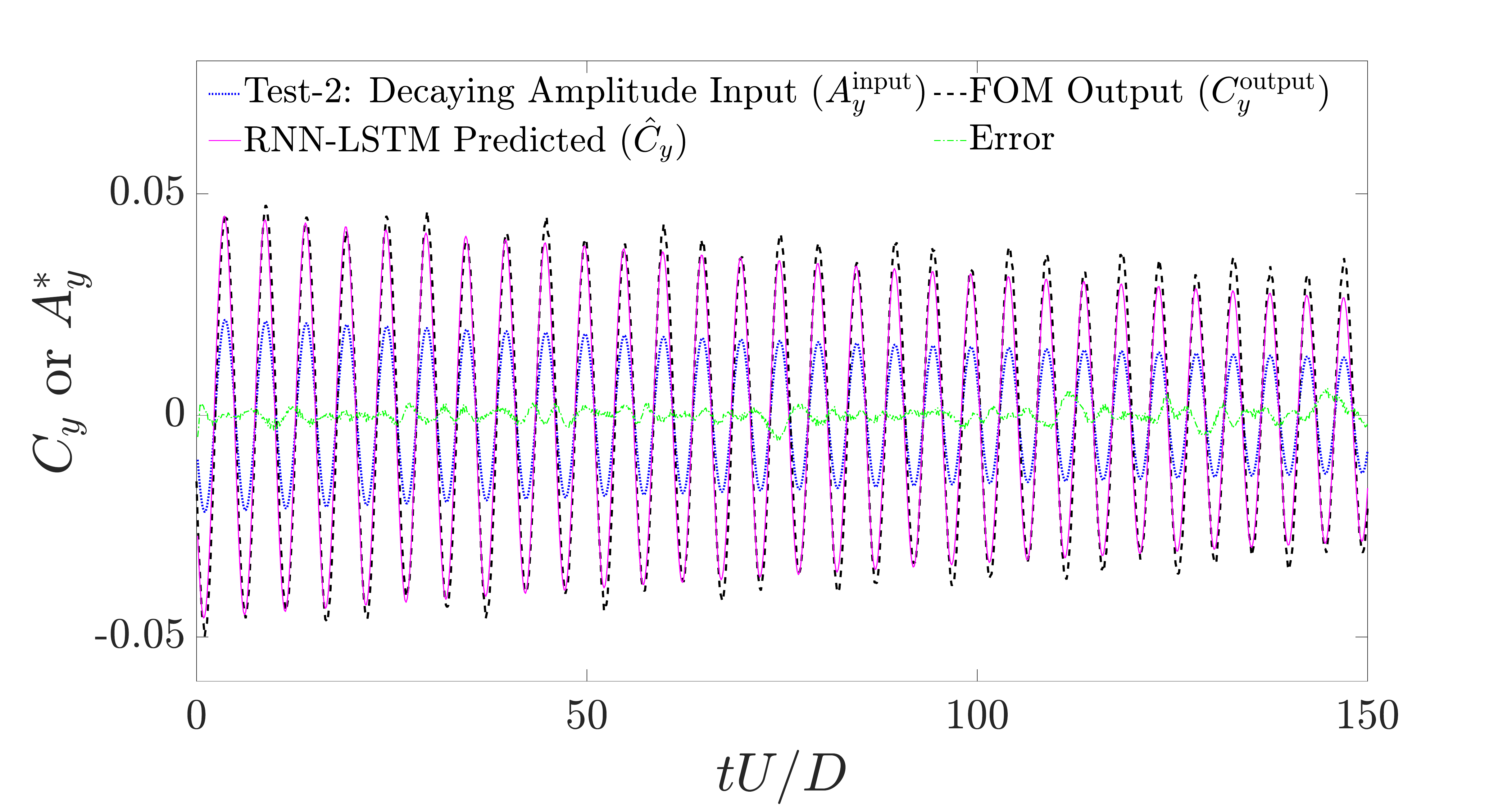}
	\end{subfigure}
	
	\caption{Time history of the normalized transverse force ($C_y^{\mathrm{output}}$) due to unstable wake behind a forced displaced sphere at $Re = 300$: (a) Train dataset: ($A_{y}^\mathrm{input}=0.3 \times e^{-0.0035 t}\mathrm{sin}(0.0142t^{0.6})t$, $tU/D\in[0-750]$), (b) Test-1 dataset: flipped $A_{y}^\mathrm{input}, (t-540)U/D\in[0-300]$, and (c) Test-2 dataset: shifted $A_{y}^\mathrm{input}, (t-750)U/D\in[0-150]$.} 
	\label{Re_low_Train}
\end{figure}

Based on the methodology described in Sec \ref{LSTM}, the LSTM network is applied to construct the nonlinear ROM based on a pure input-output dataset for a transversely vibrating sphere at $Re = 300$. The FOM data are used to fit a model. Figure \ref{Re_low_Train} (a) compares the output normalized force signal calculated from the FOM ($C_y$) and the ROM ($\hat{C}_y$), where we can see good performance ($\mathrm{Fit}=97$) for the training dataset. The model is simulated using the validation data input and the error is determined by the discrepancy between the model output and the measured validation data output as a  system identification task.

Figures \ref{Re_low_Train} (b, c) show the predicted output ($\hat{C}_y$) and the error for two different test datasets. It is worth noting that the simulation is done without any access to the measured validation output. As shown in the figures, the character of the validation dataset is different from the training datasets. The input for test dataset-1 (figure \ref{Re_low_Train} (b)) has a growing amplitude trend with a frequency range from higher to lower values, whereas the input for test dataset-2 (figure \ref{Re_low_Train} (c)) has a decaying amplitude trend with a frequency range from lower to higher values. Here for the acceptable hyper-parameter set, the main benchmark considered is the root mean square error (RMSE) and the fit for the identified model output and measured output corresponding to the non-dimensional transverse force given by
\begin{align}
	\mathrm{RMSE} = \left\lVert\frac{C_y(t)-\hat{C}_y(t|\theta)}{N}\right\rVert
\end{align}
\begin{align}
\mathrm{Fit} = 100 \times \left(1-\frac{\left\lVert C_y(t)-\hat{C}_y(t|\theta) \right\rVert}{\left\lVert C_y(t)-\bar{C}_y(t) \right\rVert}\right). 
\end{align}

We calculate the RMSE and the fit for the training set and the two aforementioned test datasets, which are shown in Table \ref{Error}. The results show good performance with the root mean squared error less than $1\%$ for all datasets and hence provide an acceptable fit.
While the model constructed using LSTM-RNN is validated by two aforementioned test datasets, the model cannot be tested with a good performance for any arbitrary input. By increasing the training dataset, one can create a relatively accurate model. Our methodology aims to construct an efficient training dataset for the VIV lock-in process.

\begin{table}
	\begin{center}
		\begin{tabular}{C{1.5cm}C{1.5cm}C{1.5cm}C{2cm}C{3cm}}
			Dataset & RMSE & Fit & Number of dataset (N) & Non-dimensional time domain  \\
			\hline
			 Train   & 0.19\% & 97.1 & 5000 & $tU/D\in[0-750]$ \\[5pt]
			 Test-1  & 1.13\% & 82.9 & 2000 & $tU/D\in[0-300]$ \\[5pt]
			 Test-2  & 0.28\% & 86.4 & 1000 & $tU/D\in[0-150]$ \\[5pt]
		\end{tabular}
		\caption{The root mean square error (RMSE) and the fit corresponding to train and test datasets.}
		\label{Error}
	\end{center}
\end{table}

\subsubsection{Hyper-parameter Sensitivity Analysis}
Extreme refining of learning rate and overuse of the hidden cells may cause the RNN-LSTM to overfit the training dataset and make it incapable of predicting the forces for perturbed geometries of the training dataset. The underutilization of hyper-parameters, on the other hand, will increase the prediction error. Here, we provide an empirical sensitivity analysis to determine the hyperparameter values for the optimum RNN-based learning performance. In particular, we study the sensitivity of the number of epochs, the number of hidden cells, and the learning rate.

\begin{figure}
	\centering
	\begin{subfigure}{0.5\textwidth}
		\centering
		\includegraphics[trim={0cm 0cm 0cm 0cm},clip,scale=0.15]{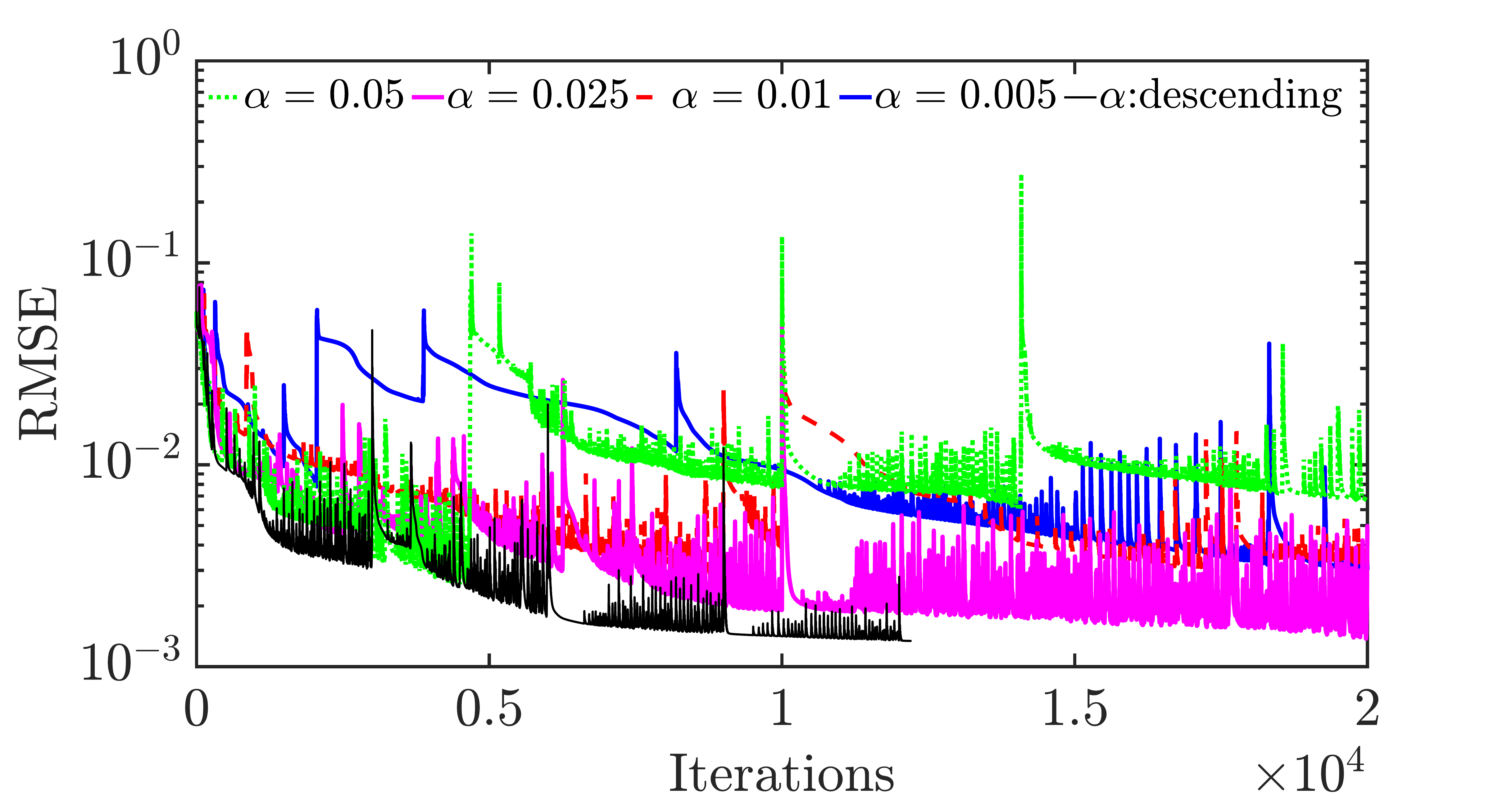}
		\caption{}
	\end{subfigure}~
	\begin{subfigure}{0.5\textwidth}
		\centering
		\includegraphics[trim={0cm 0cm 0cm 0cm},clip,scale=0.15]{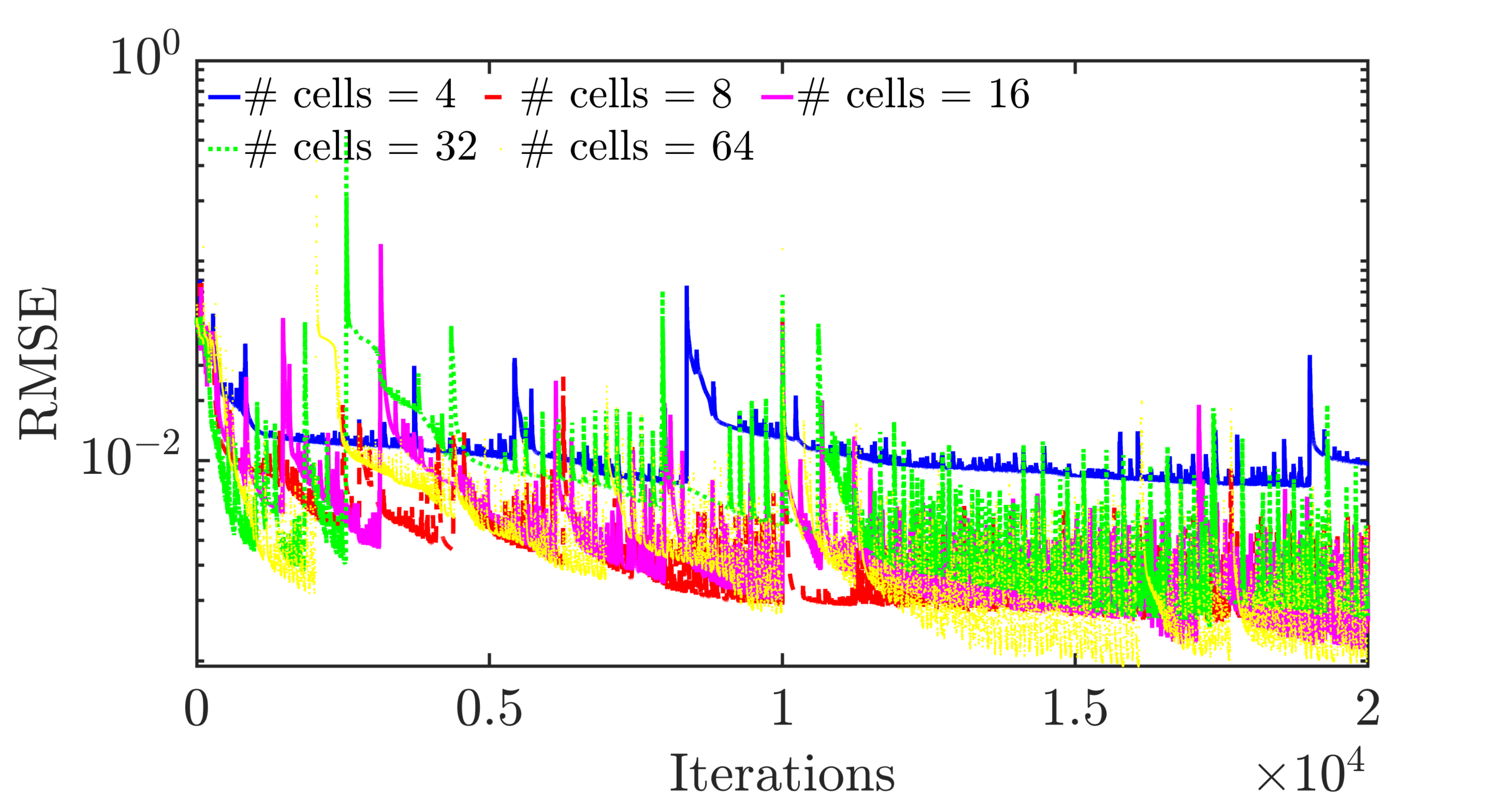}
		\caption{}
	\end{subfigure}

	\begin{subfigure}{0.51\textwidth}
		\centering
		\includegraphics[trim={0cm 0cm 0cm 0cm},clip,scale=0.15]{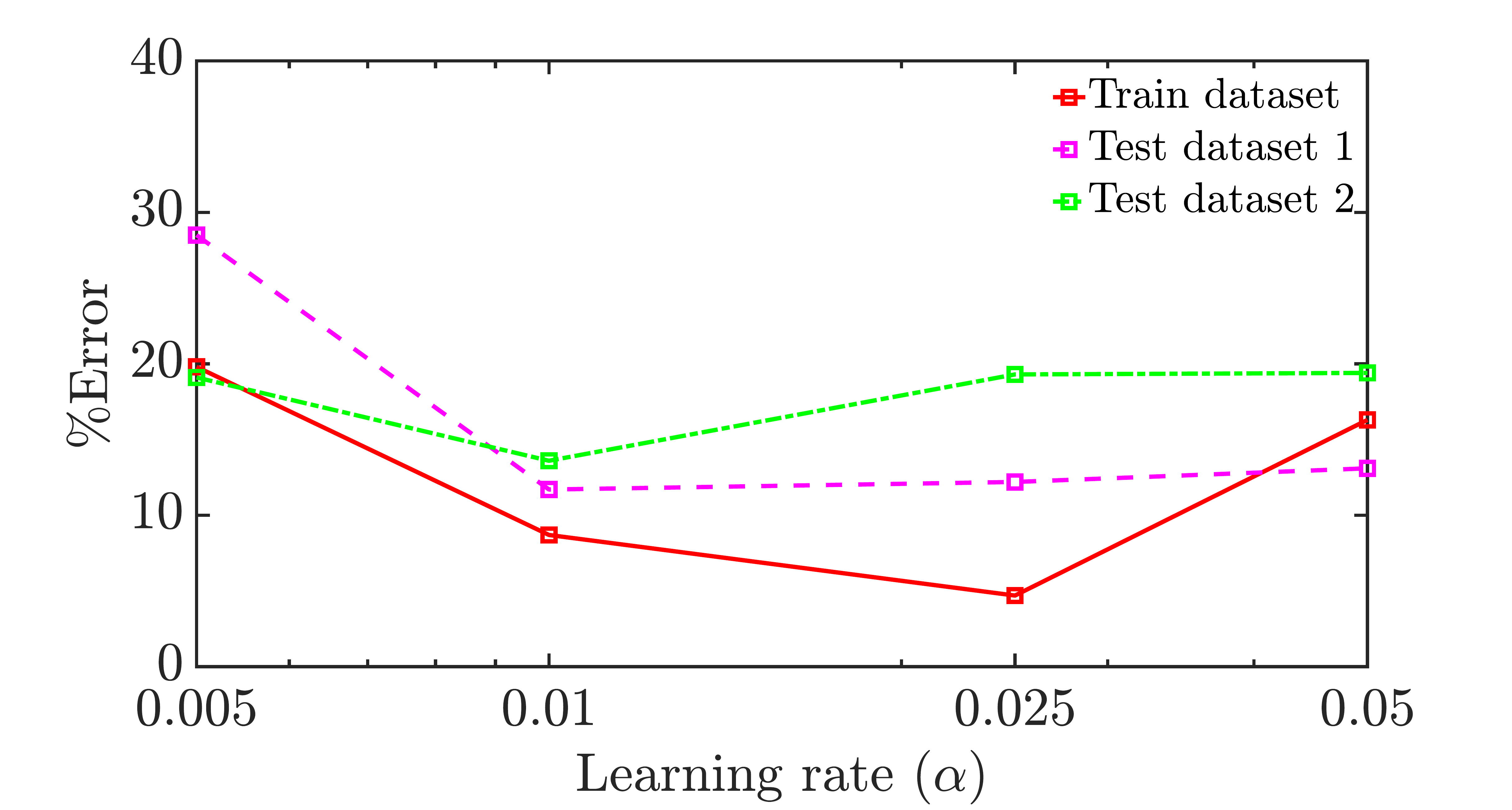}
		\caption{}
	\end{subfigure}~
	\begin{subfigure}{0.49\textwidth}
		\centering
		\includegraphics[trim={0cm 0cm 0cm 0cm},clip,scale=0.15]{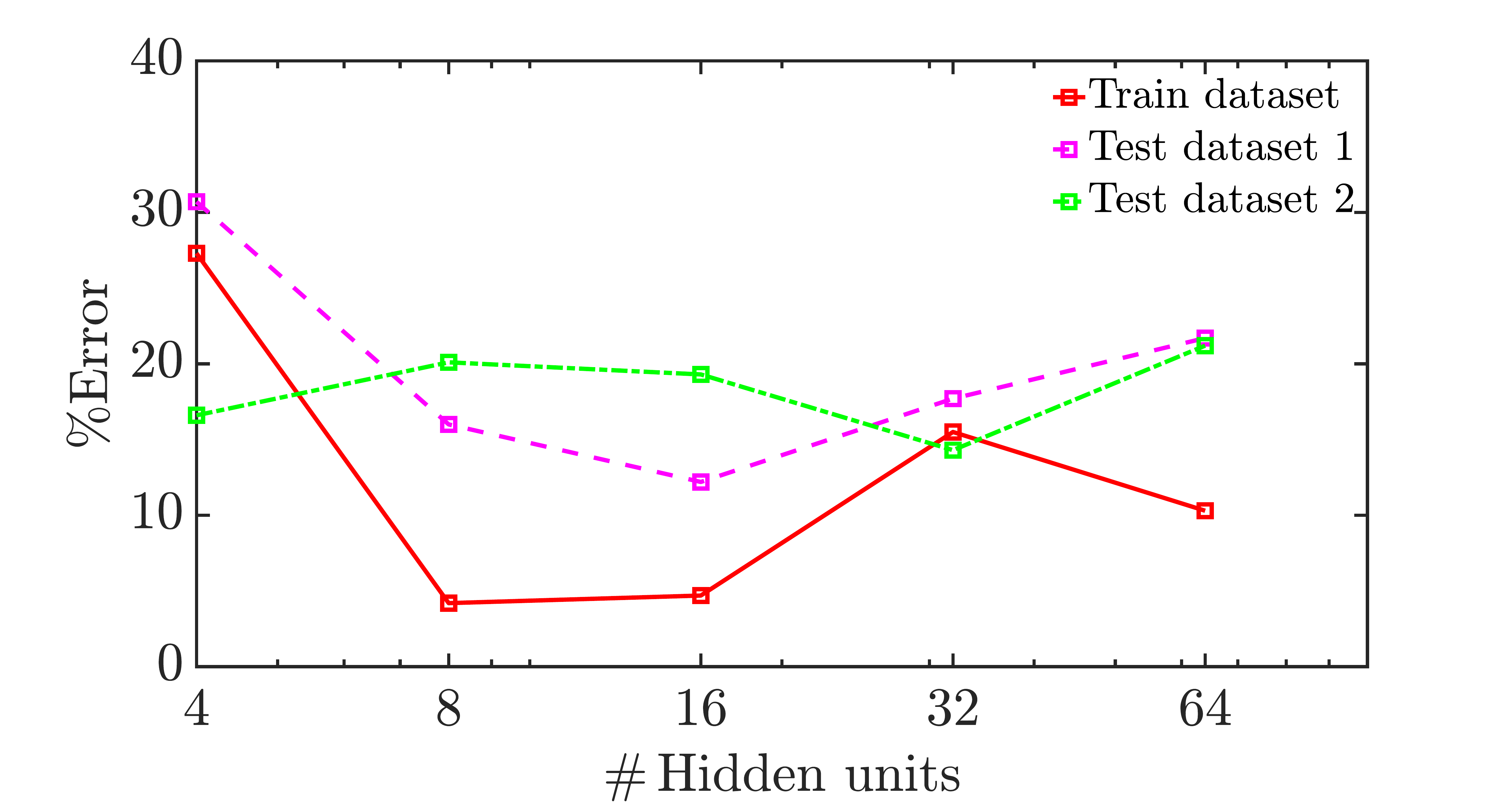}
		\caption{}
	\end{subfigure} 

	\caption{Hyper-parameter analysis: (a, b) Root mean square error (RMSE) variation with the number of epochs for different values of learning rate ($\alpha$) and hidden units. (c, d) Error variation of train and test datasets with different values of $\alpha$ and hidden units, where the error is defined as ($\mathrm{Error = 100-fit}$). }
	\label{DL_Conv}
\end{figure}

We begin by assuming that the RNN-LSTM method with the small number of hidden cells ($\# \mathrm{\ of\ Hidden\ units} = 8$) is the most appropriate for making predictions.
Figure \ref{DL_Conv} (a) shows the root mean square error (RMSE) for different learning rates ($\alpha$) while the number of hidden cells is fixed ($\# \mathrm{\ of\ Hidden\ units} = 8$) during the evolution of the neural network as it trains after each epoch. The minimum error is found when ($\alpha$=0.025) after $20\,000$ iterations.
Now, by fixing the learning rate to ($\alpha$=0.025), the variation of RMSE of the training network with a lower and higher number of hidden units is depicted in figure \ref{DL_Conv} (b). The best convergence is observed for ($\# \mathrm{\ of\ Hidden\ units} = 8$).
Figure \ref{DL_Conv} (c) and (d), show the $\%\mathrm{Error}$ with the variation of learning rate and several hidden units respectively. Figure \ref{DL_Conv} (c) shows that if the learning rate is set extremely small, the loss function might lead to a local minima issue, and if the learning rate is set too large, the loss function may exhibit undesired divergent behaviour. To overcome this issue our approach is to decrease the learning rate manually after specific iterations to obtain the global minima and fast convergence. As shown in figure \ref{DL_Conv} (a) by the black line, we can get the fastest convergence with the smallest error while decreasing the learning rate after each $3000$ iterations. Table \ref{Network_Spec} shows specifications of the selected network for our FSI problem. 
Next, we conduct the error analysis with a different number of hidden cells which is shown in figure \ref{DL_Conv} (d). We find that the RNN shows a good performance with ($\# \mathrm{\ of\ Hidden\ units} = 8$) while increasing the number of hidden units would lead to over-fitting issues and more computational cost.

\begin{table}
	\begin{center}
		\begin{tabular}{L{4cm}|L{4cm}}
			RNN-LSTM hyper-parameters 	   &  Specifications \\[5pt]
			\hline
			Number of layers  	   & 1 fully connected layer \\[5pt]
			Number of hidden unit  & 8  \\[5pt]
			Optimizer      		   & Adam  \\[5pt]
			Learning rate  		   & 0.050 $@ \mathrm{Iter}\in[0-3000]$  \\[5pt]
								   & 0.025 $@ \mathrm{Iter}\in[3001-6000]$  \\[5pt]
								   & 0.010 $@ \mathrm{Iter}\in[6001-9000]$  \\[5pt]
								   & 0.005 $@ \mathrm{Iter}\in[9001-12000]$  \\[5pt]
		\end{tabular}
		\caption{Network specifications.}
		\label{Network_Spec}
	\end{center}
\end{table}

\subsubsection{Reliability of ERA}
\begin{figure}
	\centering
	
	\begin{subfigure}{0.5\textwidth}
		\centering
		\includegraphics[trim={8cm 4.2cm 13cm 0},clip,scale=0.2]{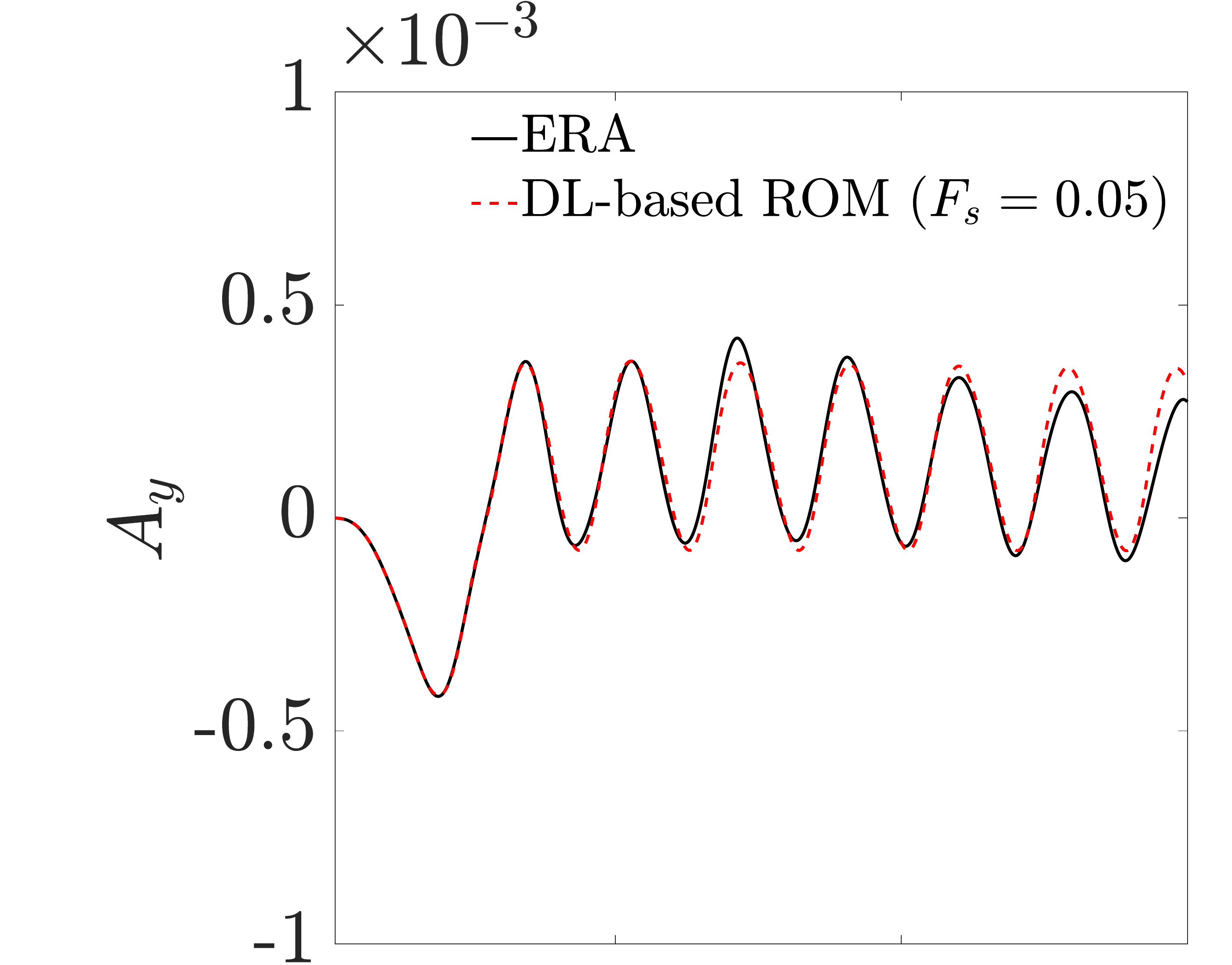}
	\end{subfigure}~
	\begin{subfigure}{0.5\textwidth}
		\centering
		\includegraphics[trim={8cm 4.2cm 8cm 0},clip,scale=0.2]{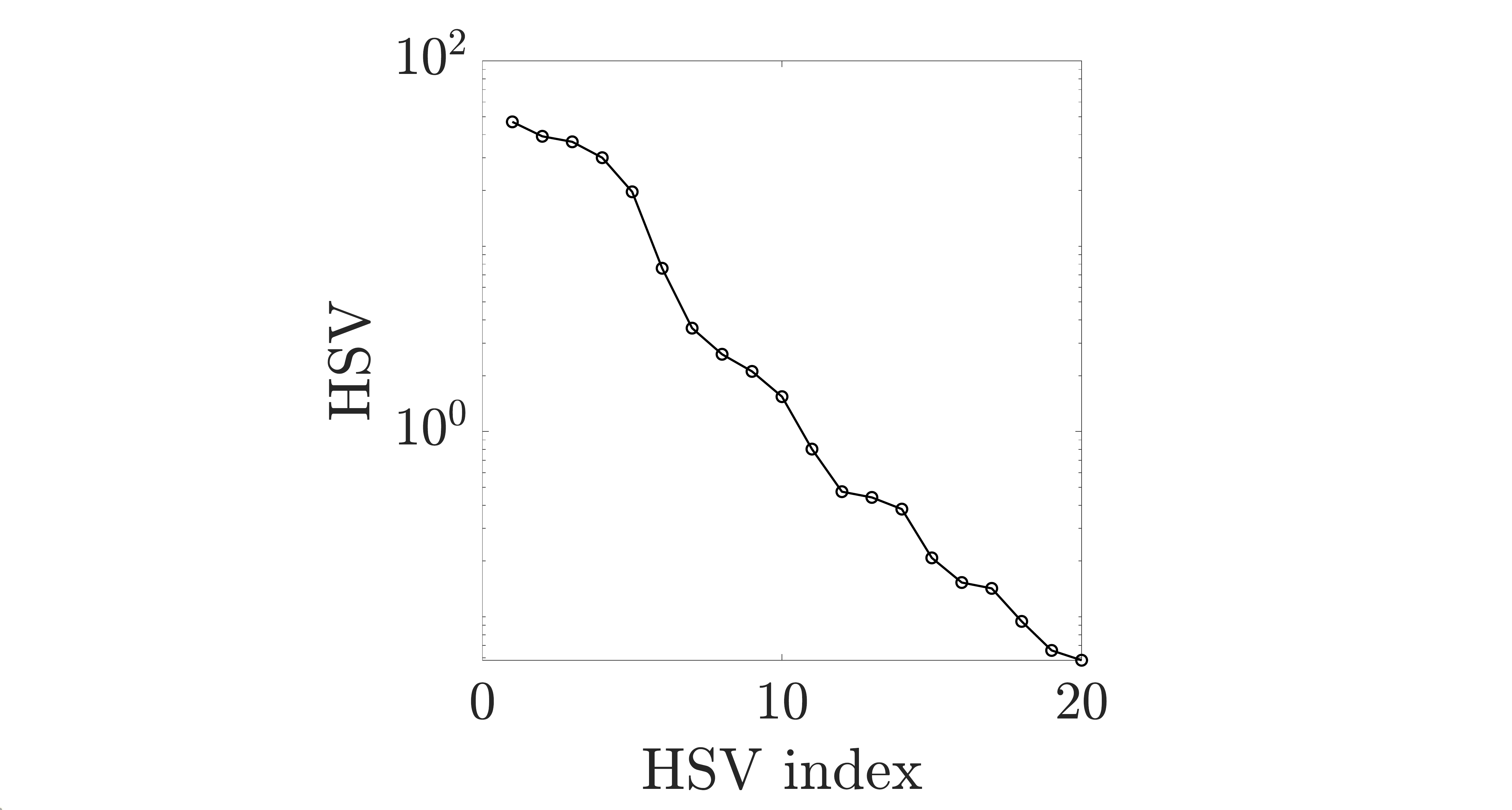}
	\end{subfigure} 
	
	\begin{subfigure}{0.5\textwidth}
		\centering
		\includegraphics[trim={8cm 4.2cm 13cm 0},clip,scale=0.2]{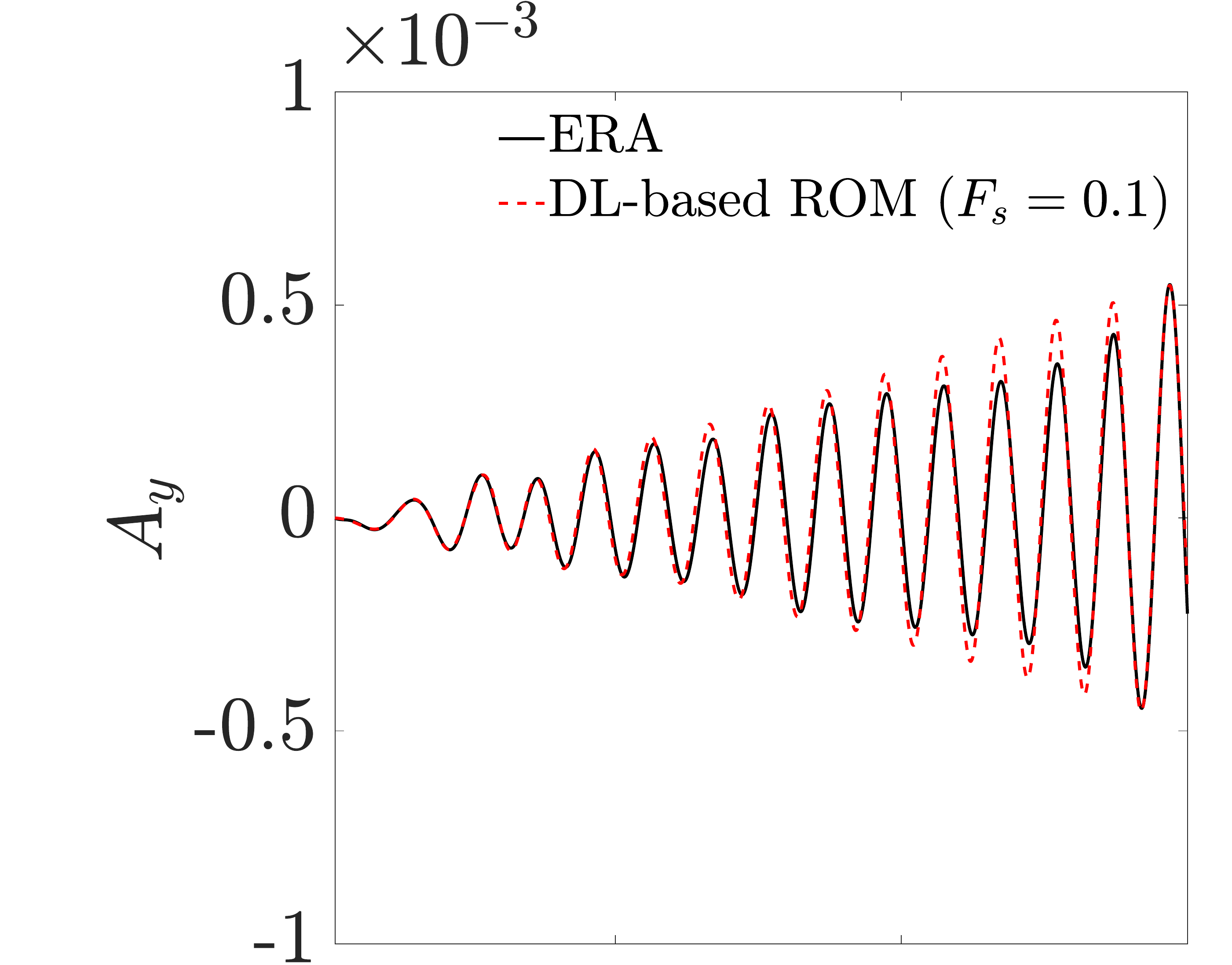}
	\end{subfigure}~
	\begin{subfigure}{0.5\textwidth}
		\centering
		\includegraphics[trim={8cm 4.2cm 8cm 0},clip,scale=0.2]{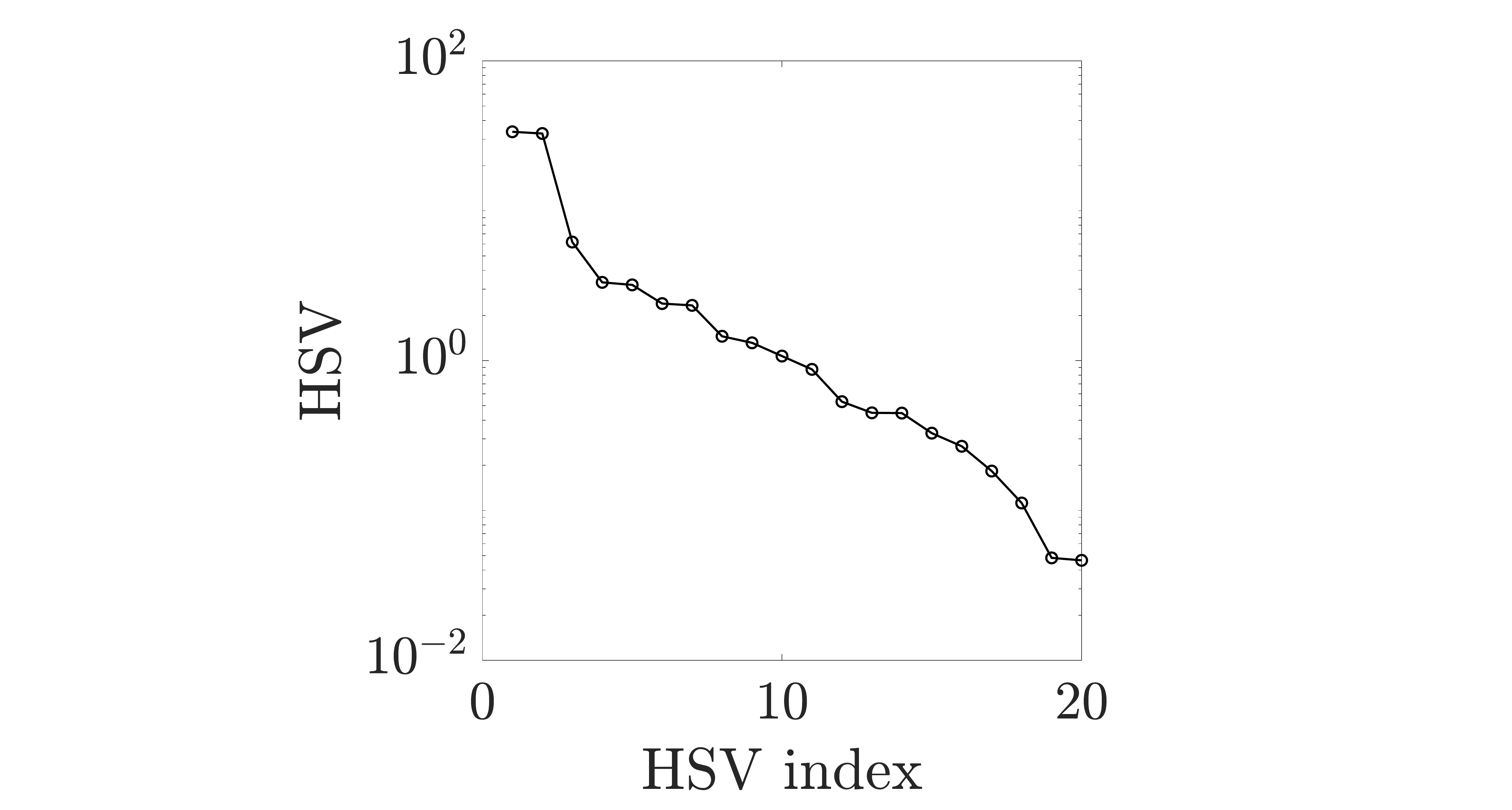}
	\end{subfigure} 
	
	\begin{subfigure}{0.5\textwidth}
		\centering
		\includegraphics[trim={8cm 4.2cm 13cm 0},clip,scale=0.2]{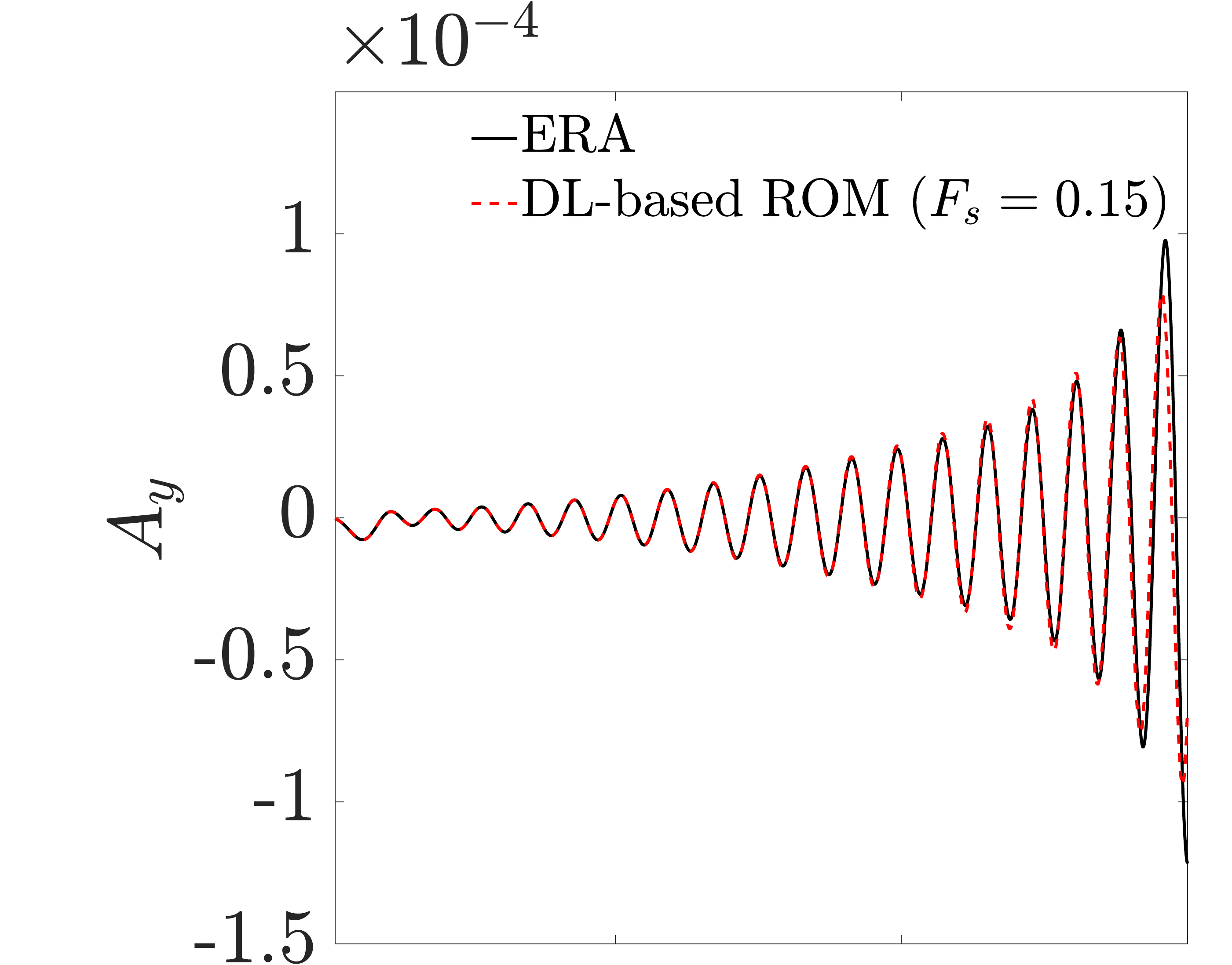}
	\end{subfigure}~
	\begin{subfigure}{0.5\textwidth}
		\centering
		\includegraphics[trim={8cm 4.2cm 8cm 0},clip,scale=0.2]{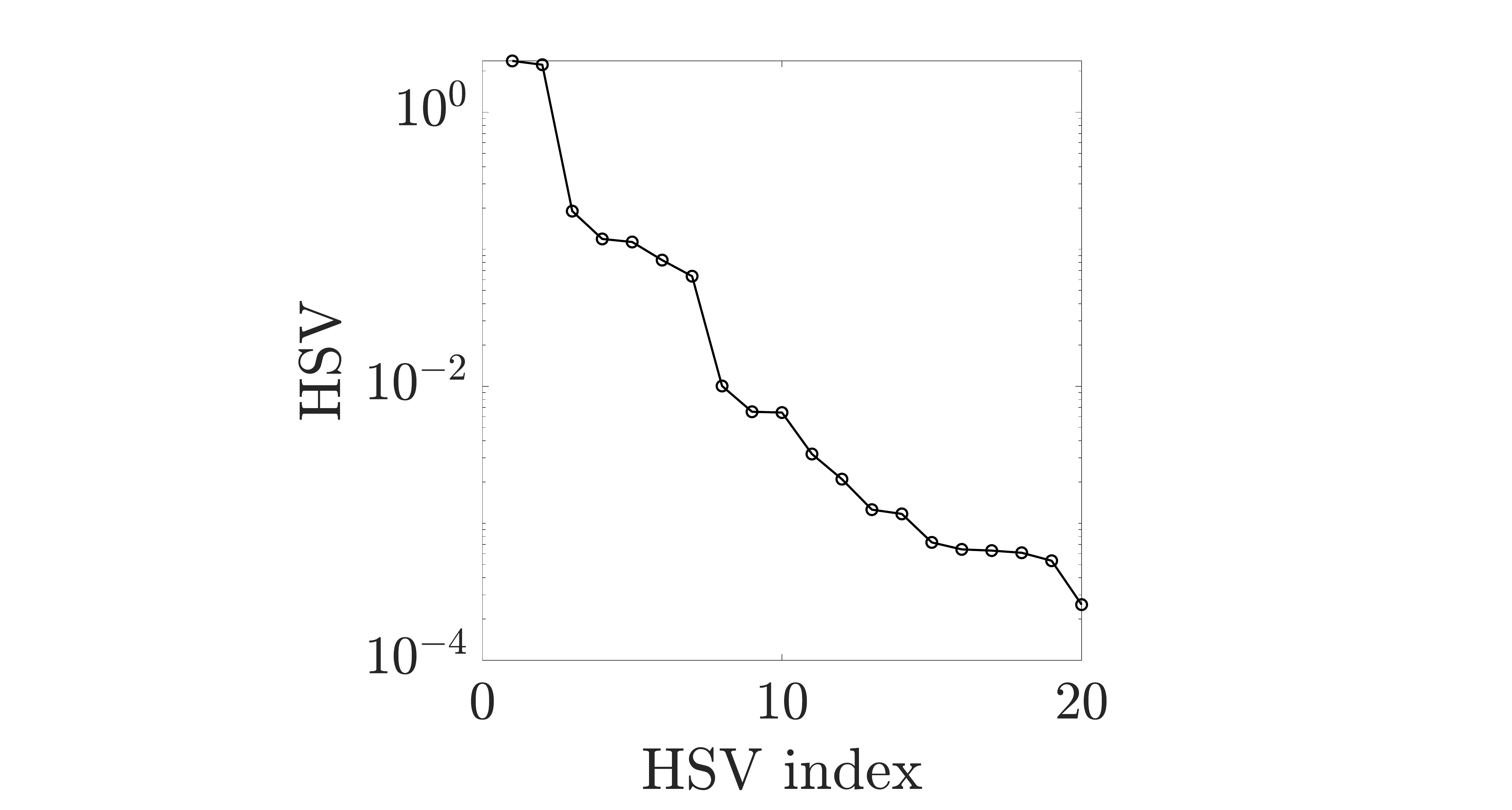}
	\end{subfigure} 
	
	\begin{subfigure}{0.5\textwidth}
		\centering
		\includegraphics[trim={8cm 0cm 13cm 0},clip,scale=0.2]{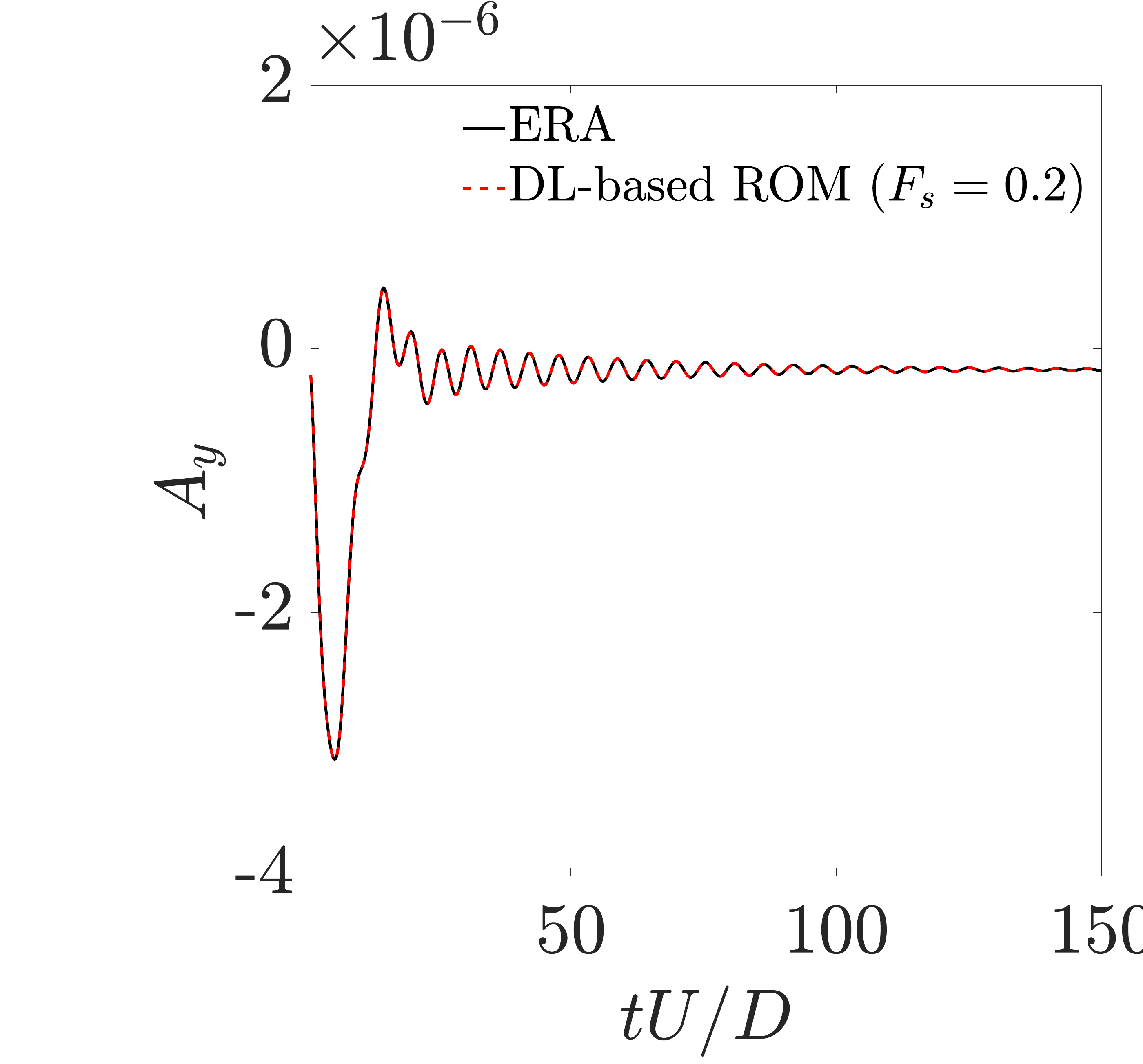}
		\caption{}
	\end{subfigure}~
	\begin{subfigure}{0.5\textwidth}
		\centering
		\includegraphics[trim={8cm 0cm 8cm 0},clip,scale=0.2]{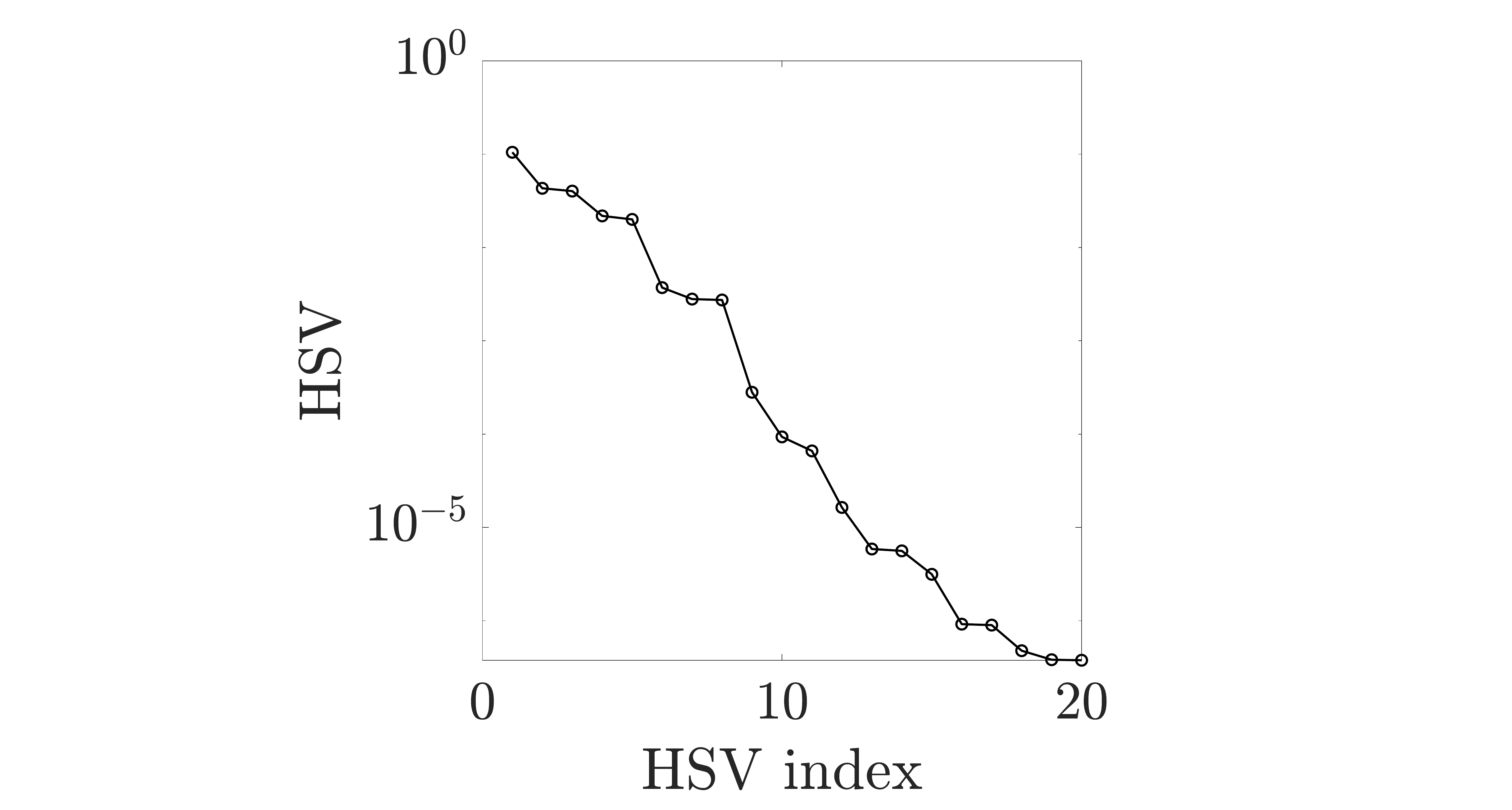}
		\caption{}
	\end{subfigure} 
	
	\caption{The DL-based ROM for the VIV of a sphere at $Re = 300$: (a) Time history of the transverse amplitude ($A_y$) due to small perturbation calculated from the DL-based ROM and the linear approximation via ERA and (b) HSV distribution corresponding to $500\times 250$ Hankel matrix.}
	\label{ERA_Cumulative}
\end{figure}

Here we aim to assess the reliability of the linearized approximation through the ERA. 
The process with the three main components forms a general procedure to be followed while constructing the ERA-based ROM as presented in Sections \ref{ERA} and \ref{Eigenvalue Selection}. It should be noted that ERA is only applied to the displacement vector $A_y$ as a desired output from the DL-based ROM for the stability predictions. 
Based on the tiny amplitude perturbation ($\delta = 10^{-3}$) given to the ERA-based ROM as an input, the transverse displacement $A_y$ is recorded for every time step $\Delta t = 0.15$. The linearity of the unstable system is confirmed by comparing the response subject to two impulse inputs with $\delta=10^{-3}$ and $\delta=10^{-4}$. A set of $750$ responses are stacked resulting in a total simulation time of $tU/D = 112.5$. Figure \ref{ERA_Cumulative} shows the output amplitude signal $A_y$ calculated from the DL-based ROM and the linearized ERA-based ROM at $Re = 300$. The Hankel matrix may not be necessarily square but can be tall, wide or square based on the problem setup. The Hankel matrix with dimension $500 \times 250$ is found to be appropriate by examining the convergence of unstable eigenvalues computed from Hankel matrices with dimensions of $500 \times 125$, $500 \times 250$, and $500 \times 500$. The order of the ERA-based ROM is selected by examining the HSV distribution. The fast-decaying singular values as depicted in figure \ref{ERA_Cumulative} (b) suggest that the ERA-based ROM with order $n_r = 12$ is sufficient. This is further confirmed by the accurate reconstruction of the impulse response by the ERA-based ROM displayed in figure \ref{ERA_Cumulative} (a). 

It should be highlighted that the computational effort associated with the development of the DL-based ROM integrated with ERA is extremely efficient compared to FOM simulations. To demarcate the lock-in range from the FOM, the long-term unsteady simulation is required to construct the relationship between the reduced natural frequency ($F_s$) and the normalized transverse amplitude ($A^*_y$) at the stationary state. The DL-based ROM integrated with ERA completely avoids these expensive simulations. Moreover, constructing the eigenspectrum is trivial and fast since it relies on the SVD procedure. 

In many situations, linear models give a reasonable approximation of the underlying process. Here we utilize the ERA methodology for the linear approximation of the coupled nonlinear model. By examining the HSV distribution, we select enough modes to obtain the error corresponding to most of the cases by less than $1\%$. However, there exist some cases for which the linear approximation cannot be made even by considering the full-rank Hankel matrix.

\section{Results and Discussion}
\label{RandD}
Here, we analyze the stability properties of a transversely vibrating sphere at $Re=300$ through the DL-based ROM integrated with ERA. For this purpose, the eigenspectrum plot for a canonical sphere is systematically examined. Consistent with the previous literature by \cite{yao2017model}, we utilize the methodology of variation of unstable modes to classify the distinct eigenvalue trajectories of the fluid-structure system governed by Eq. \ref{ERA_ROM}.
As we discussed in Sections. \ref{ERA} and \ref{Eigenvalue Selection}, ERA can be applied to the desired output of the coupled FSI system. Therefore, we can extract the structural modes (SM) by considering the transverse displacement ($A_y$) as the desired output of the coupled system. 
%



VIV lock-in may result from the instability of the unstable complex wake modes (WMs) interacting with the structural mode (SM). 
When the shedding process (unstable WMs) is dominated by the natural frequency of the body, strong coupling between the fluid and the structure is established which is called resonance-induced lock-in. 
In this work, the VIV branch is termed as resonance-induced lock-in if the imaginary part of the most unstable eigenvalue corresponding to the transverse displacement ($A_y$) gets close to the natural frequency of the structure in a vacuum. The stability of this region is determined by the value of the real part of the eigenvalues corresponding to $A_y$. 
%
%
The instability to sustain the VIV lock-in also occurs via combined mode instability of the unstable SM and WMs, indicating a flutter or galloping-type instability which is not examined in this study. 
Therefore, based on the methodology presented in Sec. \ref{Eigenvalue Selection}, we aim to identify the resonance induced lock-in regions and to predict the unseen dynamics by utilizing the eigenvalue selection process on the amplitude response of the sphere ($A_y$).

\begin{figure}
	\centering
	\begin{subfigure}{1\textwidth}
		\centering
		\includegraphics[trim={0 5cm 0 0},clip,scale=0.2]{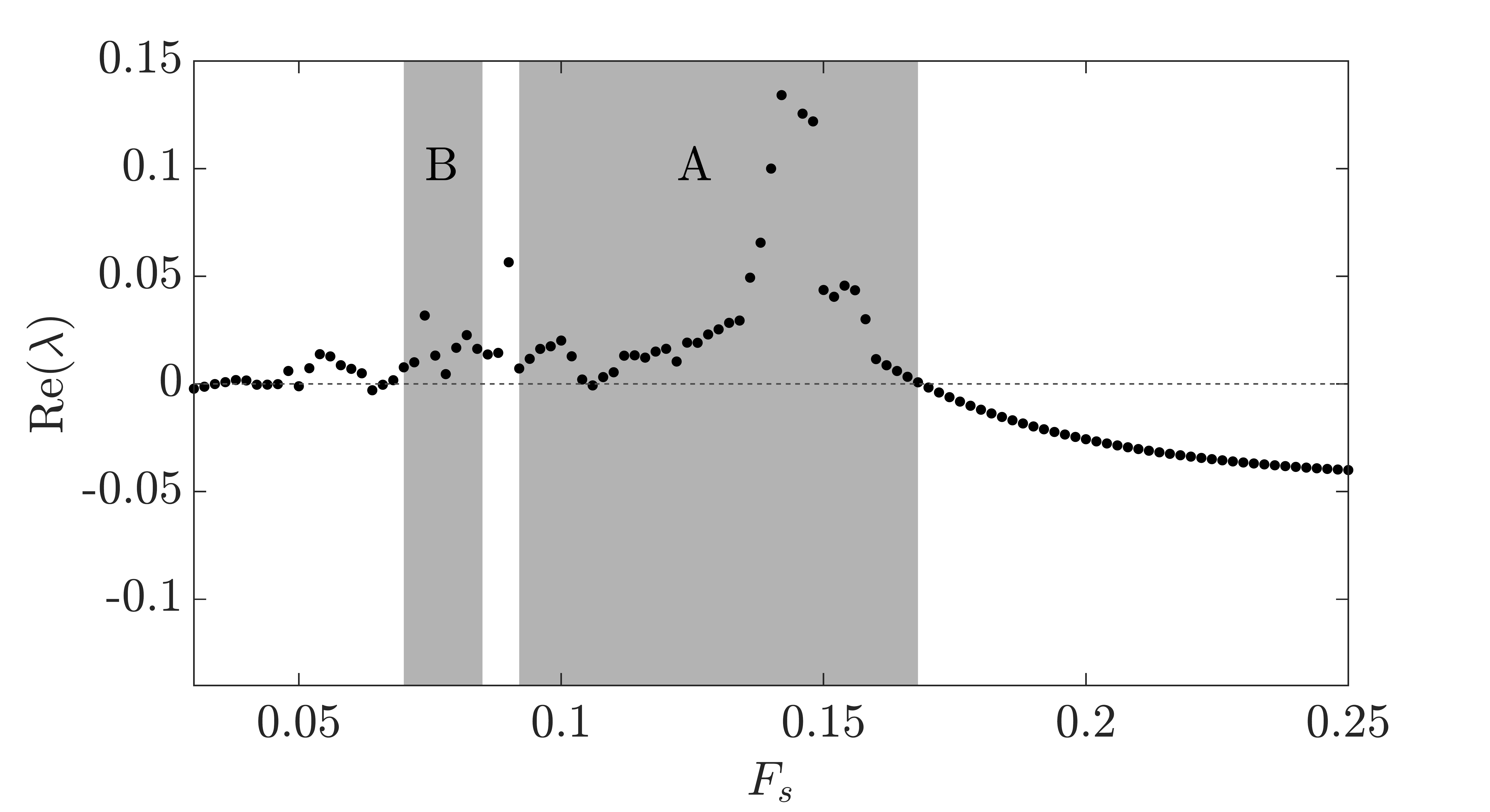}
	\end{subfigure}
	\begin{subfigure}{1\textwidth}
		\centering
		\includegraphics[trim={0 0 0 0},clip,scale=0.2]{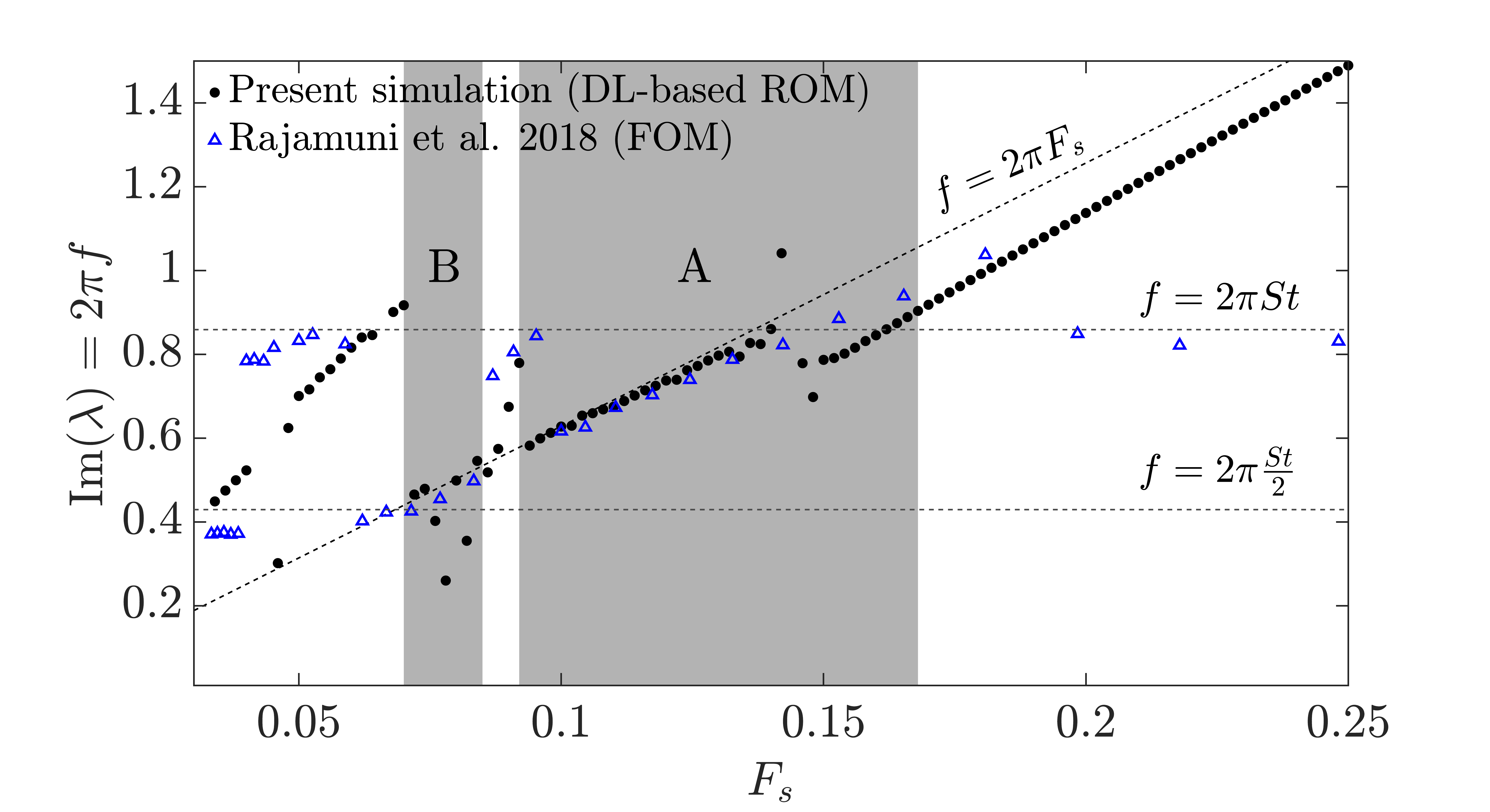}
	\end{subfigure}
	\caption{Eigenspectrum of the DL-based ROM at $(Re,m^*)=(300,2.865)$. The lock-in branches are shaded with a grey color corresponding to resonance-induced regions (branch A and B). }
	\label{Re_low_MS2}
\end{figure}

Figure \ref{Re_low_MS2} shows the selected eigenvalue trajectory corresponding to $A_y$ (SM) as a function of the reduced natural frequency ($F_s$) with $0.03<F_s<0.25$ and the increment is $\Delta F_s=0.002$. 
As elucidated in figure \ref{Re_low_MS2} (top), the SM becomes unstable for $0.0.092<F_s<0.17$, which is determined by the real part of the eigenvalues. 
To identify the resonance-based regimes, we need to consider the closeness of the SM and the natural frequency of the body. 
As shown in figure \ref{Re_low_MS2} (bottom), the imaginary part of the eigenvalue as a function of $F_s$ reveals the closeness of the SM and the natural frequency of the sphere in a vacuum
for $0.092<F_s<0.17$, which is recognized as the resonance mode (branch A). In this branch, we can also observe a change in the behaviour of the SM frequency as it intersects the $\mathrm{WM}=2\pi St$ (i.e., the vortex shedding frequency behind a stationary sphere).
Another interesting finding is that the lower-left boundary of the resonance mode can be pinpointed from the ROM at $F_s\approx0.092$. Previous studies using purely linear ERA-based approach for the stability predictions (\cite{chizfahm2021data}, \cite{yao2017model} and \cite{bukka2020stability}) were unable to predict the lower-left boundary from the ROM.

\begin{figure}
	
	\begin{subfigure}{1\textwidth}
		\centering
		\includegraphics[trim={0 4cm 0 0},clip,scale=0.25]{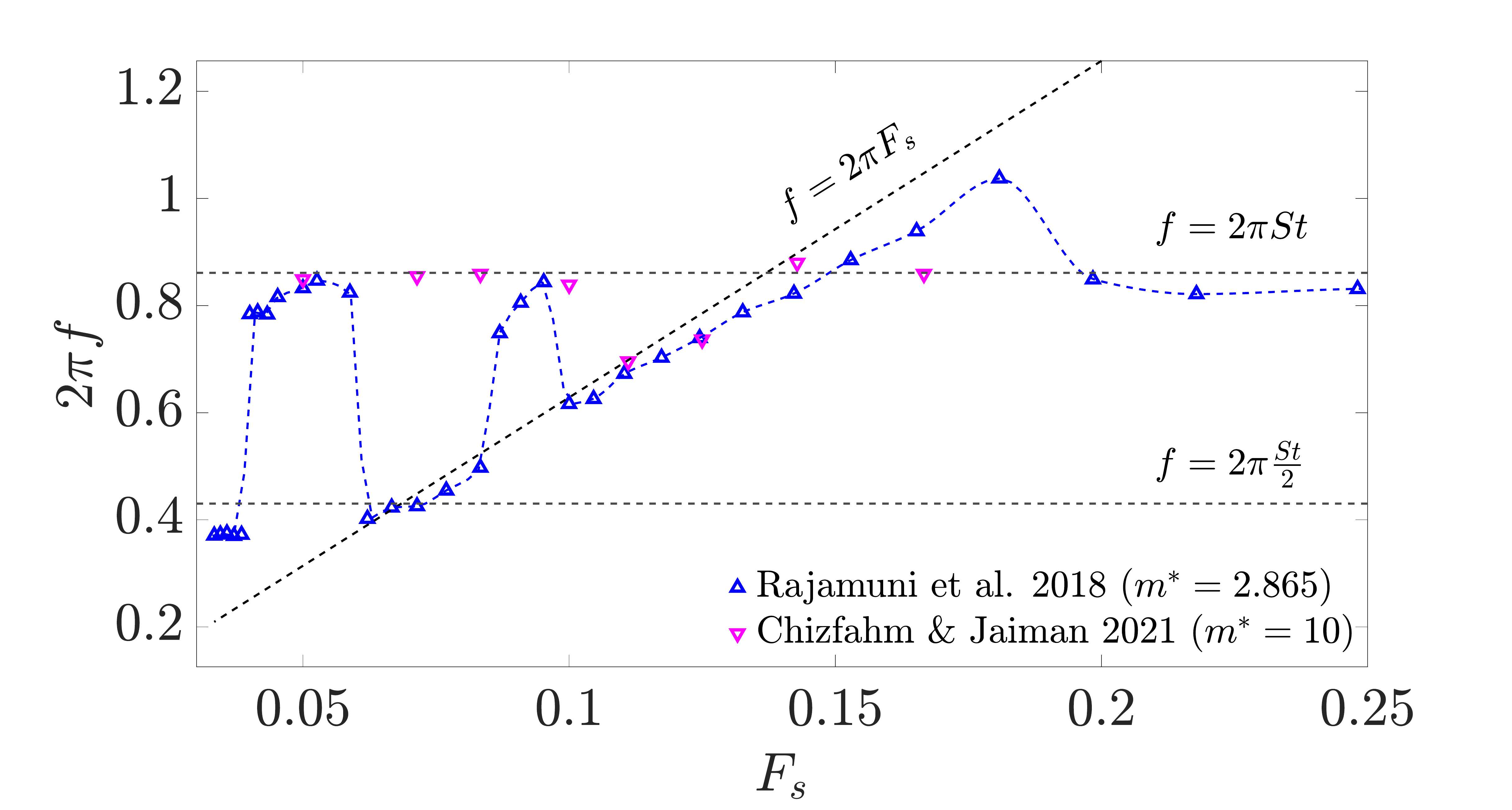}
	\end{subfigure}
	\begin{subfigure}{1\textwidth}
		\centering
		\includegraphics[trim={0 0 0 0},clip,scale=0.25]{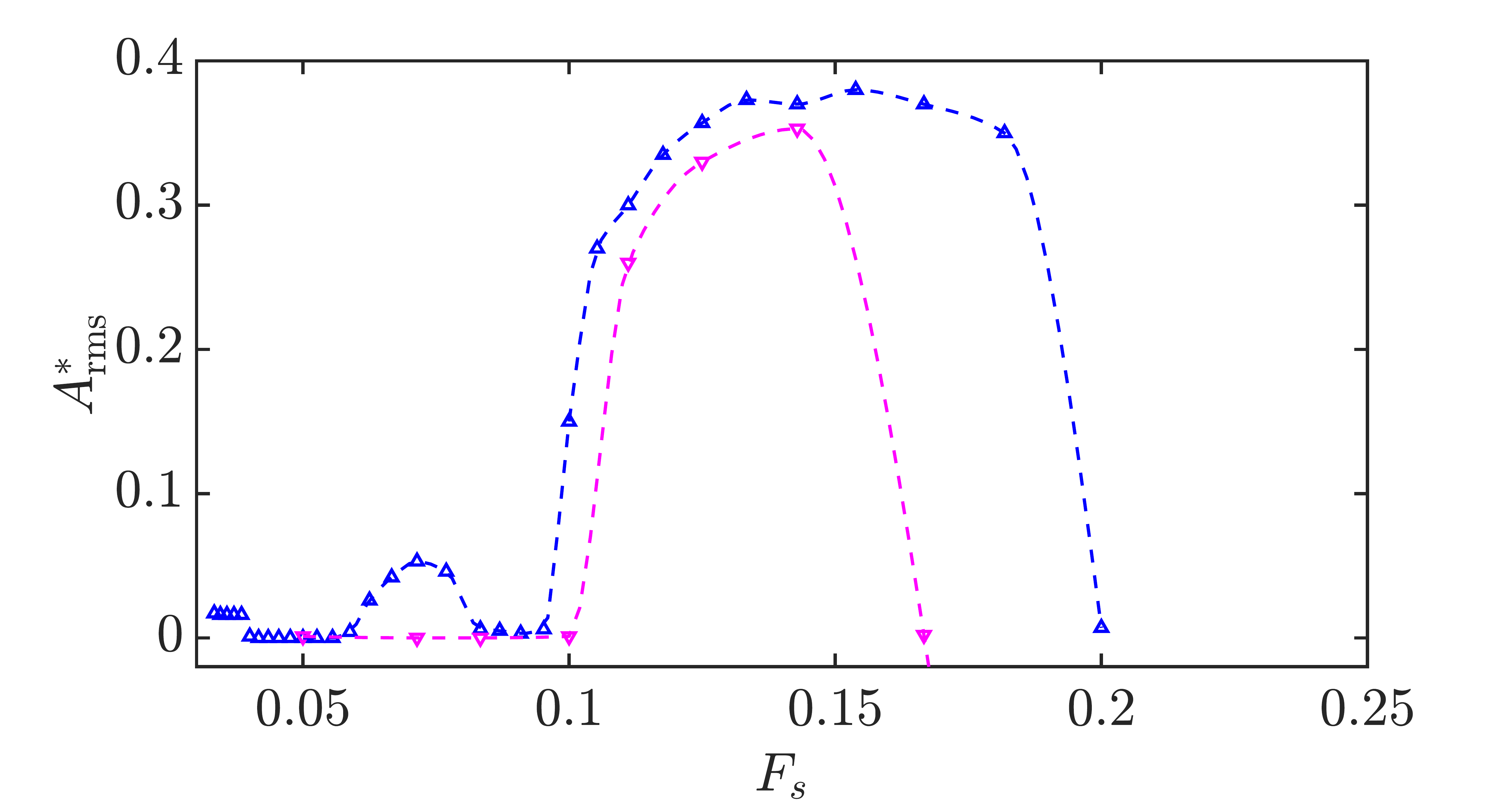}
	\end{subfigure}
	
	\caption{The FOM VIV results as a function of the reduced natural frequency $F_s$, at $Re = 300$; variation of the oscillation frequency $f$, and the r.m.s. value of the normalized amplitude ($A^*_{\mathrm{rms}}$).} 
	\label{Cross_Validate}		
\end{figure}

The eigenvalues of the SM at $0.07<F_s<0.085$, show another unstable region which is termed as branch B. We can observe that the SM frequencies are detached from the natural frequency for $0.085<F_s<0.092$ and bounce back to the intersection of the natural frequency and half of the shedding frequency ($St/2$) at $F_s\approx0.085$. The instability of this branch is recognized based on the real part of the SM trajectory, and through the imaginary part, we can confirm the closeness of the SM frequency to the natural frequency of the body as another resonance branch at lower frequencies.
%
%
To further verify the stability results predicted by the DL-based ROM integrated with ERA, the VIV response is computed by direct numerical simulation using the FOM.
Figure \ref{Cross_Validate} suggests that the VIV region with high amplitude response starts at $F_s\approx0.18$ (lock-in onset $U^*=5.5$) and ends at $F_s\approx0.095$ ($U^*=10.5$), which compares almost accurate with the resonance-induced lock-in predicted by the present DL-based ROM. In figure \ref{Cross_Validate} (bottom) the FOM simulations show small amplitude oscillations at branch B close to $St/2$ consistent with the numerical study by \cite{rajamuni2018transverse}.

\begin{figure}
	
	\begin{subfigure}{0.4\textwidth}
		\centering
		\includegraphics[trim={2mm 4.5cm 8cm 0},clip,scale=0.25]{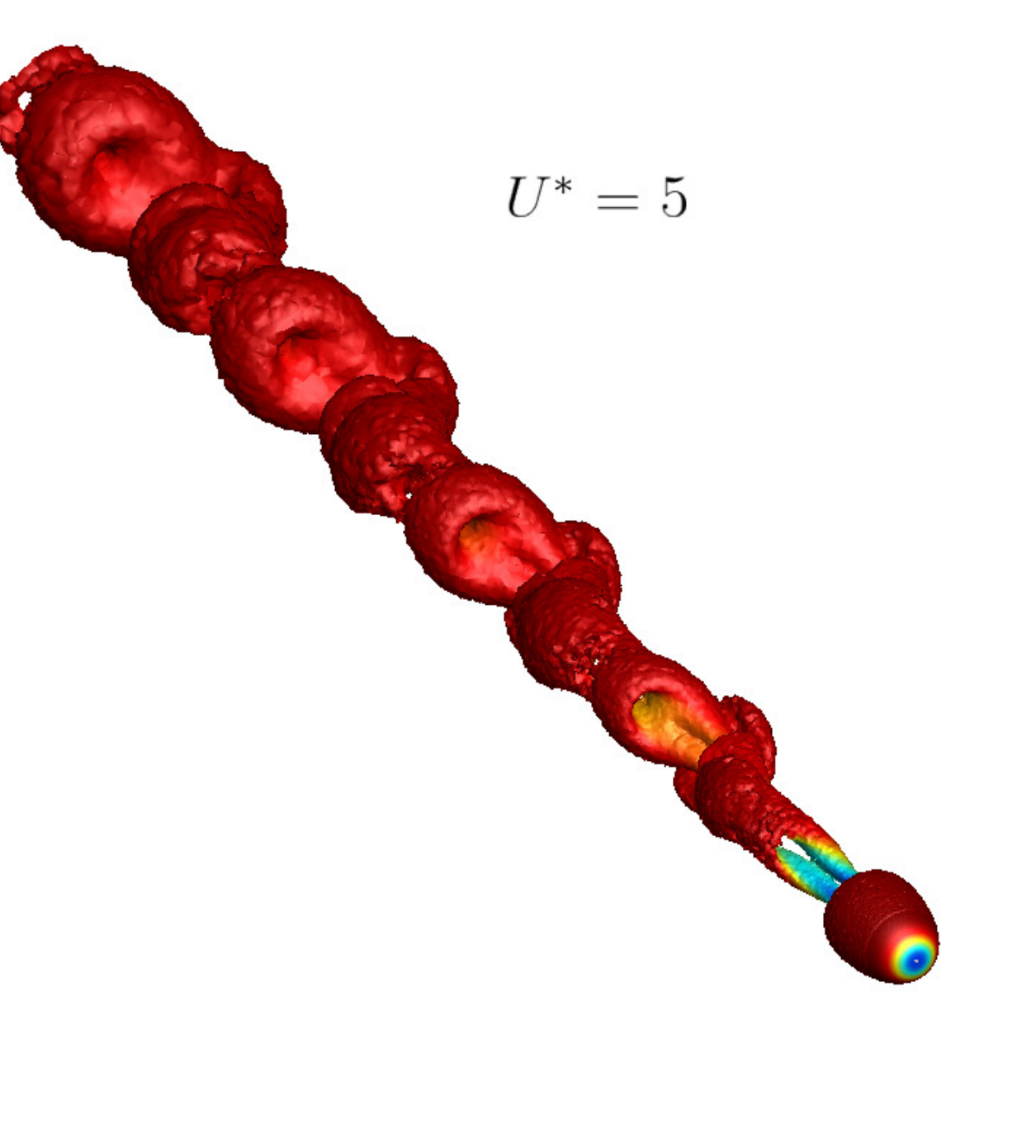}
	\end{subfigure}%
	\begin{subfigure}{0.6\textwidth}
		\centering
		\includegraphics[trim={0.5cm 4.3cm 1mm 5cm},clip,scale=0.35]{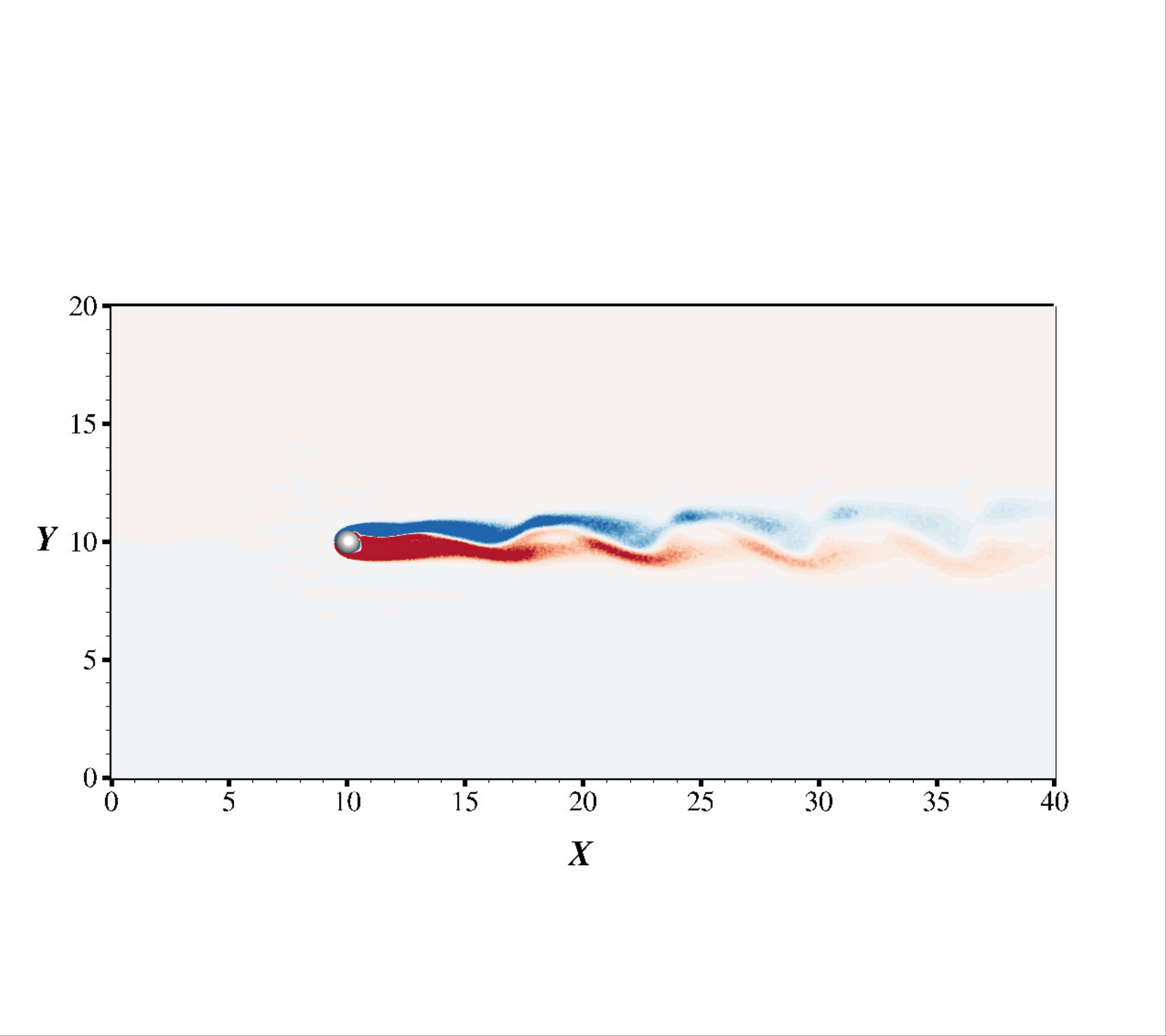}
	\end{subfigure}
	
	\begin{subfigure}{0.4\textwidth}
		\centering
		\includegraphics[trim={2mm 4.5cm 8cm 0},clip,scale=0.25]{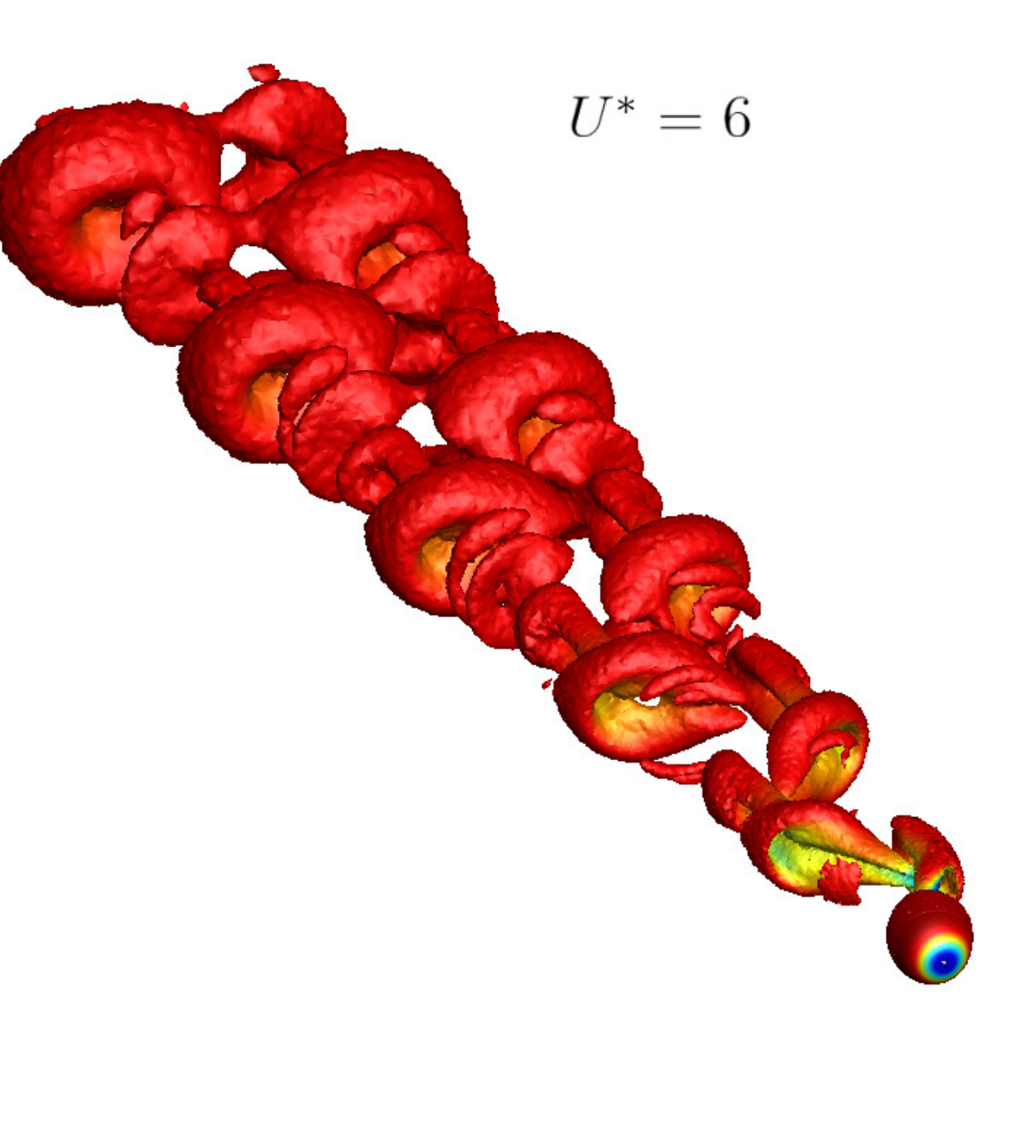}
	\end{subfigure}%
	\begin{subfigure}{0.6\textwidth}
		\centering
		\includegraphics[trim={0.5cm 4.3cm 1mm 5cm},clip,scale=0.35]{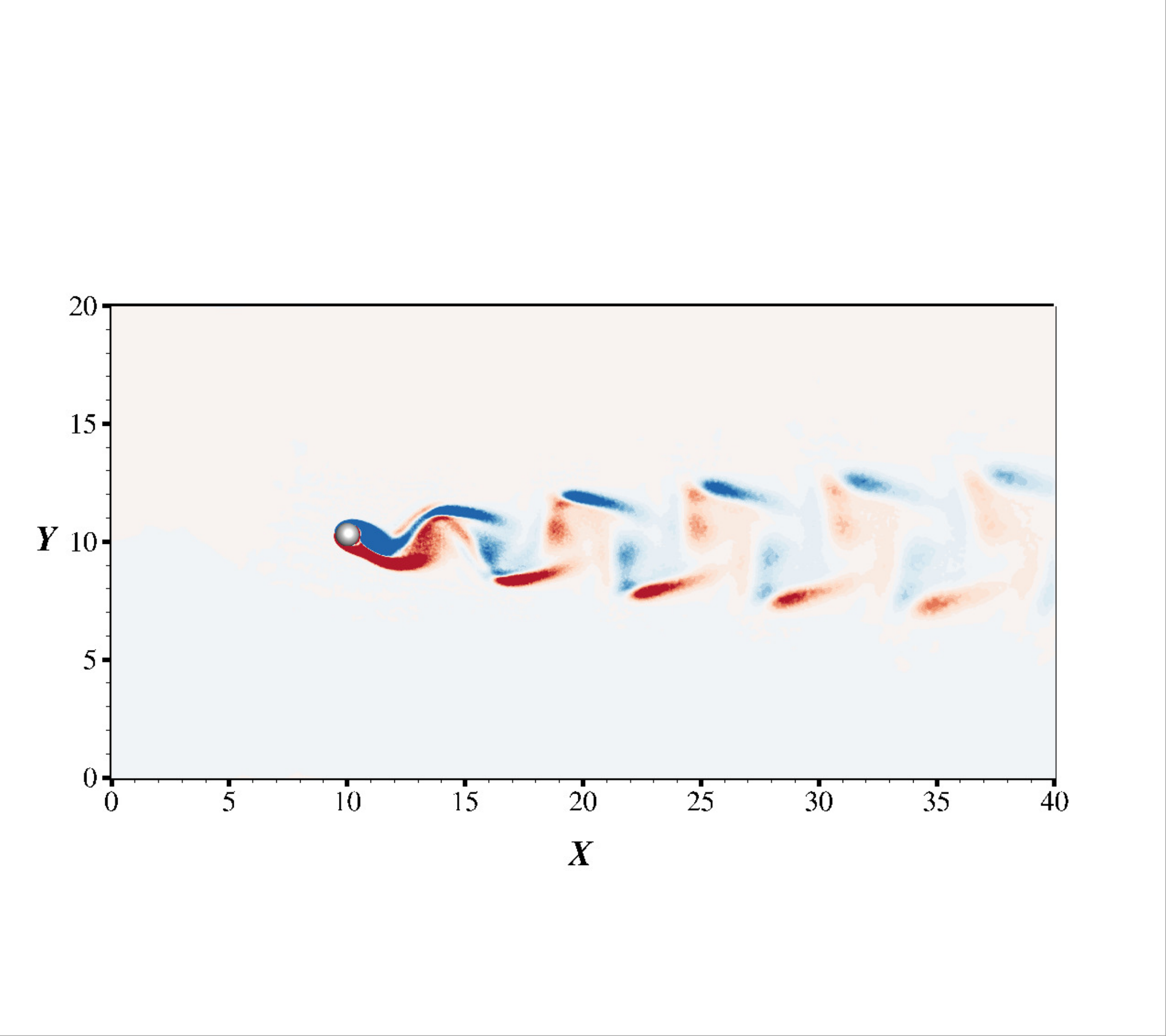}
	\end{subfigure}

	\begin{subfigure}{0.4\textwidth}
		\centering
		\includegraphics[trim={2mm 4.5cm 8cm 0},clip,scale=0.25]{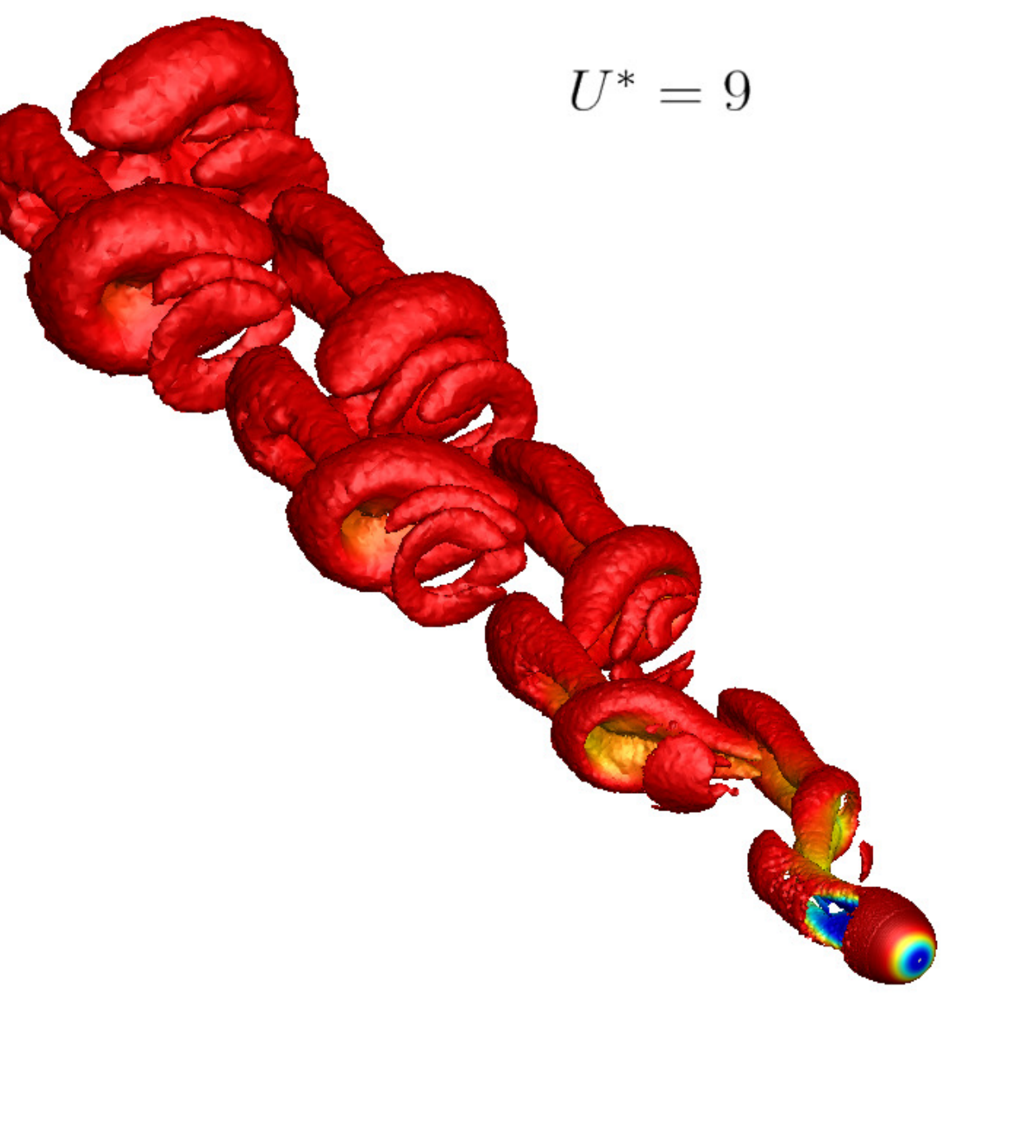}
	\end{subfigure}%
	\begin{subfigure}{0.6\textwidth}
		\centering
		\includegraphics[trim={0.5cm 4.3cm 1mm 5cm},clip,scale=0.35]{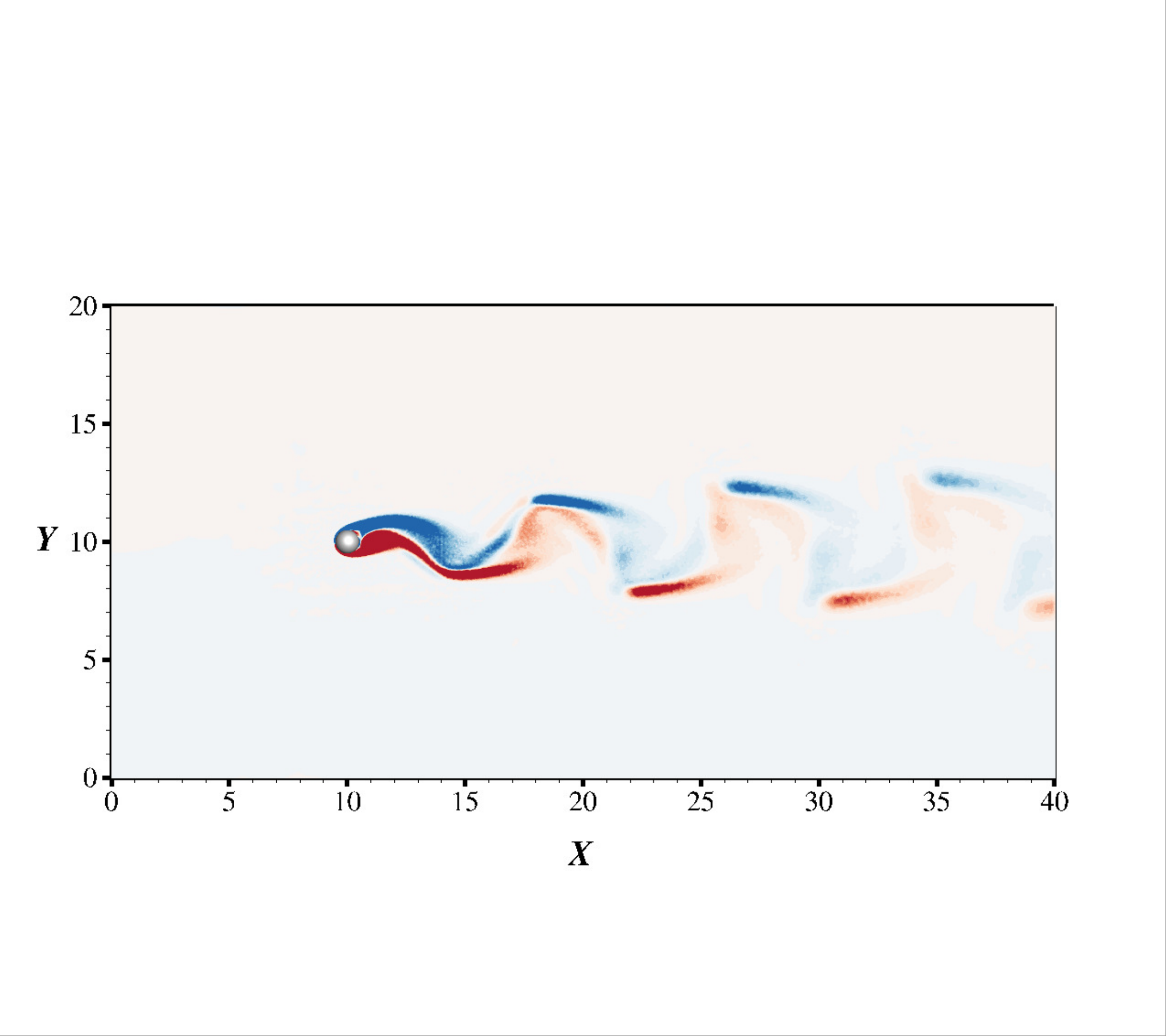}
	\end{subfigure}

	\begin{subfigure}{0.4\textwidth}
		\centering
		\includegraphics[trim={2mm 4.5cm 8cm 0},clip,scale=0.25]{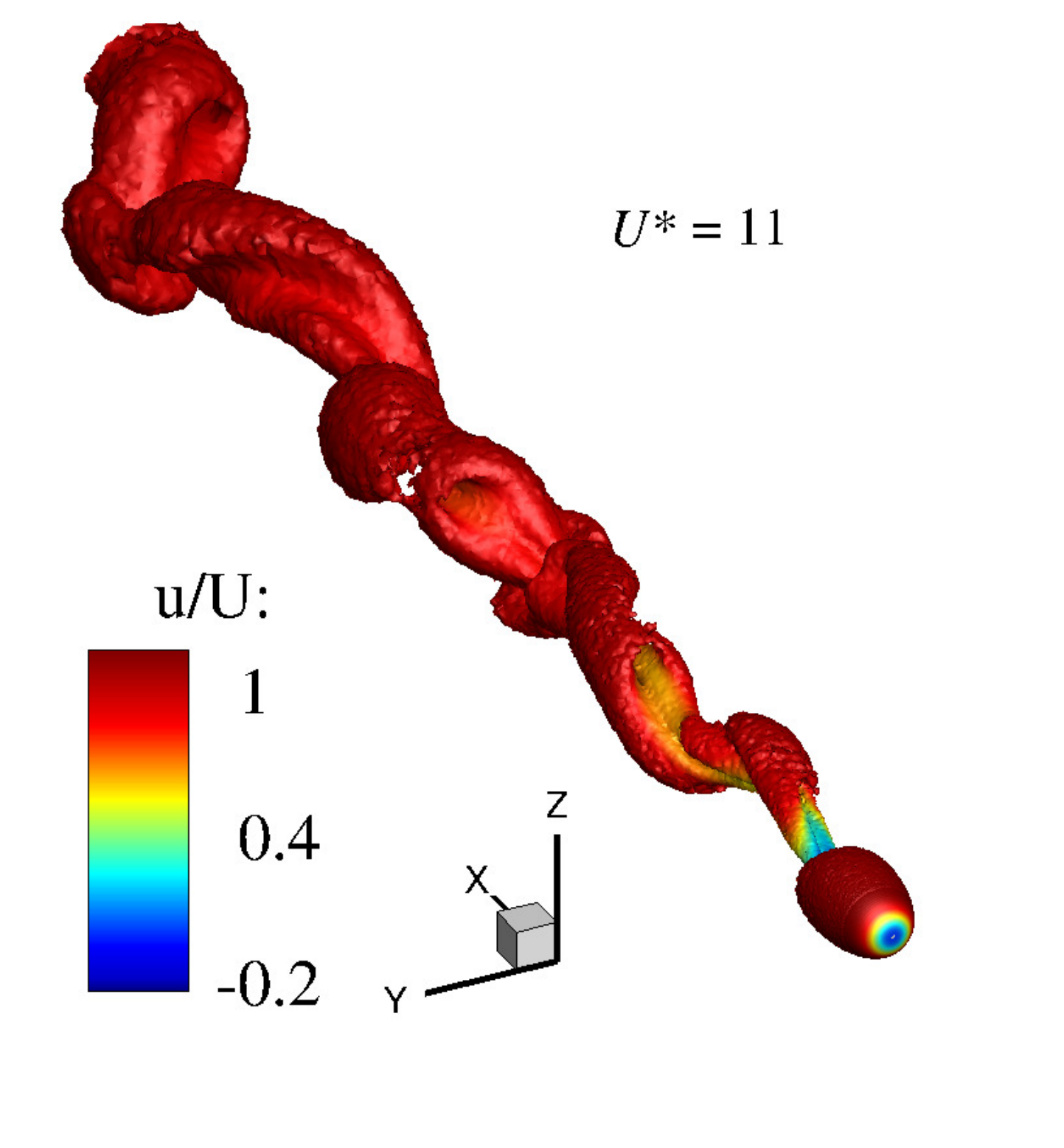}
	\end{subfigure}%
	\begin{subfigure}{0.6\textwidth}
		\centering
		\includegraphics[trim={0.5cm 3cm 1mm 5cm},clip,scale=0.35]{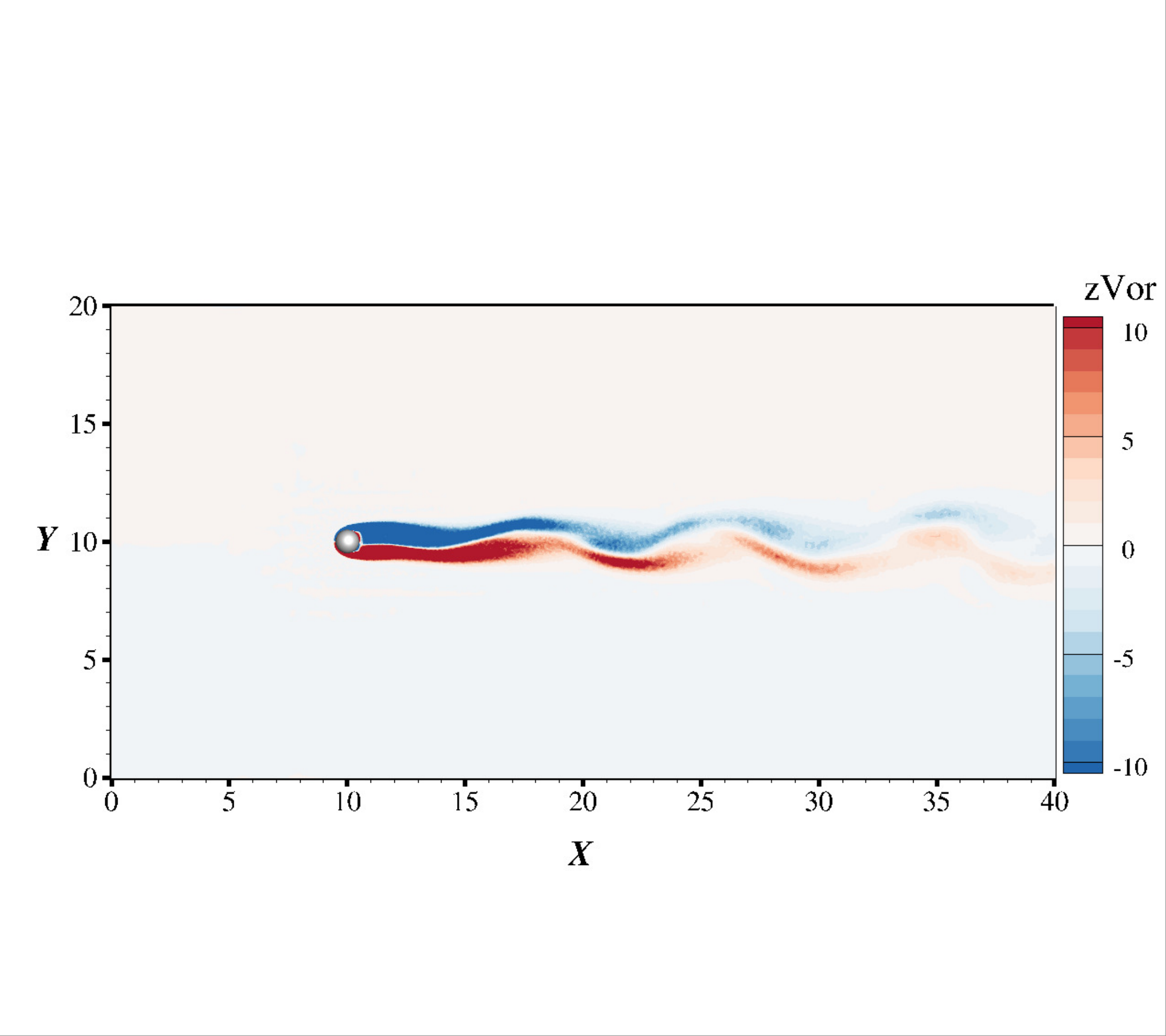}
	\end{subfigure}
	
	\caption{(a) Iso-surface of three-dimensional wake structures formed behind an elastically mounted sphere at stationary state. Iso-surfaces are plotted by the Q-criterion ($\bar{Q}=Q\frac{D^2}{U^2}=0.001$, (b) $z$-vorticity contours from the FOM  $(Re,m^*)=(300,2.865)$.} 
	\label{Q&Z}		
\end{figure}

These observations are further confirmed by our FOM calculations. The iso-surface of non-dimensional $Q$-criterion and vorticity contours from the FOM calculations for several reduced velocities are displayed in figure \ref{Q&Z} at $(Re,m^*)=(300,2.865)$. We observe the VIV instability at $U^* = 6$ which corresponds approximately close to the higher-right boundary of the branch A. This VIV instability corresponds to branch (A) sustains for $U^*=9$ as the flow becomes unsteady with completely detached hairpin-type vortices that periodically shed with a fixed plane of symmetry along the transverse motion direction ($y$-axis). At higher reduced velocity $U^*=11$, the wake lose the symmetry completely, which shows the termination of the lock-in region corresponding to branch A. The asymmetry of the wake leads to the reduction of the transverse force applied on the bluff body and consequently low amplitude response. 

\begin{figure}
	\centering
	\begin{subfigure}{1\textwidth}
		\centering
		\includegraphics[trim={0 5cm 0 0},clip,scale=0.2]{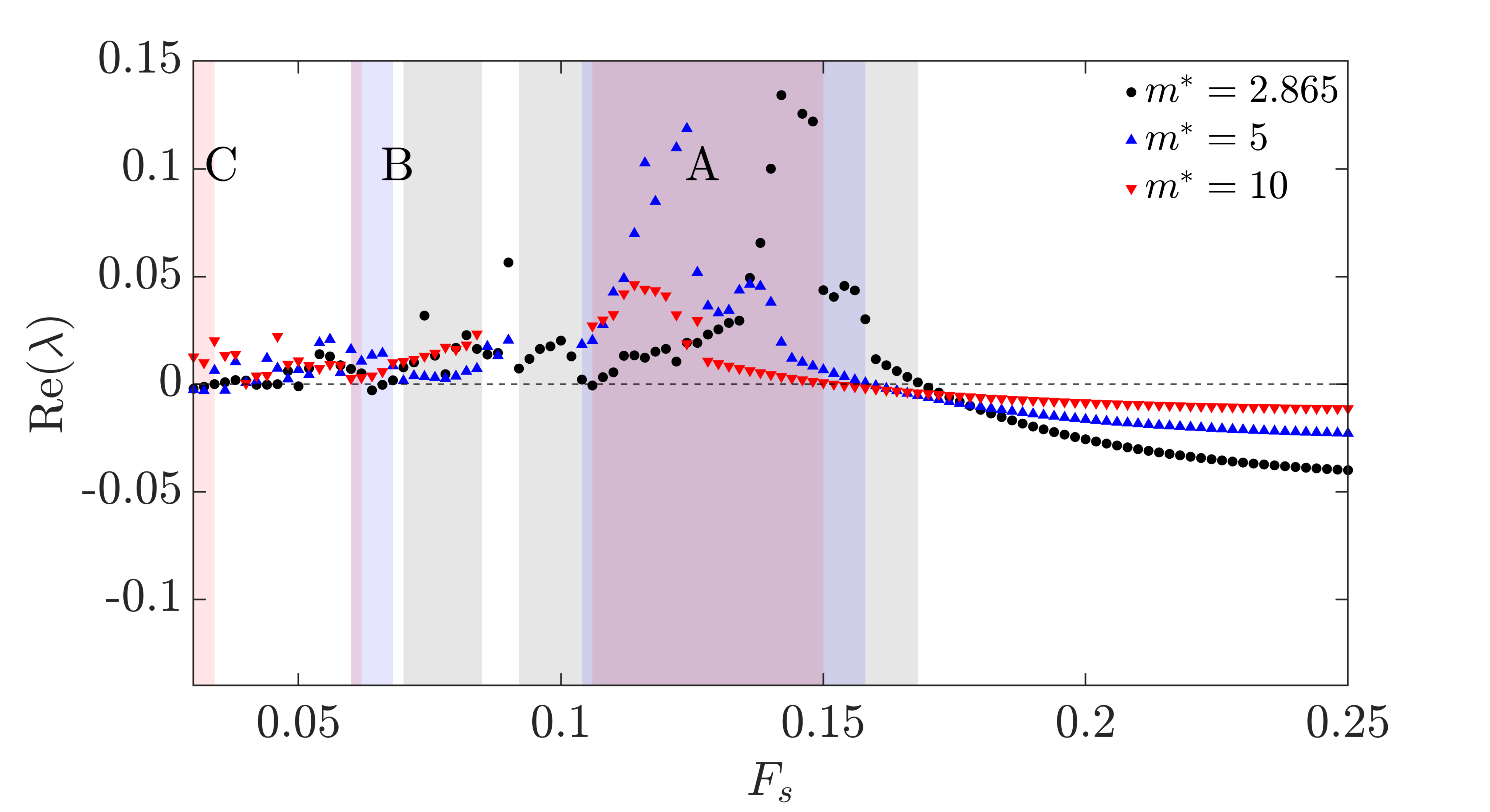}
	\end{subfigure}
	\begin{subfigure}{1\textwidth}
		\centering
		\includegraphics[trim={0 0 0 0},clip,scale=0.2]{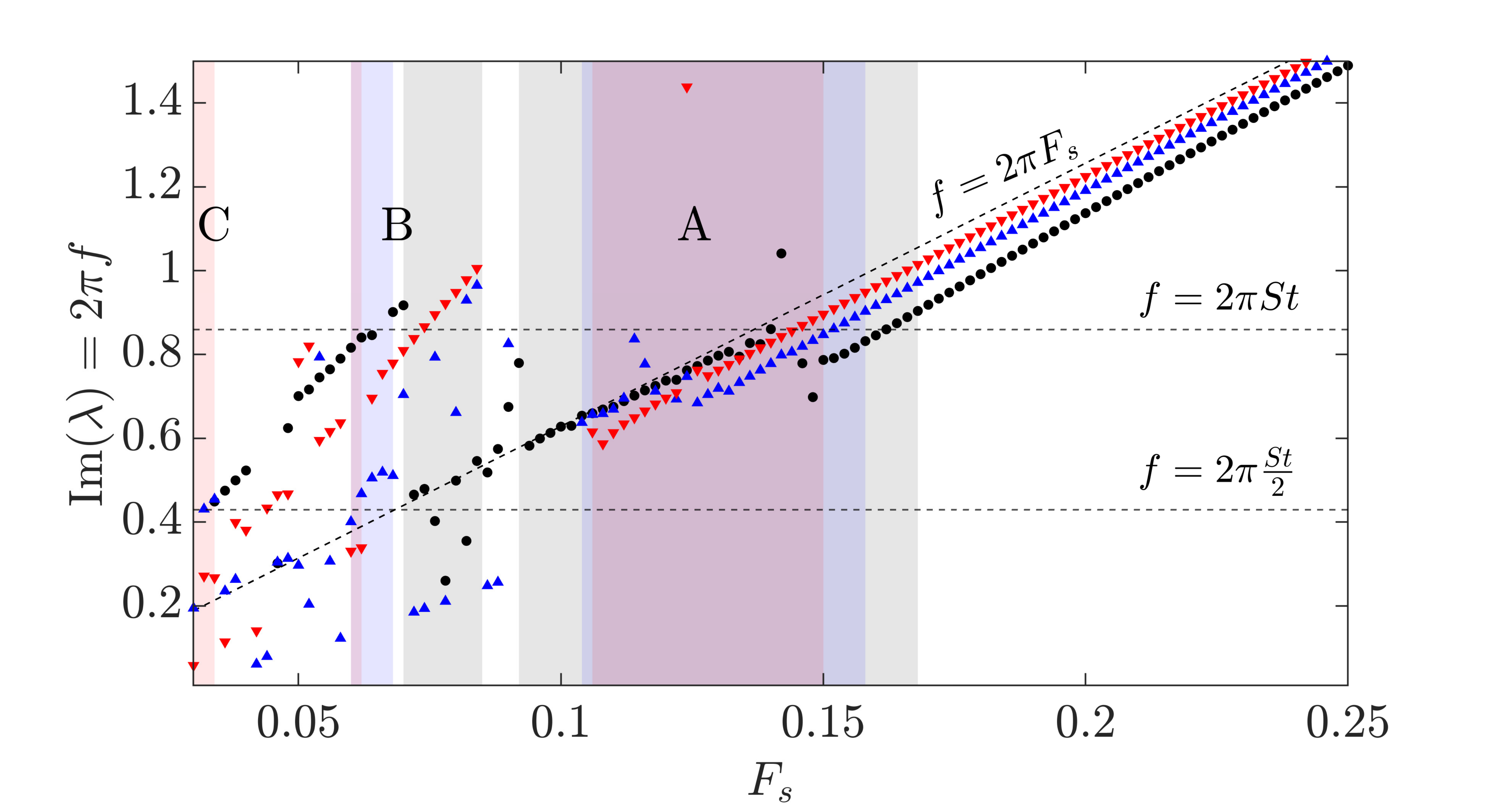}
	\end{subfigure}
	\caption{Eigenspectrum of the DL-based ROM at $Re=300$ and $m^*=2.865, 5$, $10$. The resonance-induced lock-in branches are shaded with grey, blue, and red colors corresponding to $m^*=2.865, 5$, and $10$ respectively. }
	\label{Re_low}
\end{figure}

The effectiveness of the DL-based ROM for the mass ratio effect on the VIV response is shown in figure \ref{Re_low} for ($m^*=2.865, 5, 10$) at $Re=300$. Figure \ref{Re_low} shows the real and imaginary parts of the root loci as a function of the reduced frequency $F_s$. We observe that, as the mass ratio $m^*$ increases, the lock-in onset, which is the right-higher VIV boundary corresponding to branch A, shifts to the lower reduced natural frequencies ($F_s$). In addition, the lower-left boundary of branch A shifts to higher reduced natural frequencies ($F_s$) by increasing the mass ratio $m^*$. We observe that the lock-in boundaries corresponding to branch B become narrower as the mass ratio increases and shift to the lower reduced frequency. Another unstable resonance-induced lock-in region (branch C) is identified for a higher mass ratio ($m^*=10$) at lower reduced frequency $F_s$. The accuracy of the results for higher mass ratio $m^*=10$ can be verified further through the FOM simulations in figure \ref{Cross_Validate}.

\subsection{Stability Analysis for VIV of a Sphere at $Re=2\,000$}
In this section, the performance of the DL-based ROM for stability predictions of a transversely vibrating sphere at $Re=2\,000$ is presented. The studies on the stability analysis using pure linear-type projections (\cite{yao2017model}, \cite{bukka2020stability}, \cite{chizfahm2021data}) can only be performed close to the critical Reynolds number ($Re_{cr}$) which provides a low-order representation of the unsteady flow dynamics in the neighborhood of the equilibrium steady state. However, at higher Reynolds number cases where the wake is turbulent, more complexity is added to the problem setup and linear ERA-based ROM cannot be reliable (i.e., eigenvalues do not decay exponentially).
In addition, for 3D turbulent flows at a high Reynolds number, a brute-force time integration may not be easy to converge towards the stable flow state (base flow) that is required for the stability analysis via the ERA approach. Therefore, other nonlinear-type system identification techniques are required to overcome these complexities.

\begin{figure}
	\centering
	
	\begin{subfigure}{1\textwidth}
		\centering
		\includegraphics[trim={0.5cm 0 0 0},clip,scale=0.25]{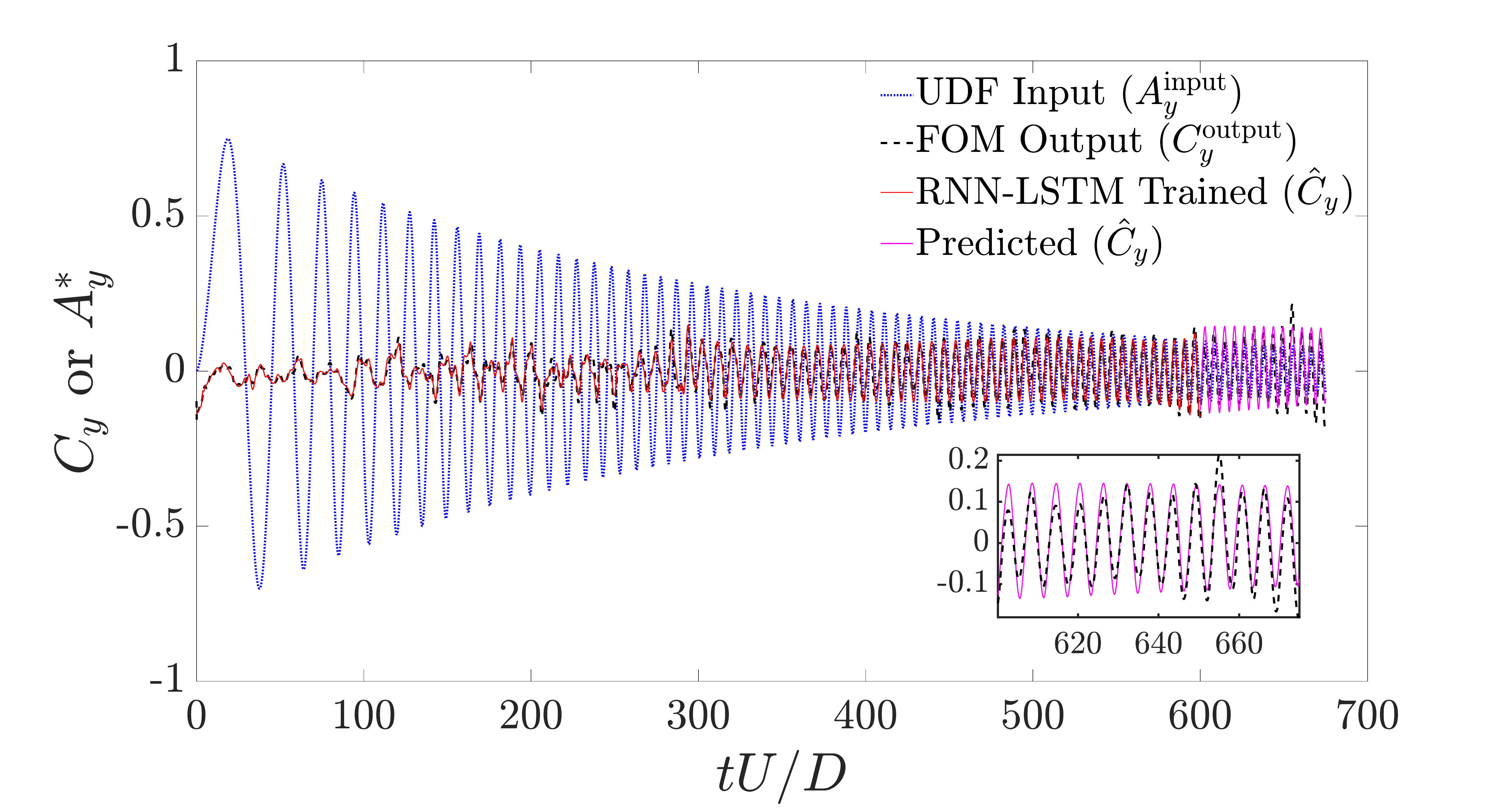}
		\caption{}
	\end{subfigure}
	\begin{subfigure}{1\textwidth}
		\centering
		\includegraphics[trim={0.5cm 0 0 0},clip,scale=0.25]{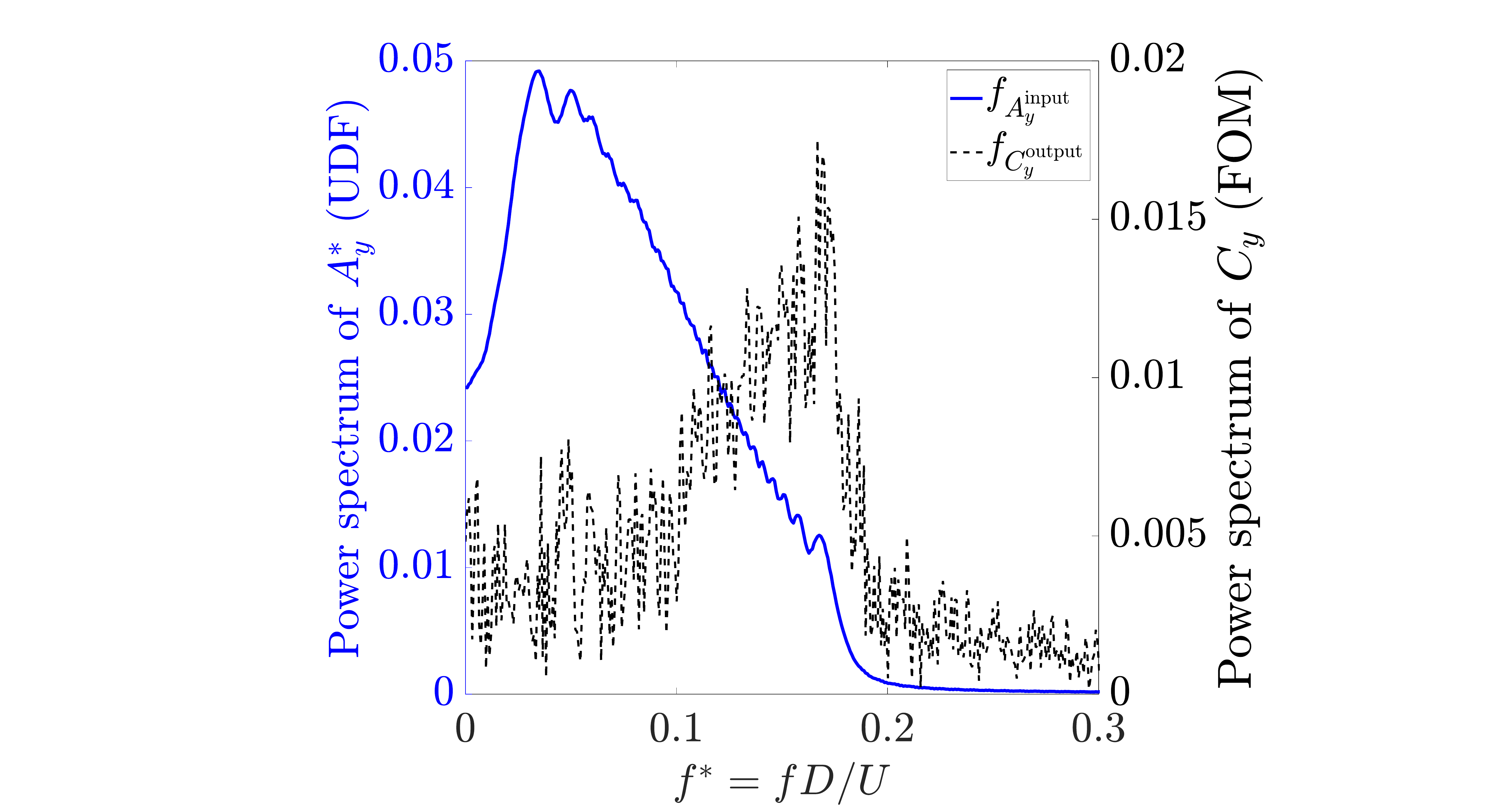}
		\caption{}
	\end{subfigure}~
	\caption{The unstable wake behind a forced displaced sphere at $Re = 2000$: (a) Time history of the normalized transverse force ($C_y$) from the FOM and the temporal prediction from the DL-based ROM LSTM network, (b) the corresponding frequency spectrum of the forced displacement of the sphere and the corresponding normalized transverse force ($C_y$) from FOM. }
	\label{Re_high_Train}
\end{figure}

For this purpose, we utilize the proposed methodology of the DL-based ROM integrated with ERA (figure \ref{Schematic_Process}), to examine the stability properties through the eigenspectrum plots for a canonical sphere at $Re=2\,000$.
A user-defined arbitrary forced transverse displacement $A_y^{\mathrm{input}}$ as a training data set is given as an input to the system which involves a range of non-dimensional frequencies $f^*\in[0.02-0.2]$ and the normalized transverse force $C_y$ is recorded for every time step $\Delta t = 0.15$.
Here we collect a set of $4\,000$ responses resulting in a total simulation time of $tU/D=600$. Figure \ref{Re_high_Train} (a) shows the output normalized force signal $C_y^{\mathrm{output}}$ calculated from the FOM at $Re=2\,000$. The FOM data are used to fit a model, and the validity of the model is checked by how well it reproduces the validation data as shown in figure \ref{Re_high_Train} (a). The simulation shows a good performance with the (Train Fit = 73, Test Fit = 62). The stability prediction is assessed via DL-based ROM integrated with ERA. The process of constructing the linearized approximation through ERA is presented in detail in Sections \ref{ERA} and \ref{Eigenvalue Selection}.

\begin{figure}
	\centering
	\begin{subfigure}{1\textwidth}
		\centering
		\includegraphics[trim={0 5cm 0 0},clip,scale=0.2]{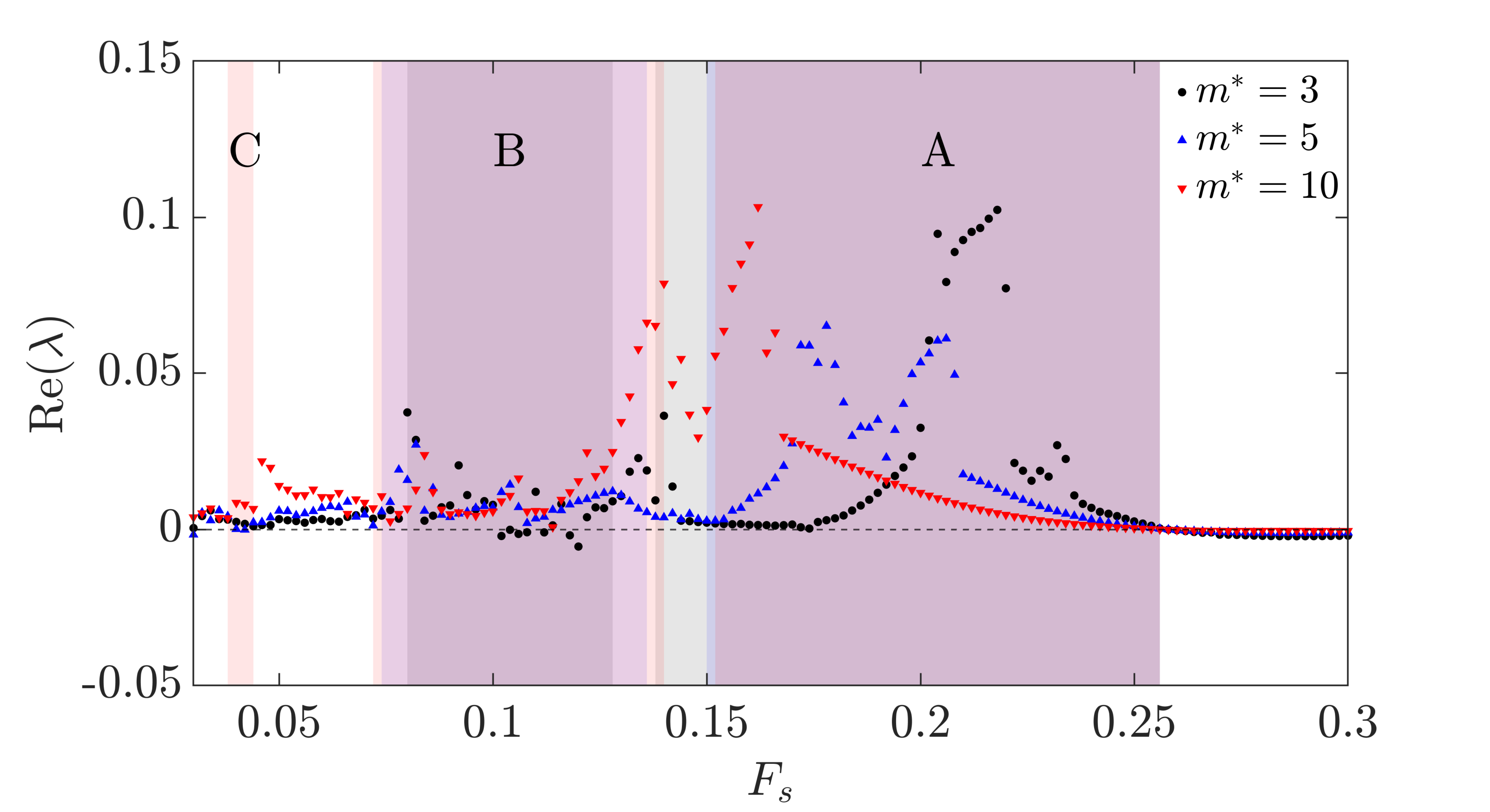}
	\end{subfigure}
	\begin{subfigure}{1\textwidth}
		\centering
		\includegraphics[trim={0 0 0 0},clip,scale=0.2]{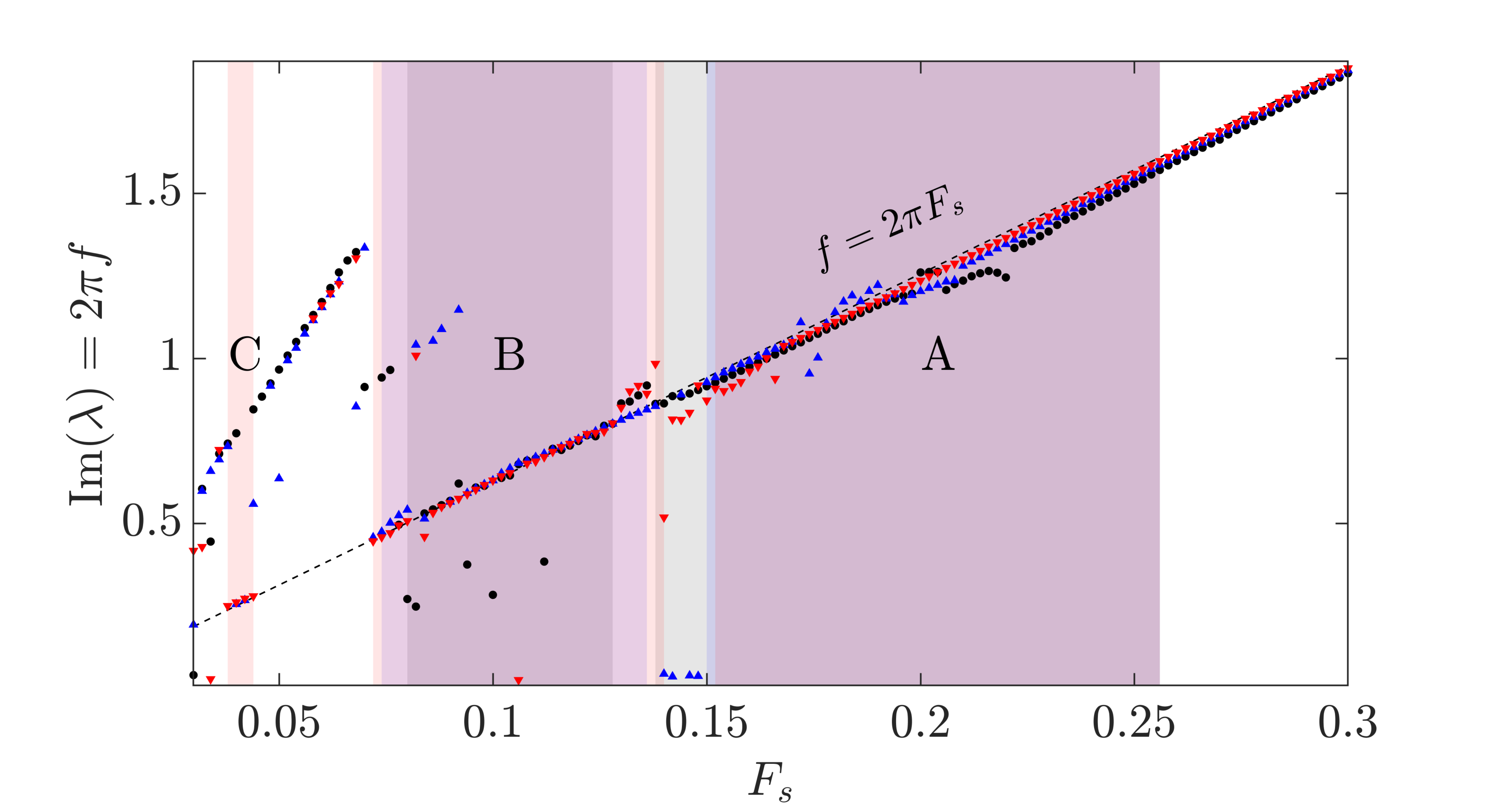}
	\end{subfigure}
	\caption{Eigenspectrum of the DL-based ROM at $Re=2000$ and $m^*=(3, 5, 10)$. }
	\label{RE_high}
\end{figure}

Figure \ref{RE_high} shows the selected eigenvalue trajectory corresponding to SM as a function of the reduced natural frequency $(0.03<F_s<0.3, \Delta F_s=0.002)$ for $m^*=(3, 5, 10)$.
By considering the imaginary part of the eigenvalues ($\mathrm{Im}(\lambda)$), we can identify resonance-induced regions (branch A and branch B) through closeness and detaching of the SM with the natural frequency of the structure in a vacuum. The instability of these branches is determined by the real part of the eigenvalues. 
These results are comparable with FOM numerical simulations provided by \cite{rajamuni2019vortex}.
The proficiency of the DL-based ROM to explore the mass ratio effects on the VIV response indicates that the lock-in region corresponding to branch A tends to become narrower as the mass ratio $m^*$ increases. In addition, the lower-left boundary of the resonance region branch A shifts to higher reduced natural frequencies ($F_s$) by increasing the mass ratio $m^*$. On the other hand, we observe that the unstable regions corresponding to branch B become wider as the mass ratio increases. Another unstable resonance-induced region (branch C) is identified for a higher mass ratio ($m^*=10$) at lower reduced natural frequencies. 

The present DL-based ROM integrated with ERA methodology has been concerned with fully submerged three-dimensional bluff-body configurations for which all three directions in space are resolved. All of the notions of the ROM can be easily extended to more complicated settings such as realistic geometries, multiphase flows, and free surface effects. Thus, the present method does not pose any theoretical limitation except that there may be numerical ones with respect to memory requirements and CPU time to obtain the FOM-trained dataset and to train the generalized neural network.

\section{Conclusions} 
\label{Conclu}
In this paper, we introduced a new deep learning-based reduced-order model for nonlinear system identification and stability predictions of 3D fluid-structure systems. The proposed DL-ROM relies on the LSTM recurrent neural network and the eigenvalue realization algorithm.
We presented a new training strategy to construct an input-output relationship for a reduced-order approximation of a fluid-structure system using full-order simulations as a temporal series of force and displacement measurements. The input function provided a range of frequencies and amplitudes based on prior knowledge of the VIV lock-in process, allowing for efficient DL-ROM without the requirement for a large training dataset for low-dimensional modeling.
We developed a methodology to integrate ERA as a linear encoder-decoder-like framework with the DL-based ROM to extract the coupled fluid-structure dynamics modes. We provided a new eigenvalue selection process to investigate the underlying mechanism and stability characteristics of VIV. Systematic stability analysis of sphere VIV has been numerically presented at $Re=300$ and $Re=2\,000$. The effectiveness of the DL-based ROM has been remarkably demonstrated for predicting the resonance lock-in and self-sustained VIV. For $Re=300$, multiple unstable regions were identified that showed strong coupling of the structural mode with shedding frequency and other low-frequency dominant wake modes resulting in different resonance-induced regimes. These coupled modes were categorized into three lock-in branches, namely A, B, and C, and cross-validated successfully with the FOM and the available literature. The effects of the mass ratio have been examined through DL-based ROM for the VIV lock-in analysis.
For $Re=2\,000$ corresponding to turbulent flow, we were able to identify unstable regions due to the strong coupling of the structural mode with complex dominant wake modes, and the results are cross-validated successfully with the available literature. We examined the effect of the mass ratio using DL-based ROM for the VIV lock-in analysis.
The simplicity and computational efficiency of the DL-based ROM allow investigation of the VIV mechanism for high Reynolds numbers and more complex problem setups and a variety of geometries and parameters and pave the way for the development of control devices.
For complex turbulent flows and fluid-structure interactions, it is worth exploring nonlinear autoencoders instead of ERA in the proposed DL-based ROM architecture.


\section*{Acknowledgement}
The authors would like to acknowledge the Natural Sciences and Engineering Research Council of Canada (NSERC) for the funding. This research was supported in part through computational resources and services provided by Advanced Research Computing at the University of British Columbia.

\appendix
\setcounter{equation}{0} 
\setcounter{figure}{0}
\renewcommand{\theequation}{A.\arabic{equation}}
\renewcommand{\thefigure}{A.\arabic{figure}}
\section*{Appendix.  Assessment with Silverbox Benchmark}
\label{Appendix}

\begin{figure}
	\centering
	
	\begin{subfigure}{1\textwidth}
		\centering
		\includegraphics[trim={0.5cm 0 0 0},clip,scale=0.25]{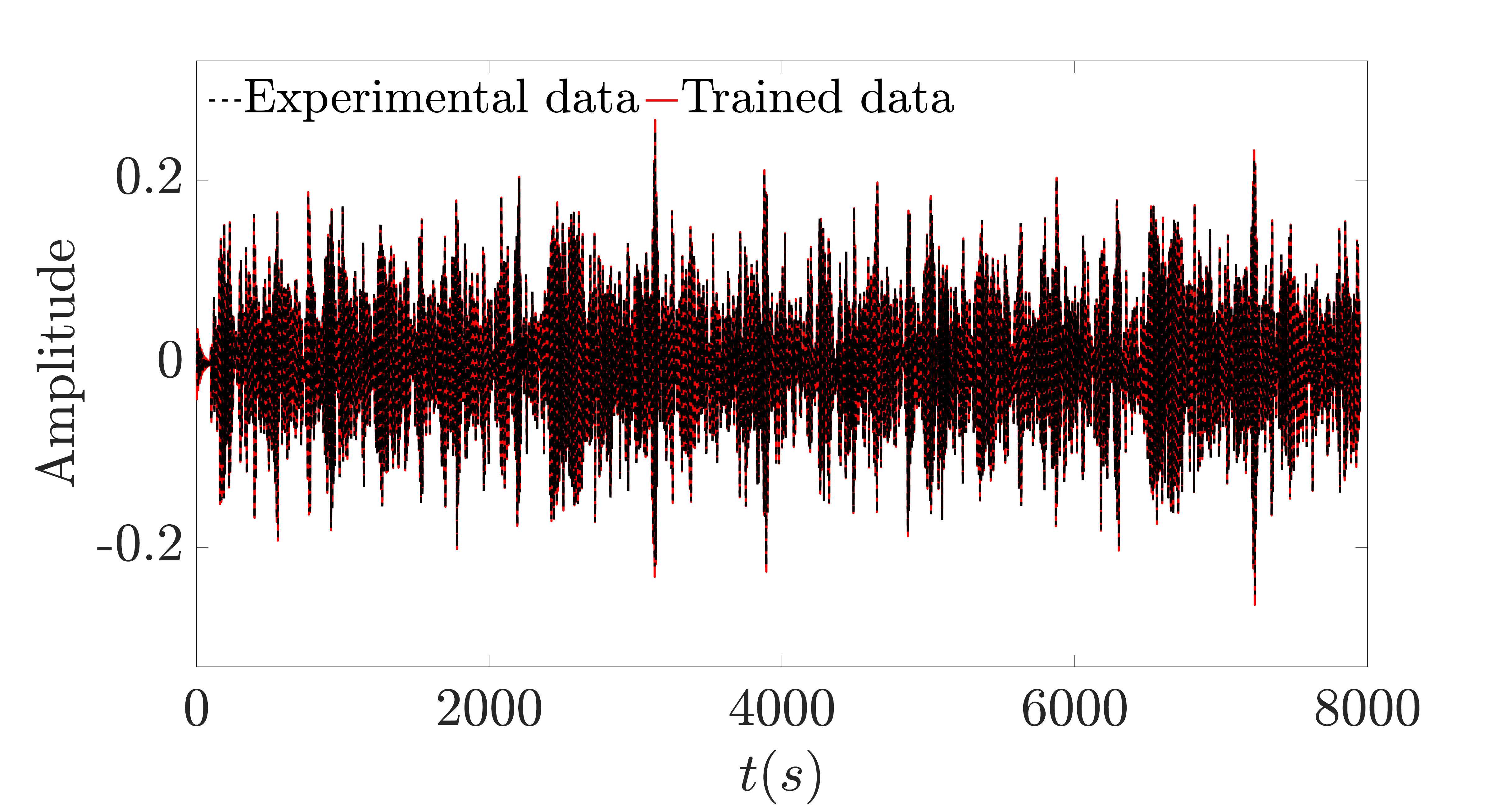}
	\end{subfigure}
	\begin{subfigure}{1\textwidth}
		\centering
		\includegraphics[trim={0.5cm 0 0 0},clip,scale=0.25]{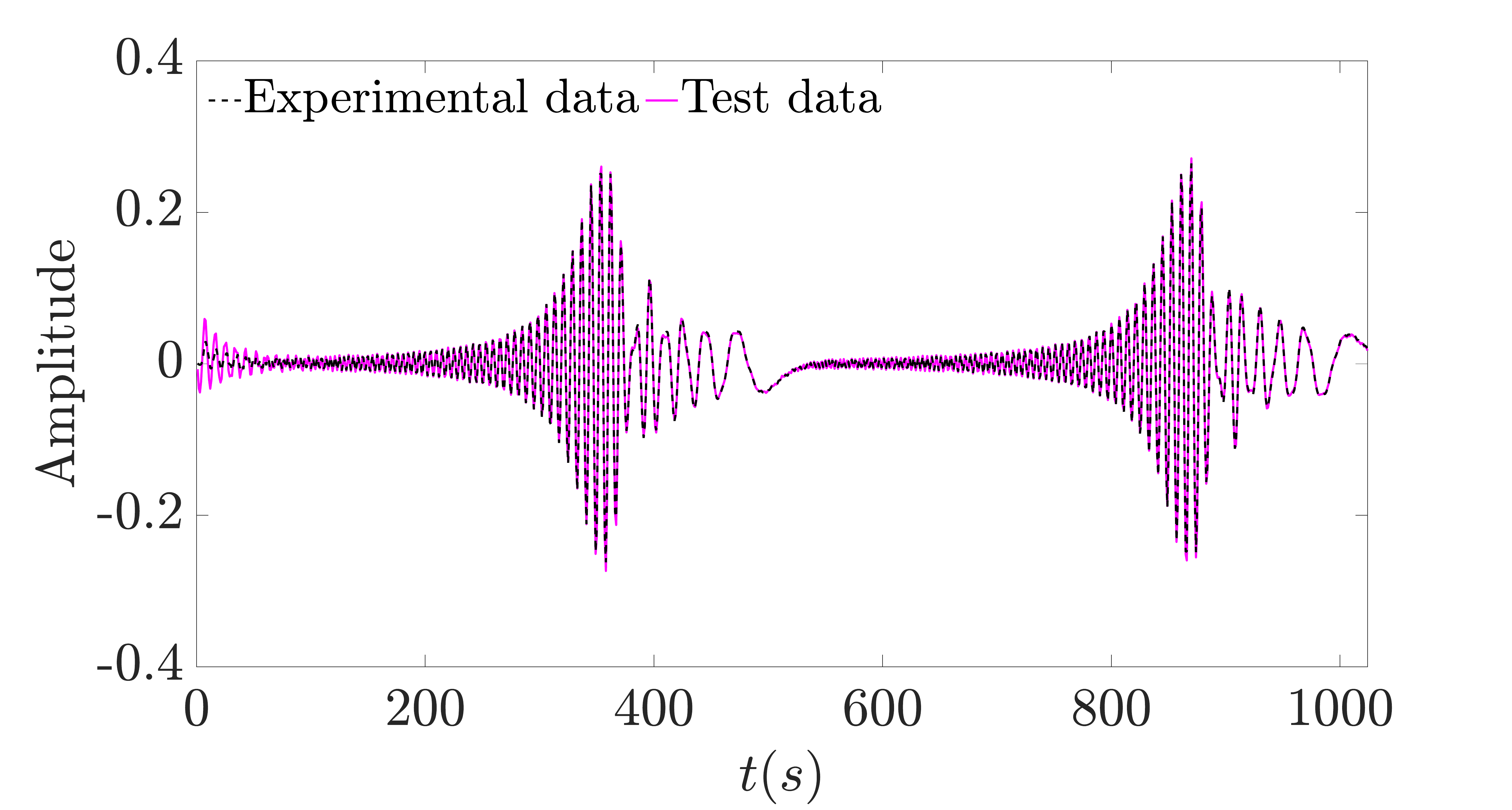}
	\end{subfigure}~
	
	\caption{Comparison of the experimental data and the LSTM model training and testing outputs. The training fit is $95.2\%$ and the testing fit is $91.6\%$}
	\label{Silver_Box}
\end{figure}

%
To demonstrate the performance and the accuracy of the LSTM network as a tool for system identification, we consider a data set of Silverbox benchmark representing a nonlinear dynamical system. The dynamical system represents a forced Duffing oscillator and includes a mechanism with a cubic hardening spring \citep{schoukens2016linear}.
Two sets of data with different responses are collected for the study. The training data set is used to construct a model, and the validity of the model is determined by how well the validation data are reproduced as a standard system identification task.
Figure \ref{Silver_Box} depicts the performance of the LSTM model while the training and testing data sets have different characteristics consistent with the simulation performed by \cite{ljung2020deep}.
The results show a good performance with the {train fit}$=95.2\%$ and the {test fit}=$91.6\%$, which corroborates the reliability of the LSTM network utilized in this study for the stability prediction of the coupled fluid-structure system.

\bibliography{JFM}
\bibliographystyle{jfm}

\end{document}